\documentclass[reprint,superscriptaddress,nofootinbib,
amsmath,amssymb,aps,prd,floatfix
]{revtex4-2}

\usepackage[utf8]{inputenc}
\usepackage{color}
\usepackage{graphicx}
\usepackage{dcolumn}
\usepackage{bm}
\usepackage{comment}

\usepackage{lineno}
\newcommand*\patchAmsMathEnvironmentForLineno[1]{%
  \expandafter\let\csname old#1\expandafter\endcsname\csname #1\endcsname
  \expandafter\let\csname oldend#1\expandafter\endcsname\csname end#1\endcsname
  \renewenvironment{#1}%
     {\linenomath\csname old#1\endcsname}%
     {\csname oldend#1\endcsname\endlinenomath}}%
\newcommand*\patchBothAmsMathEnvironmentsForLineno[1]{%
  \patchAmsMathEnvironmentForLineno{#1}%
  \patchAmsMathEnvironmentForLineno{#1*}}%
\AtBeginDocument{%
  \patchBothAmsMathEnvironmentsForLineno{equation}%
  \patchBothAmsMathEnvironmentsForLineno{align}%
  \patchBothAmsMathEnvironmentsForLineno{flalign}%
  \patchBothAmsMathEnvironmentsForLineno{alignat}%
  \patchBothAmsMathEnvironmentsForLineno{gather}%
  \patchBothAmsMathEnvironmentsForLineno{multline}%
}

\definecolor{deepgreen}{rgb}{0.2,0.8,0.2}

\definecolor{deepblue}{rgb}{0.2,0.4,0.8}

\definecolor{deepred}{rgb}{0.8,0.2,0.2}

\usepackage[colorlinks=true,linkcolor=deepblue,citecolor=deepblue,urlcolor=deepblue]{hyperref}
\usepackage{aas_macros}
\usepackage{fontawesome}
\usepackage{xspace}

\newcommand{\vect}[1]{\boldsymbol{\mathbf{#1}}}

\newcommand{\dd}{{\rm d}}

\definecolor{linkcolor}{rgb}{0.7752941176470588, 0.22078431372549023, 0.2262745098039215}
\newcommand{\nbicon}{{\color{linkcolor}\faFileCodeO}\xspace}
\newcommand{\nblink}[1]{\href{https://github.com/kenvantilburg/quaDM_lens/blob/main/#1}{\nbicon}}

\newcommand\githubicon[1]{\href{#1}{\faGithub}\xspace}

\begin{document}


\title{Discovering Substellar Dark Matter Halos\\ with Astrometric Weak Lensing of Multiply Imaged Quasars}

\author{Ken Van Tilburg}
\email{kenvt@stanford.edu}
\affiliation{Leinweber Institute for Theoretical Physics, Department of Physics, Stanford University, Stanford, CA 94305, USA}

\author{David E.~Kaplan}
 \email{david.kaplan@jhu.edu}
 \affiliation{Department of Physics and Astronomy, Johns Hopkins University, Baltimore, MD 21218, USA}
\affiliation{Kavli IPMU (WPI), UTIAS, The University of Tokyo, Kashiwa, Chiba 277-8583, Japan}

\date{\today}

\begin{abstract}
We propose time-domain astrometric weak lensing of multiply imaged quasars as a probe of substellar dark matter (DM) halos.
In $\Lambda$CDM, the photon path of each macro-image traverses numerous microhalos, which collectively produce a stochastic centroid motion with a calculable red power spectrum. 
Halos whose crossing times exceed the survey time impart a relative angular acceleration between image pairs.
The response is strongest for subhalos with scale radii of $0.005$--$2\,\mathrm{pc}$ (bound masses of $3\times10^{-6}$--$10^{2}\,M_\odot$ in $\Lambda$CDM), extending lensing sensitivity 12 to 14 orders of magnitude below the smallest DM structures detected to date, down to the free-streaming and acoustic-damping cutoffs of a canonical $100\,\mathrm{GeV}$ thermal relic.
We forecast the sensitivity of ten-year campaigns with $N=300$ epochs for two benchmark systems: the bright galaxy-lensed quadruple B1422+231 at a $0.1\,\mathrm{\mu as}$ per-epoch precision forecast for extended-path intensity correlation (EPIC), and the wide-separation cluster-lensed triple SDSS J1029+2623 at $1\,\mathrm{\mu as}$.
For the galaxy lens, the angular-acceleration channel reaches the standard $\Lambda$CDM microhalo population with a variance signal-to-noise ratio (SNR) of order ten at the best-matched halo masses, under the (optimistic) assumption that a tenth of the projected mass density at the image positions resides in surviving microhalos, modeled as tidally truncated NFW subhalos; the SNR drops to order unity after stellar microlensing background subtraction using image-shape fits.
The cluster lens, whose images sit in a far less star-rich environment, reaches a variance SNR of 0.35 at $1\,\mathrm{\mu as}$.
A positive detection would: raise universal lower bounds on the DM particle mass to $\gtrsim400\,\mathrm{keV}$ for fermions and $\gtrsim10^{-12}\,\mathrm{eV}$ for bosons; constrain the DM kinetic decoupling temperature to $\gtrsim 10\,\mathrm{MeV}$ (and thus its elastic coupling to the Standard Model); and be sensitive to the running of the spectral tilt at the $0.01$ level over 15--20 $e$-folds of the primordial curvature power spectrum.
A robust null result across well-characterized lenses would constrain microhalo survival and, once that astrophysical uncertainty is controlled, imply suppression in the primordial spectrum or the DM transfer function from nongravitational dynamics, i.e.~violations of the above inequalities.
These signatures motivate differential astrometric observations with extreme $0.1$--$1\,\mathrm{\mu as}$ light-centroiding precision on multiply imaged quasars.
\end{abstract}

\maketitle

\tableofcontents

\section{Introduction} \label{sec:intro}
Dark matter (DM) is established through its gravitational effects on galactic rotation curves, large-scale structure (LSS), and the cosmic microwave background (CMB), but its particle properties remain unknown. Viable candidates span many orders of magnitude in mass and interaction strength, and their microphysics can alter structure formation, especially on small scales.

The abundance, internal densities, and distribution of the smallest DM halos encode the primordial curvature power spectrum, the DM particle's thermal history, or its deviations from a perfect pressureless fluid due to e.g.~self-interacting or dissipative dynamics~\cite{Bechtol:2022koa}. The minimum halo mass also bounds the particle mass for both fermionic and bosonic DM. Because the relevant modes left the horizon later than those measured by the CMB and LSS, these structures probe late inflationary epochs and/or small-scale modifications of the DM transfer function.

We propose time-domain astrometric weak lensing of multiply imaged quasars to detect DM density fluctuations below one parsec, including subhalos with bound masses of a few $10^{-6}\,M_\odot$. Subhalos intersected by each macro-image's photon path produce stochastic deflections with a calculable spectrum. For the fiducial substructure in our benchmark systems, a magnified image is displaced by about a tenth of a microarcsecond over ten years. Halo crossings longer than the survey time appear as random (but statistically anisotropic) angular acceleration of order $10^{-2}\,\mathrm{\mu as\,yr^{-2}}$ (Fig.~\ref{fig:concept}) that is uncorrelated between images, while differencing image pairs suppresses common-mode systematics. At $0.1\,\mathrm{\mu as}$ per epoch, a ten-year campaign with 300 epochs on the benchmark galaxy lens reaches the fiducial $\Lambda$CDM microhalo population with a peak variance SNR of order ten, assuming that a tenth of the projected mass density at the image positions resides in surviving, tidally truncated microhalos. That reach is for data \emph{certified} free of stellar microlensing: intensity interferometry can establish, image by image and epoch by epoch, whether any star lies close enough to the sightline to matter. This certification of clean images is a significantly easier task than subtracting a star's astrometric deflection after inferring its mass and trajectory, the subject of a companion work~\cite{companionstars}. The typical stellar microlensing residuals reduce the SNR to order unity (after subtraction). The cluster lens, whose images sit in a far sparser stellar environment, reaches a variance SNR of 0.35 at $1\,\mathrm{\mu as}$ light-centroiding precision and otherwise similar survey parameters, and unity at $0.59\,\mathrm{\mu as}$. Throughout, these two per-epoch precisions serve as benchmarks for what is \emph{required} to see the $\Lambda$CDM signal. Similarly, a nondetection at these capabilities would constrain the substructure content of the lens halos to be different from that in standard cosmology, thereby suggesting the presence of new physics.

No existing facility can currently provide this required sub-microarcsecond \emph{relative} astrometry across the arcsecond-to-arcminute separations of lensed-quasar images.  Phase-referenced radio interferometry reaches $\sim10\,\mathrm{\mu as}$ at best, and optical imagers are less precise~\cite{Malbet:2021}, although more ambitious concepts have been proposed~\cite{2017arXiv170701348T,2004NewAR..48.1473F,2019arXiv191013086R,2025arXiv251018920M}. Extended-path intensity correlation (EPIC), a variant of intensity interferometry, is forecast to bridge this gap and may attain $0.1\,\mathrm{\mu as}$ precision when deployed at scale~\cite{VanTilburg:2023tkl,Galanis:2023gef}. The same measurements would resolve image-shape distortions sufficiently well to identify and subtract stellar microlenses individually~\cite{companionstars}. The science case developed in this paper provides, in our view, a central motivation for building such instruments.

Our work complements other probes of substructure in overlapping mass ranges: astrometric weak lensing by Milky Way subhalos~\cite{VanTilburg:2018ykj, Mishra-Sharma:2020ynk}, photometric microlensing of highly magnified stars~\cite{2020AJ....159...49D,Kelly:2018,Diego:2018}, arrival-time modulation of fast radio bursts (FRBs)~\cite{2024PhRvD.110b3516X}, and Doppler and Shapiro perturbations in pulsar timing arrays~\cite{Baghram:2011is,Dror:2019twh,Ramani:2020hdo,Lee:2020wfn}. We compare these methods with the proposed signatures in Sec.~\ref{sec:discussion}. At larger masses, dynamical heating of stars in ultrafaint dwarf galaxies constrains compact substructure for halos heavier than $10\,M_\odot$ and for comoving wavenumbers $k \lesssim 10^{3}\,\mathrm{Mpc}^{-1}$~\cite{Graham:2024hah}, while perturbations of cold stellar streams probe subhalos above $10^{6}\,M_\odot$~\cite{Banik:2019smi}. 

At substellar masses, \emph{photometric} microlensing surveys tightly constrain extremely dense dark objects such as primordial black holes over $10^{-11}$--$10\,M_\odot$~\cite{Tisserand:2006zx,Niikura:2017zjd,Niikura:2019kqi}. Those bounds do not apply to diffuse NFW-like halos or prompt cusps in this mass range~\cite{Croon:2020ouk}, which remain observationally unconstrained. 

Strongly lensed quasars also probe more massive ($\gtrsim 10^{6}\,M_\odot$) substructure through anomalous image flux ratios~\cite{MaoSchneider:1998,DalalKochanek:2002,Nierenberg:2017,Gilman:2020,Hsueh:2020}; the current frontier bounds the minimum halo mass at around $10^{8}\,M_\odot$~\cite{Nierenberg:2026}. Gravitational imaging of extended arcs has also detected individual perturbers directly, most recently one at $10^{6}\,M_\odot$ resolved with milliarcsecond-resolution VLBI~\cite{Powell:2025}. Our proposal for \emph{time-domain} astrometric weak lensing targets the same systems but can potentially reach structures orders of magnitude lighter than these analyses.
Previous studies of astrometric quasar microlensing focused mainly on baryonic or stellar lenses~\cite{williams1995,1998ApJ...501..478L,2004A&A...416...19T,ForesToribio:2024a,2026arXiv260502181M}. In a companion paper~\cite{companionstars}, we show that intensity interferometry can identify and subtract this dominant stellar background star by star. 

Section~\ref{sec:theory} derives the stochastic astrometric signal and its observable summary statistics. Section~\ref{sec:sensitivity} evaluates instrumental, stellar, and source backgrounds for two systems that are promising testbeds for the signal under consideration: B1422+231, one of the brightest lensed quasars in the sky, imaged into a highly magnified quadruple by a foreground galaxy whose stellar surface density happens to be low at the image positions; and SDSS J1029+2623, a wide-separation triple imaged by a galaxy cluster, whose images fall far from any cluster member and are correspondingly star-poor, at the cost of being too faint for intensity interferometry. Section~\ref{sec:structures} compares the forecast with the small-scale structure predicted by purely cold DM and derives implications for DM microphysics and the primordial spectrum, should a discovery or nondetection of structures around $10^{-6}\,M_\odot$ occur. Section~\ref{sec:discussion} summarizes the observational requirements, compares our technique to complementary approaches, and outlines the remaining theoretical and numerical work. Appendices~\ref{app:psd}--\ref{app:nl-power} contain the derivations and modeling details.

Where needed, we employ a cosmic expansion rate of $H_0 = 70 \, \mathrm{km\,s^{-1}\,Mpc^{-1}}$, fractional matter (dark energy) densities of $\Omega_M = 0.3$ ($\Omega_\Lambda = 0.7$), and vanishing spatial curvature ($\Omega_K = 0$). We use units wherein $c = 1$ throughout. All code used in this work is publicly available at \githubicon{https://github.com/kenvantilburg/quaDM_lens}\href{https://github.com/kenvantilburg/quaDM_lens}{\texttt{github.com/kenvantilburg/quaDM\_lens}}; the \nbicon icon in each caption links to the notebook that produces the corresponding Figure or Table.

\section{Lensing Signal Theory} \label{sec:theory}
\begin{figure}
    \centering
    \includegraphics[width=0.48\textwidth]{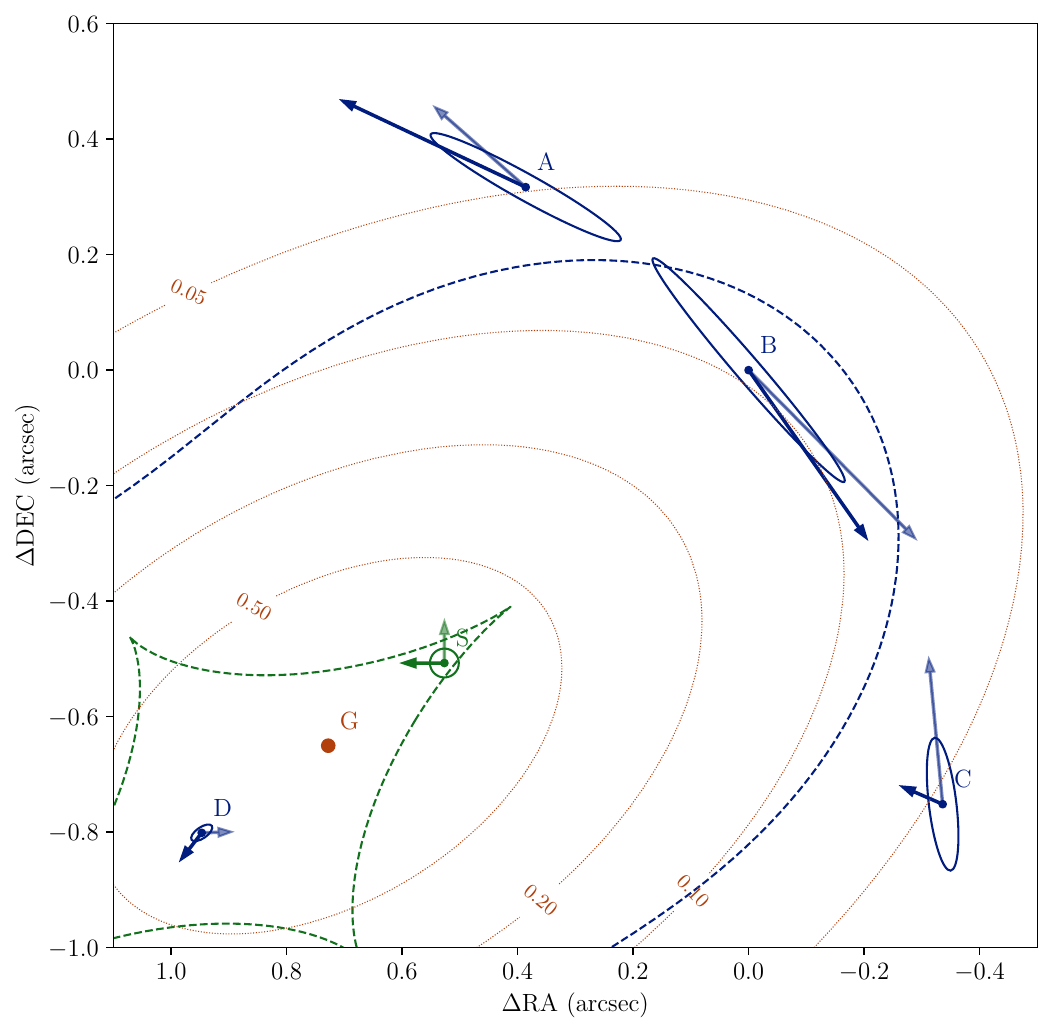}
    \caption{\nblink{code/macro_lens.ipynb} Diagram of the macrolensing setup in the B1422+231 system. The foreground galaxy G at $z_\mathrm{L} = 0.34$ lenses a quasar source S at $z_\mathrm{S} = 3.62$, whose source-plane location ($\vect{\beta}$) is shown in green and four image locations ($\vect{\theta}^\mathrm{I}$) for I = A,B,C,D in blue. The change in light and dark vectors (or equivalently, the ellipses) indicate the lensing distortion given by the inverse Jacobian $B_{ij}^\mathrm{I}$ relative to the (assumed) azimuthally symmetric source. Source and image sizes are shown larger than their true values by about a factor of a million in linear dimension. The critical curve (caustic) is shown as the blue (green) dashed line. Orange contours display the modeled stellar convergence $\kappa_*$.}
    \label{fig:diagram}
\end{figure}

This section derives the power spectral density (PSD) of the stochastic astrometric fluctuations produced by (sub)structure along the line of sight and connects it to two time-domain observables.

\subsection{Setup}
Let $\vect{\varphi}$ be the macrolensing map that takes an image-plane location $\vect{\theta}$ to a source-plane location $\vect{\beta}$, both 2D coordinates on the celestial sphere. The map $\vect{\varphi}$ only accounts for the macroscopic lensing (hereafter, ``macrolensing'') of the foreground galaxy with reduced deflection angle $\vect{\alpha}_\mathrm{G}$. The signal of interest comes from the fine-grained structure of the foreground lens (or elsewhere along the line of sight), in the form of DM microhalos, which collectively give a reduced deflection angle of $\vect{\alpha}(\vect{\theta})$. Hence the total lensing map is:
\begin{align}
    \vect{\beta} = \vect{\theta} - \vect{\alpha}_\mathrm{G}(\vect{\theta}) - \vect{\alpha}(\vect{\theta}) \equiv \vect{\varphi}(\vect{\theta}) - \vect{\alpha}(\vect{\theta}).
\end{align}
The separation between $\vect{\alpha}_\mathrm{G}$ and $\vect{\alpha}$ depends on the smoothing scale. We require only that $\vect{\alpha}$ contain the fine-grained structure that varies over the survey time, while $\vect{\alpha}_\mathrm{G}$ describes the effectively static macrolens, which moves only a tiny fraction of its own size. Time variation is predominantly induced by the proper tangential velocities $\vect{v}_\mathrm{S}$, $\vect{v}_\mathrm{L}$, and $\vect{v}_\mathrm{o}$ of the quasar, lens, and observer, respectively, leading to a source-plane angular motion $\vect{\mu} \equiv \dd \vect{\beta}/\dd t$ of 
\begin{align}
    \vect{\mu} =  \frac{\vect{v}_\mathrm{S}}{1+z_\mathrm{S}} \frac{1}{D_\mathrm{S}} - \frac{\vect{v}_\mathrm{L}}{1+z_\mathrm{L}}\frac{1}{D_\mathrm{L}} + \frac{\vect{v}_\mathrm{o}}{1+z_\mathrm{L}} \frac{D_\mathrm{LS}}{D_\mathrm{L}D_\mathrm{S}}, \label{eq:mu}
\end{align}
where $D_\mathrm{S}$ and $D_\mathrm{L}$ ($z_\mathrm{S}$ and $z_\mathrm{L}$) are the angular diameter distances (redshifts) of the quasar and lens, and $D_\mathrm{LS}$ is the lens--source angular diameter distance~\cite{kayser1986astrophysical}. All symbols $D$ denote \emph{physical} angular diameter distances throughout this work; they are not additive, $D_\mathrm{LS} \neq D_\mathrm{S} - D_\mathrm{L}$, and $D(z)$ is not even monotonic in redshift $z$. Comoving distances, which are additive in a flat cosmology, are written $\chi(z) \equiv (1+z) D(z)$ and are used only in the cosmological sightline integrals of Eqs.~\ref{eq:alpha_los} and~\ref{eq:C_tilde_1} (see App.~\ref{app:power-spectrum}). In those variables, Eq.~\ref{eq:mu} takes the compact form $\vect{\mu} = \vect{v}_\mathrm{S}/\chi_\mathrm{S} - \vect{v}_\mathrm{L}/\chi_\mathrm{L} + \vect{v}_\mathrm{o} (\chi_\mathrm{S}-\chi_\mathrm{L})/(\chi_\mathrm{L}\chi_\mathrm{S})$.

The inverse lensing map $\vect{\varphi}^{-1}$ is generically multi-valued. Denote by $\vect{\theta}^\mathrm{I} = \vect{\varphi}^{-1}(\vect{\beta}_\mathrm{S})|_\mathrm{I}$ the multiple images, labeled by $\mathrm{I} = \mathrm{A}, \mathrm{B}, \ldots$, of the quasar's location $\vect{\beta}_\mathrm{S}$ on the source plane. Given the lensing map's Jacobian $J_{ij}(\vect{\theta}) \equiv \partial \varphi_i(\vect{\theta}) / \partial \theta_j$, the quasar's images are distorted (relative to the unlensed image) by the inverse local Jacobian $B^\mathrm{I}_{ij} \equiv [J(\vect{\theta}^\mathrm{I})^{-1}]_{ij}$. The (signed) magnification of each image is the scalar $A^\mathrm{I} \equiv \det B^\mathrm{I}$.

\subsection{Power Spectrum}
For perturbatively small proper motions $\vect{\mu}$ in the source plane and even smaller (stochastic) deflections $\vect{\alpha}$, 
we can solve for the image trajectories:
\begin{align}
\theta^\mathrm{I}_i(t) - \theta^\mathrm{I}_i(t_0) \simeq B^\mathrm{I}_{ij} \left\lbrace \mu_j (t - t_0) + \alpha_j\big[\vect{\theta}^\mathrm{I}(t)\big] - \alpha_j\big[\vect{\theta}^\mathrm{I}(t_0)\big] \right\rbrace \label{eq:thetaI}
\end{align}
with Einstein summation convention assumed on repeated lowercase Cartesian indices. 
After fitting the location and magnified proper motion (the latter first proposed as a realistic observable in Ref.~\cite{Kochanek:1996}),
\begin{align}
\dot{\theta}^\mathrm{I}_i \equiv \widetilde{\mu}^\mathrm{I}_i = B^\mathrm{I}_{ij} \mu_j, \label{eq:mu_tilde}
\end{align}
each image will have an excess astrometric noise of $\delta \theta^\mathrm{I}_i \equiv B^\mathrm{I}_{ij} \alpha_j\big[\vect{\theta}^\mathrm{I}(t) \big]$ with vanishing expectation value $\langle \delta \theta_i^\mathrm{I} \rangle = 0$ but nonzero two-point function $\langle \delta \theta_p^\mathrm{I}(t') \delta \theta_q^\mathrm{I}(t'+t) \rangle$:
\begin{align}
    C^\mathrm{I}_{pq}(t) &= B_{pi}^\mathrm{I} B_{qj}^\mathrm{I} \left \langle \alpha_i\big(\vect{\theta}^\mathrm{I}(t')\big) \alpha_j\big(\vect{\theta}^\mathrm{I}(t'+t)\big)  \right\rangle \label{eq:2ptC_1} \\
    &\equiv \int \frac{\dd \omega}{2\pi} \, e^{-i\omega t} \widetilde{C}_{pq}^\mathrm{I}(\omega). \label{eq:2ptC_2}
\end{align}
The reduced lensing deflection angle $\vect{\alpha}$ is the standard line-of-sight integral of the perpendicular gradient of the gravitational potential:
\begin{align}
    \alpha_i(\vect{\theta}) = 2 \int_0^{\chi_\mathrm{S}} \dd \chi \, \frac{\chi_\mathrm{S} - \chi}{\chi_\mathrm{S} \chi} \partial_{\theta_i} \Phi(\chi,\chi \vect{\theta}). \label{eq:alpha_los}
\end{align}
Small-scale fractional DM density fluctuations $\delta(\vect{x}) \equiv \rho(\vect{x})/\overline{\rho} - 1$ with power spectrum $P_\delta(k)$ source stochastic potential gradients. Their astrometric PSD, as defined in Eq.~\ref{eq:2ptC_2}, is
\begin{align}
    \widetilde{C}^\mathrm{I}_{pq}(\omega)
    &=B_{pi}^\mathrm{I} B_{qj}^\mathrm{I} \frac{4}{|\omega|} \int_0^{\chi_\mathrm{S}} \dd \chi \, \left(\frac{\chi_\mathrm{S} - \chi}{\chi_\mathrm{S}}\right)^2 (4\pi G \overline{\rho} a^2)^2 \nonumber \\
    &\phantom{=}\times \int_{-\frac{\pi}{2}}^{\frac{\pi}{2}}  \frac{\dd \phi}{2\pi}
    \mathcal{M}_{ij}(\phi,\hat{\widetilde{\vect{\mu}}}{}^\mathrm{I})
    P_\delta \left(\frac{|\omega|}{ \widetilde{\mu}^\mathrm{I} \mathcal{X}(\chi) \cos(\phi)} \right). \label{eq:C_tilde_1}
\end{align} 
Equation~\ref{eq:C_tilde_1} is derived in App.~\ref{app:power-spectrum}. The matrix $\mathcal{M}_{ij}$, defined in Eq.~\ref{eq:matM}, depends on the integration angle $\phi$ and the image proper-motion direction $\hat{\widetilde{\vect{\mu}}}{}^\mathrm{I}$ of the image. Here $a = 1/(1+z)$ is the scale factor at comoving distance $\chi$ along the path, and $\overline{\rho}a^2 = \overline{\rho}_{m,0}(1+z)$ with $\overline{\rho}_{m,0}$ the present-day mean matter density, so that the wavenumber argument of $P_\delta$ is comoving. The comoving \emph{sweep length}
\begin{align}
    \mathcal{X}(\chi) = \chi_\mathrm{L} \, \min\!\left[ \frac{\chi}{\chi_\mathrm{L}} , \, \frac{\chi_\mathrm{S}-\chi}{\chi_\mathrm{S}-\chi_\mathrm{L}} \right] \label{eq:sweep}
\end{align}
sets the velocity $\widetilde{\mu}^\mathrm{I} \mathcal{X}(\chi)$ at which the light beam crosses the plane at $\chi$. We assume the photon path is straight between deflections and pinned at both ends (on the observer and on the quasar), so the magnified sweep velocity grows as $\chi$ out to the lens plane, where it reaches $\chi_\mathrm{L}\widetilde{\mu}^\mathrm{I}$, and then falls back to zero on the source, where only the (unmagnified, and here neglected) relative motion of source, lens, and observer remains. 
Note that $P_\delta(k)$ is the nonlinear (dark) matter power spectrum specific to the line of sight, heavily biased toward overdense regions by its trajectory through the lensing galaxy or cluster.

\subsection{One-Halo Terms} \label{sec:one_halo}
If just \emph{one} small lens of mass $M_\mathrm{L}$ and characteristic radius $r_\mathrm{L}$ were located (within the larger lens galaxy) at $(\vect{\theta}_\mathrm{L},D_\mathrm{L})$ with proper motion $\vect{\mu}_\mathrm{L}$, then we obtain the single-lens deflection:
\begin{align}
    \vect{\alpha}^{(1)}(\vect{\theta}) &= \int \frac{\dd^2k}{(2\pi)^2} \, \widetilde{\vect{\alpha}}(\vect{k}) e^{- i \vect{k} \cdot (\vect{\theta} - \vect{\theta}_\mathrm{L})}; \label{eq:alpha_FT1}\\ 
    \widetilde{\vect{\alpha}}(\vect{k}) &= \frac{D_\mathrm{LS}}{D_\mathrm{L} D_\mathrm{S}} 4 G M_\mathrm{L} \, \frac{2 \pi i}{k} \hat{\vect{k}} \, F(k \gamma_\mathrm{L}), \label{eq:alpha_FT2}
\end{align}
where $\gamma_\mathrm{L} \equiv r_\mathrm{L} / D_\mathrm{L}$ is the angular radius of the lens. The form factor is 
\begin{align}
    F(k \gamma_\mathrm{L}) = \int_0^\infty \dd x \, J_1(x) M_\mathrm{L}^{-1} M\left(\frac{x r_\mathrm{L}}{k \gamma_\mathrm{L}}\right) \label{eq:form}
\end{align}
with $M(\theta D_\mathrm{L})$ the enclosed lens mass as a function of impact parameter. This form factor goes to unity as $k \gamma_\mathrm{L} \ll 1$ and is suppressed for $k \gamma_\mathrm{L} \gg 1$, with the dependence encoding the microhalo's internal density profile. We model the collective, stochastic effect from \emph{all} small-scale subhalos with the two-point function:
\begin{align}
    \left \langle \alpha_i(\vect{\theta}) \alpha_j(\vect{\theta}')  \right\rangle  = \int \dd D_\mathrm{L} \, D_\mathrm{L}^2 \int \dd^2 \theta_\mathrm{L} \, n_\mathrm{L} \, \alpha^{(1)}_i(\vect{\theta}) \alpha^{(1)}_j(\vect{\theta}'), \label{eq:2ptalpha}
\end{align}
effectively assuming uncorrelated positions of the small-scale lenses in the host galaxy, distributed with mean number density $n_\mathrm{L}$, i.e. neglecting the two-halo term from correlations between distinct lenses. Here, $n_\mathrm{L}$ is a physical number density and $\dd D_\mathrm{L}\,D_\mathrm{L}^2 \,\dd^2\theta_\mathrm{L}$ a proper volume element: the integral runs only over the thin slab occupied by the host, across which $D_\mathrm{L}$ is (fractionally) constant to excellent approximation.

The small lenses also have their own (only partially correlated) proper motion $\vect{\mu}_\mathrm{L}$ \emph{within} the lens galaxy/cluster, which we omit from Eq.~\ref{eq:2ptalpha}. For our benchmark systems, these unmagnified internal dispersions contribute negligibly ($\lesssim1\%$ in quadrature) to the ``sweep rate'' (the rate at which the image moves across the subhalo's deflection field). Where needed, they can be incorporated exactly by promoting $n_\mathrm{L}$ to the subhalo phase-space distribution and letting each deflector drift, as in Eq.~\ref{eq:Cder_0}, which amounts to averaging the frequency-selecting delta function of the derivation (App.~\ref{app:1-halo}) over the subhalo velocity distribution.\footnote{\label{fn:mu}The kinematics here deserve care: the rate at which image $\mathrm{I}$ sweeps across a given microhalo's deflection field is $\dd [\vect{\theta}^\mathrm{I} - \vect{\theta}_\mathrm{L}]/\dd t = B^\mathrm{I}\vect{\mu} - \vect{\mu}_\mathrm{L}$, so the bulk source--lens--observer motion is magnified by $B^\mathrm{I}$ (Eq.~\ref{eq:thetaI}), whereas a subhalo's own orbital motion displaces the deflector, not the image, and enters \emph{unmagnified} to leading order in the perturbation. When distinct \emph{macroscopic} components $c$ of the overall lensing potential move relative to one another, the magnified term generalizes to $B^\mathrm{I}[\vect{\mu} - \sum_c H_c\,\delta\vect{\mu}_c]$, with $H_c$ each component's contribution to the local deflection gradient and $\delta\vect{\mu}_c$ its peculiar motion. This effect could be significant for unvirialized cluster lenses (Sec.~\ref{sec:cluster}).}

The subhalo population therefore produces a stochastic image deflection with two-point function $C^\mathrm{I}_{pq}(t)$ and PSD $\widetilde{C}_{pq}^\mathrm{I}(\omega)$ defined in Eqs.~\ref{eq:2ptC_1} and~\ref{eq:2ptC_2}.
Define the projected surface mass density of small lenses $\Sigma^\mathrm{I}_\mathrm{L} = \int \dd D_\mathrm{L} \, n_\mathrm{L} M_\mathrm{L}$ with associated convergence $\kappa^\mathrm{I}_\mathrm{L} = \Sigma^\mathrm{I}_\mathrm{L}/\Sigma_\mathrm{cr}$, where the critical surface mass density is defined as $\Sigma_\mathrm{cr} \equiv (4\pi G D_\mathrm{L})^{-1} D_\mathrm{S}/(D_\mathrm{LS})$. In the thin-lens approximation, the signal power is
\begin{align}
    \hspace{-0.4em}
    \widetilde{C}^\mathrm{I}_{pq}(\omega) &= B_{pi}^\mathrm{I} B_{qj}^\mathrm{I} \frac{\kappa^\mathrm{I}_\mathrm{L} \theta_\mathrm{E,L}^2}{|\omega|} \int_{-\frac{\pi}{2}}^{\frac{\pi}{2}} \dd \phi \, Q^\mathrm{I}_{ij} F\left[\frac{|\omega| \gamma_\mathrm{L}}{\widetilde{\mu}^\mathrm{I} c_\phi}\right]^2; \label{eq:C}\\
    Q^\mathrm{I} &\equiv 
    \begin{pmatrix}
    1 + c_{2\zeta^\mathrm{I}} c_{2\phi} & s_{2\zeta^\mathrm{I}} c_{2\phi} \\
    s_{2\zeta^\mathrm{I}} c_{2 \phi} & 1 - c_{2\zeta^\mathrm{I}} c_{2\phi}
    \end{pmatrix}, \label{eq:Q}
\end{align}
where $c_\phi \equiv \cos \phi$ and $s_\phi \equiv \sin \phi$.
We derive Eq.~\ref{eq:C} in App.~\ref{app:1-halo}. Here $\widetilde{\mu}^\mathrm{I}$ is the magnitude of the image proper motion, with components $\widetilde{\mu}^\mathrm{I}_i \equiv B^\mathrm{I}_{ij} \mu_j \equiv \widetilde{\mu}^\mathrm{I} (\cos \zeta^\mathrm{I}, \sin \zeta^\mathrm{I})_i$ as in Eq.~\ref{eq:mu_tilde}. We have also introduced the angular Einstein radius of the small lenses:
\begin{align}
    \theta_\mathrm{E,L} 
    &\equiv \sqrt{\frac{4 G M_\mathrm{L}}{D_\mathrm{L}} \frac{D_\mathrm{LS}}{D_\mathrm{S}}} 
    \approx 2.85 \, \mathrm{\mu as} \, \sqrt{\frac{M_\mathrm{L}}{M_\odot} \frac{\mathrm{Gpc}}{D_\mathrm{L}}\frac{D_\mathrm{LS}}{D_\mathrm{S}}}. \label{eq:thetaE}
\end{align}
In the low-frequency limit $|\omega| \ll \widetilde{\mu}^\mathrm{I}/\gamma_\mathrm{L}$, the form factor approaches unity and Eq.~\ref{eq:C} reduces to the red-noise spectrum 
\begin{align}
\widetilde{C}^\mathrm{I}_{pq}(\omega) \approx B_{pi}^\mathrm{I} B_{qi}^\mathrm{I} \frac{\pi \kappa^\mathrm{I}_\mathrm{L} \theta_\mathrm{E,L}^2}{|\omega|}, \label{eq:C_low}
\end{align}
which is isotropic save for the lensing distortions parametrized by $B^\mathrm{I}$, and independent of $\zeta^\mathrm{I}$ and $\gamma_\mathrm{L}$.
For a population of small lenses, Eq.~\ref{eq:C} is precisely the one-halo-term evaluation of the general matter-power-spectrum expression of Eq.~\ref{eq:C_tilde_1} (App.~\ref{app:power-spectrum}).

Equations~\ref{eq:C} and~\ref{eq:C_low} have a simple parametric interpretation. Consider a column of mass parallel to the line of sight with a transverse size equal to the lens radius $r_\mathrm{L}$ and a survey time $\tau$ such that the transverse image movement $\widetilde{\mu} D_\mathrm{L} \tau$ also matches this size (i.e.~$\tau$ equal to the crossing time). The total mass in this column is $M_\mathrm{col} \sim \Sigma_\mathrm{L} r_\mathrm{L}^2$, within which one expects $N_\mathrm{col} = M_\mathrm{col}/M_\mathrm{L} = \kappa_\mathrm{L}\Sigma_\mathrm{cr}r_\mathrm{L}^2/M_\mathrm{L}$ lenses on average. For our fiducial substructure assumptions of $\rho_s \sim 1\,M_\odot\,\mathrm{pc}^{-3}$, $\kappa_\mathrm{L}\approx 0.04$ for B1422+231 (a tenth of the total matter convergence, cf.~Tab.~\ref{tab:macro}; Sec.~\ref{sec:signals}), and a bound halo mass $M_\mathrm{L} = 10^{-3}\,M_\odot$ (truncation radius $r_\mathrm{L} = r_\mathrm{t}\approx 0.07\,\mathrm{pc}$), this occupancy is large, $N_\mathrm{col}\sim 5\times 10^{2}$. However, Poisson fluctuations in this total mass from column to column are of order $\delta M_\mathrm{col} \sim M_\mathrm{col}/\sqrt{N_\mathrm{col}}$, which lead to stochastic lensing deflections by typical angles $\delta \alpha \sim G \delta M_\mathrm{col} / r_\mathrm{L} \sim G \sqrt{\Sigma_\mathrm{L} M_\mathrm{L}} \sim \sqrt{\kappa_\mathrm{L}} \theta_\mathrm{E,L}$. After amplification by the inverse lensing Jacobian $B^\mathrm{I}$, this estimate agrees with Eq.~\ref{eq:C}, considering the integral of Eq.~\ref{eq:2ptC_2} in this case has dominant support at $\omega \sim \tau^{-1} \sim \widetilde{\mu}^\mathrm{I}/\gamma_\mathrm{L}$.

The ``redness'' of the spectrum follows from the same picture, extended to columns of transverse size $b \gtrsim r_\mathrm{L}$. The number of lenses within a distance $b$ of the image trajectory grows as $N(b)\sim \Sigma_\mathrm{L} b^2/M_\mathrm{L}$, so their net Poisson deflection $\alpha(b) \sim G\sqrt{N(b)}M_\mathrm{L}/b \sim G\sqrt{\Sigma_\mathrm{L}M_\mathrm{L}} \sim \sqrt{\kappa_\mathrm{L}}\,\theta_\mathrm{E,L}$ is \emph{independent} of $b$: the $1/\theta$ deflection tail is scale free, and every $e$-fold in impact parameter contributes equally. Since a scale $b$ is swept in a time $b/(\widetilde{\mu}^\mathrm{I} D_\mathrm{L})$, i.e.~appears at $|\omega| \sim \widetilde{\mu}^\mathrm{I} D_\mathrm{L}/b$, equal variance per $e$-fold of $b$ means equal variance per $e$-fold of frequency, $|\omega|\widetilde{C}^\mathrm{I} \sim B^2 \kappa_\mathrm{L}\theta_\mathrm{E,L}^2 = \mathrm{const}$, which is the $1/|\omega|$ law of Eq.~\ref{eq:C_low}. Scale invariance is broken only at $b \lesssim r_\mathrm{L}$, where the lens is resolved and the enclosed mass drops, an effect encapsulated by the form factor $F$, which suppresses the spectrum for $|\omega|\gtrsim\widetilde{\mu}^\mathrm{I}/\gamma_\mathrm{L}$.

The magnification enters Eq.~\ref{eq:C} twice, both instances amplifying the signal. The first is the explicit prefactor $B^\mathrm{I}_{pi}B^\mathrm{I}_{qj}$, coming from the mapping of a small geodesic deflection $\vect{\alpha}_i$ to a magnified image displacement $B^\mathrm{I}_{pi}\alpha_i$ (Eq.~\ref{eq:thetaI}), so the power carries two factors of the inverse lensing Jacobian $B^\mathrm{I}$. The second is implicit, through the magnified image proper motion $\widetilde{\mu}^\mathrm{I} = |B^\mathrm{I}_{pi}\mu_i|$ in the argument of the form factor: at a fixed frequency $\omega$ within the survey band, the image sweeps across a transverse scale $b \sim \widetilde{\mu}^\mathrm{I} D_\mathrm{L}/|\omega|$ that is itself proportional to the magnification. A more magnified (and thus faster-moving) image keeps \emph{larger} halos unresolved within the band, with the roll-off scale $r_\mathrm{L} \approx D_\mathrm{L}\widetilde{\mu}^\mathrm{I}/|\omega| \propto B^\mathrm{I}$. These larger halos are more massive ($M_\mathrm{L}\propto  r_\mathrm{L}^3$ at fixed scale density) and hence lens more strongly ($\theta_\mathrm{E,L}^2 \propto M_\mathrm{L}$). For a monochromatic population inside the band, the two effects compound to $\widetilde{C}{}^\mathrm{I} \propto (B^\mathrm{I})^2\theta_\mathrm{E,L}^2 \propto (B^\mathrm{I})^5$ at fixed $\omega$.

The large occupancy $N_\mathrm{col} \gg 1$ also underpins the Gaussian approximation. Using $M_\mathrm{L} \sim \rho_s r_\mathrm{L}^3$, the count scales as $N_\mathrm{col}\propto \kappa_\mathrm{L}\Sigma_\mathrm{cr}\,\rho_s^{-2/3}M_\mathrm{L}^{-1/3}$ and stays well above unity across the entire mass range of interest ($N_\mathrm{col} \sim 50$ at $M_\mathrm{L} = 1\,M_\odot$). Because each column's deflection comes from the incoherent sum of many independent perturbers, the central limit theorem renders the stochastic deflection field Gaussian to good approximation. Under these assumptions, the PSD $\widetilde{C}^\mathrm{I}_{pq}(\omega)$ of Eq.~\ref{eq:C} is a complete statistical observable, with information from higher-point functions fractionally suppressed by powers of $N_\mathrm{col}^{-1/2}$. The effective number of perturbers can be smaller for cuspier profiles or enhanced internal microhalo densities; in extreme cases, the Gaussian approximation could break down. Such microhalo populations would lead to stronger deflections at small impact parameters, so the signal only grows, with the excess appearing in nearest-neighbor-dominated non-Gaussian tails that the PSD of Eq.~\ref{eq:C} does not capture. Restricting ourselves to Gaussian statistics is thus a conservative choice; extracting the additional information in those tails would require Monte Carlo realizations of the deflection field. In particular, the above reasoning fails for stellar microlenses, which are far rarer (only $N_\mathrm{fit}\sim\mathcal{O}(\mathrm{few})$ within the star-fitting region; Sec.~\ref{sec:sensitivity}) and point-like: their background is neither Gaussian nor perturbative, and can be measured and subtracted star by star using methods described in a companion paper~\cite{companionstars}.

\subsection{Finite Source Size}\label{sec:finite_source}
The preceding calculations treat the quasar as a point source, whereas the base observable is the \emph{light centroid} of a macro-image of finite size. The angular extent $\theta^\mathrm{I}_\mathrm{src}$ of each image weights the astrometric deflection by flux across the image and thus averages over structures that are smaller (in projection) than the image itself. This introduces a second form factor, the squared Fourier transform $|\widetilde{W}^\mathrm{I}|^2$ of the image surface-brightness profile~\cite{Galanis:2023gef}, multiplying the integrands of Eqs.~\ref{eq:C_tilde_1} and~\ref{eq:C} in analogy with the lens form factor $F$. In the one-halo-term form of Eq.~\ref{eq:C}, this amounts to the replacement
\begin{align}
    F\!\left[\frac{|\omega| \gamma_\mathrm{L}}{\widetilde{\mu}^\mathrm{I} c_\phi}\right]^2 \longrightarrow F\!\left[\frac{|\omega| \gamma_\mathrm{L}}{\widetilde{\mu}^\mathrm{I} c_\phi}\right]^2 \left|\widetilde{W}^\mathrm{I}\!\left[\frac{|\omega| \theta^\mathrm{I}_\mathrm{src}}{\widetilde{\mu}^\mathrm{I} c_\phi}\right]\right|^2 , \label{eq:finite_source}
\end{align}
so the finite size of the source imposes a high-frequency cutoff at $|\omega| \gtrsim \widetilde{\mu}^\mathrm{I}/\theta^\mathrm{I}_\mathrm{src}$ that washes out substructures projecting on angular scales below the source image size, $\gamma_\mathrm{L} \lesssim \theta^\mathrm{I}_\mathrm{src}$. For the compact UV/optical continuum region of a luminous quasar, with physical size of a few light-days, the magnified image size is typically $\theta^\mathrm{I}_\mathrm{src} \sim \mathcal{O}(1)\,\mathrm{\mu as}$. For example, for image A of B1422+231, the flux-normalized disk size is $R_\mathrm{src}\approx6.9\times10^{15}\,\mathrm{cm}$ (App.~\ref{app:B1422}), corresponding to an unlensed angular size of $\theta_\mathrm{src}\approx0.04\,\mathrm{\mu as}$, stretched tangentially by a factor $7.5$ to $\approx0.3\,\mathrm{\mu as}$. The intensity-interferometric observations advocated here would also determine the source size to percent-level precision (in addition to other aspects of the image morphology), so this irreducible cutoff scale can be measured directly. This is safely below the optimal $\gamma_\mathrm{L} \sim \widetilde{\mu}^\mathrm{I}\tau \sim 10\,\mathrm{\mu as}$ that maximizes the stochastic signal.
Essentially, because an optical quasar image moves several times its own size during a decade-long survey, subhalos larger than the image dominate the signal and the finite source size only mildly suppresses contributions from smaller subhalos: including it lowers the peak acceleration SNR of Sec.~\ref{sec:prospects} at the percent level. Therefore, $|W^\mathrm{I}| \sim 1$ over most of the relevant range of $\gamma_\mathrm{L}$ and $\omega$ in the sensitivity calculations of Sec.~\ref{sec:sensitivity}.\footnote{Significant image shape distortions can be caused by \emph{stellar} microlenses. In fact, their effects on $\big| \widetilde{W}^\mathrm{I} \big|^2$ can be used to mitigate the stellar microlensing background~\cite{companionstars}.}
In the radio/microwave band, in which the angular size of the quasar emission is much larger, this form factor is more important, and sets the lower limit of the subhalo mass range that can be probed by astrometric lensing. We include it in the cluster-lens analysis of Sec.~\ref{sec:cluster} (App.~\ref{app:J1029}), where it would significantly impact the discovery reach of very-long-baseline interferometry (VLBI).

\subsection{Observables} \label{sec:observables}
\begin{figure}
    \centering
    \includegraphics[width=0.5\textwidth]{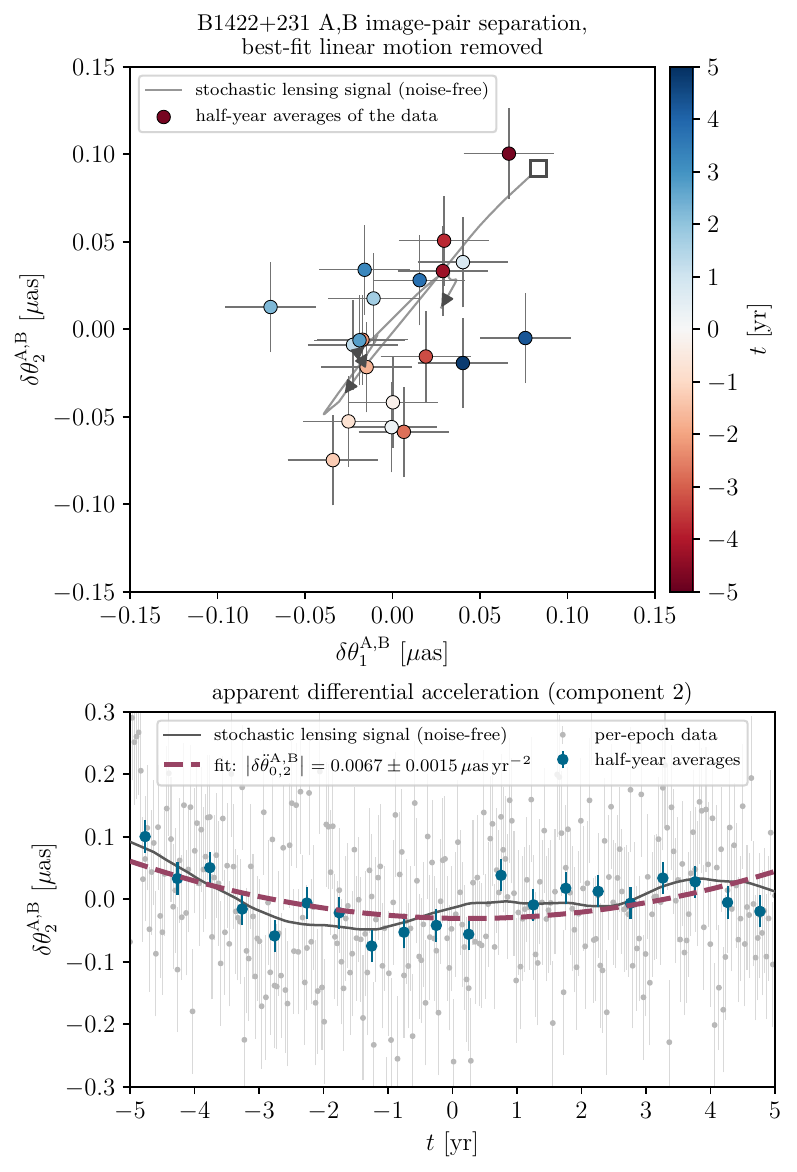}
    \caption{\nblink{code/concept.ipynb} Simulated realization of the astrometric weak-lensing signal for the A,B image pair of B1422+231 (Sec.~\ref{sec:sensitivity}), for truncated NFW subhalos of bound mass $M_\mathrm{L}=10^{-4}\,M_\odot$ with $\rho_s = 10\,M_\odot\,\mathrm{pc^{-3}}$ and $f_\mathrm{sub}=0.1$, observed with the fiducial EPIC campaign ($\tau=10\,\mathrm{yr}$, $N=300$, $\sigma_{\delta\theta}=0.1\,\mathrm{\mu as}$). \emph{Top:} sky-plane trajectory of the image-pair separation after removing the best-fit relative proper motion; the gray curve is the noise-free lensing random walk, running from the open square at $t=-\tau/2$ in the direction of the arrowheads, and the points are half-year averages of the simulated data, colored by time. \emph{Bottom:} the vertical (declination) component versus time. The contributions to the random walk from subhalo crossings that are not time resolved manifest as an anomalous differential acceleration (dashed quadratic fit, reduced chi-squared $\chi^2_\nu = 1.17$), here $\big|\delta\ddot\theta_{0,2}^{\mathrm{A,B}}\big| = 0.0067 \pm 0.0015\,\mathrm{\mu as\,yr^{-2}}$ (one component of the observable in Eq.~\ref{eq:observable_2}).}
    \label{fig:concept}
\end{figure}
Our base observable is the time-dependent difference between the light centroids of any two macro-images $\mathrm{I}$ and $\mathrm{I'}$ in a strongly lensed quasar system, namely the separation vector $\delta \vect{\theta}^\mathrm{I,I'} \equiv \vect{\theta}^\mathrm{I} - \vect{\theta}^\mathrm{I'}$, as a function of time. The small-scale structures along the two lines of sight are independent, so the lensing contribution to the covariance of the image separation vector is additive in PSD, $\widetilde{C}^\mathrm{I,I'}_{pq} \equiv \widetilde{C}^\mathrm{I}_{pq}+\widetilde{C}^\mathrm{I'}_{pq}$:
\begin{align}
    &\left\langle \delta \theta_p^\mathrm{I,I'}(t) \delta \theta_q^\mathrm{I,I'}(0) \right\rangle =\int \frac{\dd \omega}{2\pi} e^{-i \omega t} \widetilde{C}^\mathrm{I,I'}_{pq}(\omega). \label{eq:observable_1}
\end{align}
Our first summary statistic is the discrete Fourier transform (DFT) of the $\delta\vect{\theta}^\mathrm{I,I'}$ time series, which estimates $\widetilde{C}^\mathrm{I,I'}_{pq}(\omega)$. We call the excess variance carried by these DFT modes the \emph{stochastic lensing signal}.

If the survey time $\tau$ is shorter than the lens crossing time $\gamma_\mathrm{L} / \widetilde{\mu}$ of larger DM halos, then their stochastic fluctuations will not be time resolved. In this case, one can Taylor expand the motion as 
$\theta^\mathrm{I}_i(t) = \theta^\mathrm{I}_{0,i} + \dot{\theta}^\mathrm{I}_{0,i} t + \ddot{\theta}^\mathrm{I}_{0,i} t^2/2 + \mathcal{O}(t^3)$. Any small-scale lensing contributions to the image's location $\theta^\mathrm{I}_{0,i}$ and proper motion $\dot{\theta}^\mathrm{I}_{0,i}$ are degenerate with uncertainties in the macrolens model and the large-scale proper motion, cf.~Eqs.~\ref{eq:mu} \&~\ref{eq:thetaI}, and not observable in practice. The acceleration from the macrolens and other large-scale structures is negligibly small---it is the first term in the Taylor expansion dominated by small-scale structures (assuming approximate scale invariance of the halo spectrum). The induced angular acceleration $\ddot{\theta}^\mathrm{I}_{0,i}$ is thus the leading observable for these ``large'' microhalos. In App.~\ref{app:mu-alpha-est}, we derive that for a survey over a time interval $t \in [-\tau/2,\tau/2]$ with equally spaced observations, the expected covariance matrix for an optimal estimator $\hat{\ddot{\theta}}^\mathrm{I}_{0,i}$ of the angular acceleration is: 
\begin{alignat}{2}
    \big\langle \hat{\ddot{\theta}}^\mathrm{I}_{0,p} \hat{\ddot{\theta}}^\mathrm{I}_{0,q} \big\rangle
    &= \int \frac{\dd \omega}{2\pi} \widetilde{C}^{\mathrm{I}}_{pq}(\omega) \omega^4 \mathcal{F}_2(\omega \tau) \label{eq:est_2_var_main}
\end{alignat}
with $\widetilde{C}^{\mathrm{I}}_{pq}$ from Eq.~\ref{eq:C}, and where the smearing factor $\mathcal{F}_2(\omega \tau)$, defined in Eq.~\ref{eq:F_2_smearing}, has asymptotic limits of 1 as $\omega \tau \to 0$ and $14400 \sin^2(\omega \tau/2)/(\omega \tau)^6$ for $|\omega \tau| \gg 1$.

Our second summary statistic is the differential angular-acceleration covariance, averaged over equally spaced measurements during a survey time $\tau$ and additive between the two sightlines:
\begin{align}
\left\langle \delta \ddot \theta_{0,p}^\mathrm{I,I'} \delta \ddot \theta_{0,q}^\mathrm{I,I'} \right\rangle &= \int \frac{\dd \omega}{2\pi} \widetilde{C}^\mathrm{I,I'}_{pq}(\omega) \omega^4 \mathcal{F}_2(\omega \tau) \label{eq:observable_2}
\end{align}
equal to the sum of Eq.~\ref{eq:est_2_var_main} for images $\mathrm{I}$ and $\mathrm{I'}$.

Figure~\ref{fig:concept} shows a simulated realization of the signal, illustrating the essence of both observables (Eqs.~\ref{eq:observable_1}~\&~\ref{eq:observable_2}) for the B1422+231 system described in Sec.~\ref{sec:B1422}. The two observables draw on complementary parts of the DFT spectrum of the $\delta\vect{\theta}^\mathrm{I,I'}$ time series, split at the band edge $2\pi/\tau$. The modes with $\omega \geq 2\pi/\tau$ complete at least one cycle over the survey and are individually accessible. Each is a zero-mean complex Gaussian whose expected power is set by $\widetilde{C}^\mathrm{I,I'}_{pq}(\omega)$. The stochastic lensing signal of Eq.~\ref{eq:observable_1} is the excess power in these modes \emph{above} the instrumental noise. Hence, the optimal detection statistic is an appropriately weighted sum of $\chi^2$ variables testing for an excess red-tilted variance on top of the (white) instrumental noise floor.

Modes with $\omega < 2\pi/\tau$---dominated by halos with crossing times longer than the survey time $\tau$---do not complete a full cycle over the observations and instead produce a slowly varying deflection. The constant and linear terms are absorbed into the image position and bulk proper motion, leaving a random angular acceleration as the leading observable, with covariance given by Eq.~\ref{eq:est_2_var_main} for each image. For a scale-invariant power spectrum, we derive $\widetilde{C} \propto \omega^{-4}$ in Eq.~\ref{eq:C_tilde_scale_invariant}; combined with the kernel $\omega^4\mathcal{F}_2(\omega\tau)$, this places $93\%$ of the integral of Eq.~\ref{eq:est_2_var_main} below $2\pi/\tau$. In the bottom panel of Fig.~\ref{fig:concept}, this contribution appears as the dashed quadratic fit to the image-pair trajectory. 

Both observables are excess variances. The signal from DM substructure lensing is essentially an inflated scatter of the data, above the expectation from instrumental noise only. The acceleration signal covariance $\mathcal{S}_{pq}$, Eq.~\ref{eq:observable_2}, is a $2\times2$ matrix in the sky-plane indices, as is the (assumed isotropic) instrumental noise of Eq.~\ref{eq:noise_2}: $\mathcal{N}_{pq}=\delta_{pq}\mathcal{N}$. Measuring the two acceleration components in units of the noise rms $\sqrt{\mathcal{N}}$ therefore reduces the noise covariance to the identity and the signal covariance to $\mathcal{S}_{pq}/\mathcal{N}$, whose eigenvalues $\lambda_1\geq\lambda_2$ are the fractional excess variances along the two principal directions on the sky. Throughout this work, ``SNR'' denotes the larger of the two, $\lambda_1$: the sensitivities quoted in Secs.~\ref{sec:sensitivity}~and~\ref{sec:structures} use that single eigenvalue and, for simplicity, (conservatively) discard the information in $\lambda_2$. The stochastic channel of Eq.~\ref{eq:observable_1} is treated identically, one DFT frequency at a time, and merely supplies a longer list of eigenvalues $\lambda_a$ and rescaled amplitudes $x_a$ (the observed power measured in units of expected noise power). The optimal search is then the quadratic statistic $\sum_a [\lambda_a/(1+\lambda_a)]\,x_a^2$, which counts each mode with a weight that approaches unity where the signal exceeds the noise and is downweighted elsewhere. The expected log-likelihood ratio separating the signal-plus-noise and noise-only hypotheses is $\langle 2\Delta \ln \mathcal{L}\rangle = \sum_a [\lambda_a - \ln(1+\lambda_a)]$ when the data do contain the signal, and the smaller $\sum_a [\ln(1+\lambda_a) - \lambda_a/(1+\lambda_a)]$ when they do not.

The relative acceleration of a single image pair consists of two modes, and even $\lambda_a\sim10$ in both amounts to no more than a $4\sigma$ detection (or $2\sigma$ exclusion)---an excess variance inferred from two numbers is intrinsically hard to establish. The stochastic channel of Eq.~\ref{eq:observable_1} contributes four real modes per resolved DFT frequency (two sky eigendirections times the two quadratures of the complex Fourier coefficient) and a system of $N_\mathrm{I}$ images offers $N_\mathrm{I}-1$ linearly independent pair separations. For three images, the closure $\delta\vect{\theta}^\mathrm{A,B}+\delta\vect{\theta}^\mathrm{B,C}+\delta\vect{\theta}^\mathrm{C,A}=0$ serves as a null channel, in which any excess variance signals systematics rather than substructure. Pairs sharing an image remain correlated, so their mutual covariance must enter a joint fit. The two expectations above also show that detection and exclusion are not symmetric for a variance signal. They coincide for $\lambda_a \ll 1$, where both reduce to $\sum_a \lambda_a^2/2$, but differ for $\lambda_a \gtrsim 1$. For example, for two acceleration components at $\lambda_a \sim 10$, $\sqrt{\langle 2\Delta\ln\mathcal{L}\rangle}\approx3.9$ on signal-containing data but only $\approx1.7$ on noise-only data (the two headline numbers above). The asymmetry originates in the different scaling of the two expectations with $\lambda_a$: the detection expectation grows linearly, $\lambda_a - \ln(1+\lambda_a) \to \lambda_a$, whereas the exclusion expectation grows only logarithmically, $\ln(1+\lambda_a) - \lambda_a/(1+\lambda_a) \to \ln \lambda_a - 1$. Noise alone is exponentially unlikely to fluctuate up to signal-sized amplitudes, so loud data are strongly incompatible with the noise-only hypothesis, provided the latter is well characterized. A signal realization that lands as ``quiet'' as pure noise is only power-law unlikely (its probability density is suppressed by $(1+\lambda_a)^{-1/2}$ per mode), so quiet data are compatible with substantial substructure along the line of sight. This asymmetry disappears when repeated over many image pairs and/or systems, since a conspiracy in which many independent substructure realizations all fluctuate low is exceedingly unlikely. 

A complete analysis should entail a time-domain likelihood over all epochs, images, and sky coordinates, with covariance $C = C_\mathrm{inst}+C_\mathrm{src}+C_*+C_\mathrm{DM}$ and with image positions, proper motions, the macrolensing model, and (potentially time-dependent) source morphology as nuisance parameters. We expect the additional modes and information to offset part of the loss from nuisance marginalization, but defer such an end-to-end likelihood analysis to future work. We will simply use the variance SNR as a forecasting proxy. 

The signal PSD falls steeply with frequency, so the information contained in the in-band stochastic lensing signal is highly compressible. Of the several hundred sampled modes in the realization of Fig.~\ref{fig:concept}, only the lowest-frequency ones carry lensing information---those whose deflection power rises above the astrometric floor $\sigma_{\delta\theta}$, i.e.~$\lambda_a\gtrsim1$. The redness of the spectrum implies that the effective number of modes is small---$N_\mathrm{eff}\equiv\sum_a \lambda_a/(1+\lambda_a)\approx6$ for the parameters of Fig.~\ref{fig:concept}---with the lowest few being the most informative (populated mostly by the acceleration). For the realization displayed in Fig.~\ref{fig:concept}, the estimator rejects the noise-only hypothesis at $6.4\sigma$ ($5.8\sigma$ expected). Even after fitting out the acceleration (in addition to the linear motion), a resolved-band excess ($\omega\ge2\pi/\tau$) remains, here at $3.1\sigma$ ($3.6\sigma$ expected; see bottom panel), indicative of anomalous low-frequency noise from stochastic lensing.

Section~\ref{sec:sensitivity} evaluates the signal and backgrounds for Eqs.~\ref{eq:observable_1} and~\ref{eq:observable_2} in more detail.

\section{Sensitivity and Noise} \label{sec:sensitivity}
\subsection{Galaxy Lens: B1422+231} \label{sec:B1422}
The quadruply imaged quasar B1422+231~\cite{Patnaik:1992} is a well-studied strong lens and a promising testbed for the stochastic, temporally resolved astrometric fluctuations (Eq.~\ref{eq:observable_1}) and angular-acceleration signatures (Eq.~\ref{eq:observable_2}) of this work. At redshift $z_\mathrm{S} = 3.62$, the quasar is lensed by a foreground galaxy~G at $z_\mathrm{L} = 0.34$, producing four macro-images, of which three (A, B, C) are exceptionally bright and highly magnified. We plot the macrolensing configuration of B1422+231 in Fig.~\ref{fig:diagram}, and describe our modeling in App.~\ref{app:B1422}.

B1422+231 has four properties favorable for detecting DM substructure. First, the stellar surface mass density at the positions of the bright images is relatively low (due to the external shear by the other member galaxies in the group to which G belongs~\cite{1998AJ....115....1T}), reducing microlensing contamination from the stellar population in the lens galaxy, as discussed below. Second, the strong magnification of the three bright images amplifies the astrometric lensing signal, directly through $B_{ij}^\mathrm{I}$, and indirectly through the magnified image motion $\widetilde{\mu}_i^\mathrm{I} = B_{ij}^\mathrm{I} \mu_j$ (cf.~Eq.~\ref{eq:C}). Third, the shear axes of images A--C are mostly aligned, nearly doubling (in power) the effects under consideration. Finally, B1422+231 is one of the brightest (lensed) quasars in the sky, which puts it in reach of future intensity interferometers.

\begin{figure}
    \centering
    \includegraphics[width=0.5\textwidth]{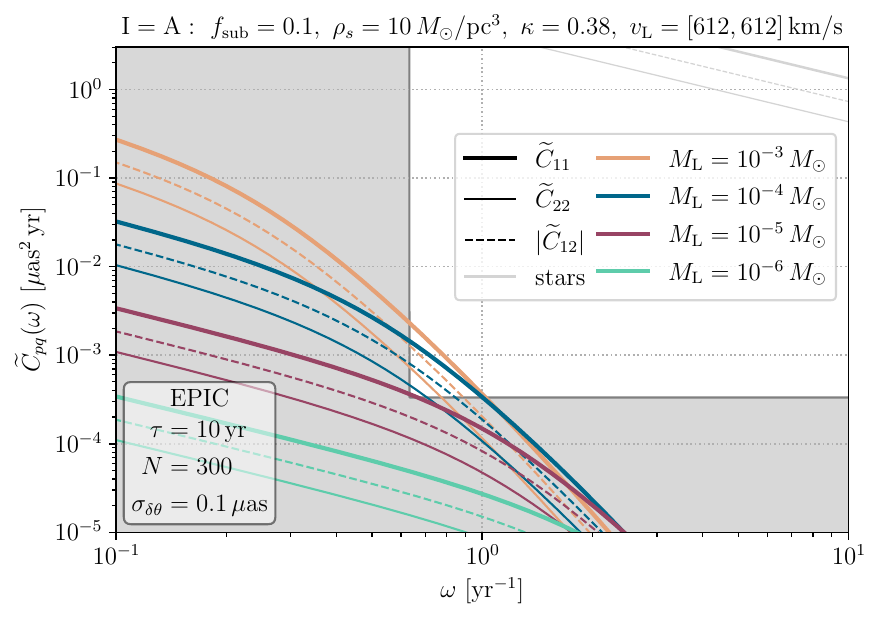}
    \caption{\nblink{code/sensitivity.ipynb} Power spectral density $\widetilde{C}^\mathrm{I}_{pq}$ of Eq.~\ref{eq:C} for image $\mathrm{I} = \mathrm{A}$ of B1422+231, as a function of angular frequency $\omega$. Different colors signify different bound halo masses $M_\mathrm{L}$ of truncated NFW subhalos ($r_\mathrm{t} = 2.3\,r_s$), all with a fixed scale density $\rho_s = 10\,M_\odot/\mathrm{pc^3}$ and substructure fraction of $f_\mathrm{sub} = 10\%$ of the convergence $\kappa = 0.38$. The spatial components $pq$ are shown as thick (11), thin (22), and dashed (12) lines, in the (RA, DEC) basis on the sky. The thin gray lines indicate the \emph{naive} power from stellar microlensing, but we caution that this contribution is not Gaussian; it should be subtracted star by star through intensity-interferometric image-shape fits~\cite{companionstars} (Sec.~\ref{sec:noise}). Gray regions depict inaccessible parameter space for the stochastic signal in a $\tau = 10\,\mathrm{yr}$ EPIC survey with $N=300$ observations with (relative) astrometric precision of $\sigma_{\delta \theta} = 0.1\,\mathrm{\mu as}$ due to band limitations ($\omega < 2\pi/\tau$) or astrometric precision (Eq.~\ref{eq:noise_1}).}
    \label{fig:Ctilde}
\end{figure}

\paragraph*{Expected motion.} The relative proper motion of Eq.~\ref{eq:mu} is dominated by the peculiar velocity of the lens galaxy: G orbits within a rich compact group whose line-of-sight velocity dispersion is measured at $\sigma_\mathrm{grp}\approx550\,\mathrm{km\,s^{-1}}$~\cite{Kundic:1997}, or $\approx 470\,\mathrm{km\,s^{-1}}$ with expanded group membership~\cite{Momcheva:2006}. We adopt $500\,\mathrm{km\,s^{-1}}$ per axis, combined in quadrature with a $\sim300\,\mathrm{km\,s^{-1}}$ (per axis) large-scale bulk flow of the group itself. The observer term is known: the CMB dipole velocity of $369.8\,\mathrm{km\,s^{-1}}$~\cite{Planck:2018} has a transverse component of $306\,\mathrm{km\,s^{-1}}$ at this line of sight (the dipole axis is misaligned by $56^\circ$). The source term, suppressed by $1/[(1+z_\mathrm{S})D_\mathrm{S}]$, contributes negligibly for a quasar-host peculiar velocity of $\sim300\,\mathrm{km\,s^{-1}}$ per axis. Together these give an expected $\langle|\vect{\mu}|^2\rangle^{1/2}\approx0.14\,\mathrm{\mu as\,yr^{-1}}$, i.e.~an equivalent bulk transverse velocity $(1+z_\mathrm{L})D_\mathrm{L}\langle|\vect{\mu}|^2\rangle^{1/2}\approx865\,\mathrm{km\,s^{-1}}$, which is our fiducial $|\vect{v}_\mathrm{L}|$. For the chosen direction in Fig.~\ref{fig:Ctilde} of $\vect{v}_\mathrm{L} = [612,612]\,\mathrm{km/s}$, it yields \emph{magnified} image motions $\widetilde{\mu}^\mathrm{I}\approx\lbrace1.0,1.4,0.5\rbrace\,\mathrm{\mu as\,yr^{-1}}$ for $\mathrm{I}=\mathrm{A,B,C}$. Averaging instead over the unknown direction of $\vect{\mu}$ at fixed $|\vect{v}_\mathrm{L}|$, i.e.~$\langle|\widetilde{\vect{\mu}}{}^\mathrm{I}|^2\rangle^{1/2} = |\vect{\mu}| \big[\mathrm{Tr}\big(B^\mathrm{I} B^{\mathrm{I}\mathsf{T}}\big)/2\big]^{1/2}$, gives rms values of $\lbrace0.7,1.0,0.5\rbrace\,\mathrm{\mu as\,yr^{-1}}$: the chosen direction is somewhat favorable for A and B, whose shear axes are mostly aligned with it. At present, there are large forecasting uncertainties associated with the unknown image proper motions $\widetilde{\vect{\mu}}{}^\mathrm{I}$, but these can be measured directly from the data, as we will argue below Eq.~\ref{eq:observable_2_parametric}.

We estimate an internal galaxy halo dispersion of $\sigma_\mathrm{int}\approx150\,\mathrm{km\,s^{-1}}$ (per axis). Yielding an expected relative proper motion of $\approx0.03\,\mathrm{\mu as\,yr^{-1}}$ in the lens, this contribution is negligible against $|B^\mathrm{I}_{pi}\mu_i|$, so the sweep rate is set by the magnified bulk motion. (The group members other than G, which enter the local lensing field only through the external shear $\gamma_\mathrm{ext}\approx0.17$, further add a macro-magnified moving-clump term $\sim\sqrt{2}\,\gamma_\mathrm{ext}\,\sigma_\mathrm{grp}\approx120\,\mathrm{km\,s^{-1}}$, expected to be roughly a $1\%$ quadrature correction.)

\subsection{Stochastic and Acceleration Signals}\label{sec:signals}
In Fig.~\ref{fig:Ctilde}, we plot $\widetilde{C}^\mathrm{I}_{pq}$ of Eq.~\ref{eq:C} for image $\mathrm{I} = \mathrm{A}$ and the fiducial bulk velocity of Sec.~\ref{sec:B1422}, implemented as $\vect{v}_\mathrm{L} = [612,612]\,\mathrm{km/s}$, i.e.~$|\vect{v}_\mathrm{L}| \approx 865\,\mathrm{km/s}$ (the known observer term and the residual source term having been folded into this effective lens-frame velocity, with $\vect{v}_\mathrm{S} = \vect{v}_\mathrm{o} = 0$). For illustrative purposes, we assume that a fraction $f_\mathrm{sub} = 0.1$ of the convergence is contained in halos of a single mass, i.e.~$\kappa^\mathrm{I}_\mathrm{L} = f_\mathrm{sub} \kappa^\mathrm{I}$ with $\kappa^\mathrm{I}$ the total convergence ($\kappa^\mathrm{A} = 0.38$, see App.~\ref{app:B1422}), for bound masses $M_\mathrm{L} = \lbrace 10^{-3}, 10^{-4}, 10^{-5}, 10^{-6} \rbrace \, M_\odot$, shown in $\lbrace \text{orange, blue, red, cyan} \rbrace$, respectively. Each such curve is a \emph{monochromatic} response characterizing the instrument's sensitivity to halos of one mass carrying the full substructure convergence, and is not by itself a $\Lambda$CDM population forecast (see Sec.~\ref{sec:pure_CDM} for the convolution over an extended mass spectrum). 

We model the subhalos as NFW profiles with a Gaussian cutoff at the tidal radius $r_\mathrm{t} \equiv x_\mathrm{t}\, r_s$, i.e.~with a profile $\rho(r) = 4\rho_s\,(r/r_s)^{-1}(1+r/r_s)^{-2}\,e^{-r^2/2r_\mathrm{t}^2}$ and scale density $\rho_s$ equal to the density of the untruncated profile at $r_s$. The bound mass is $M_\mathrm{L} = 16\pi\rho_s r_s^3\, m_\mathrm{t}(x_\mathrm{t})$, with $m_\mathrm{t}(x_\mathrm{t}) \equiv \int_0^\infty \dd x\, x\,(1+x)^{-2}\,e^{-x^2/2x_\mathrm{t}^2}$; the untruncated profile encloses $16\pi\rho_s r_s^3\, m(x)$ inside $x\,r_s$, with $m(x) \equiv \ln(1+x) - x/(1+x)$. The form factor of Eq.~\ref{eq:form}, with $\gamma_\mathrm{L} = r_s/D_\mathrm{L}$ and $y \equiv k\gamma_\mathrm{L}$, is
\begin{align}
    F(y) = \frac{1}{m_\mathrm{t}(x_\mathrm{t})\,y}\int_0^\infty \dd x\, \frac{\sin(xy)\,e^{-x^2/2x_\mathrm{t}^2}}{(1+x)^2}. \label{eq:F_tNFW}
\end{align}
It tends to unity for $x_\mathrm{t}y \ll 1$ and falls as $F \simeq [m_\mathrm{t}(x_\mathrm{t})\,y^2]^{-1}$ for $y \gg 1$, the transform of the $1/r$ cusp; the $\phi$ integral of Eq.~\ref{eq:C} then has no closed form, and we evaluate it numerically. For reference, a 3D Gaussian profile $\rho \propto \exp\lbrace-r^2/2r_\mathrm{L}^2\rbrace$ has $F(k\gamma_\mathrm{L}) = \exp\lbrace-(k\gamma_\mathrm{L})^2/2\rbrace$ with $\gamma_\mathrm{L} = r_\mathrm{L}/D_\mathrm{L}$, for which the isotropic part of that integral is $\pi\,\mathrm{erfc}(|\omega|\gamma_\mathrm{L}/\widetilde{\mu}^\mathrm{I})$. This exponential cutoff removes the small-scale power of the NFW cusp: at the same bound mass, convergence, and $r_\mathrm{L} = r_s$, 3D Gaussian halos with $r_\mathrm{L} = r_s$ but without a cusp would lower the peak acceleration SNR of Sec.~\ref{sec:prospects} by a factor of 1.8.

We estimate the truncation radius from the King radius~\cite[Eq.~17]{Moline:2017}, where the mean density of the subhalo's untruncated profile equals three times that of its host inside the subhalo's orbital radius $R$,
\begin{align}
    \overline{\rho}_\mathrm{NFW}(<r_\mathrm{t}) = \frac{12\,\rho_s\, m(x_\mathrm{t})}{x_\mathrm{t}^3} = 3\,\overline{\rho}_\mathrm{host}(<R). \label{eq:King}
\end{align}
This condition only depends on $\rho_s$, as denser subhalos are more resistant to tides and are thus truncated farther out. Reference~\cite{Moline:2017} applies the same truncation radius to the concentration relation that we use in Sec.~\ref{sec:pure_CDM}, which describes the inner profiles of present-day subhalos. For B1422+231, we take the smallest radius $R$ that a subhalo on the image sightlines can have, the projected distance $R = 4.7\,\mathrm{kpc}$ of image B from G, and the isothermal model of G, for which $\overline{\rho}_\mathrm{host}(<R) \approx 0.17\,M_\odot\,\mathrm{pc^{-3}}$ (App.~\ref{app:B1422}).\footnote{In general, there will be a distribution of tidal radii, mainly depending on the distribution of subhalo pericenters along the line of sight. This is a simple choice and, at fixed $f_\mathrm{sub}$, an optimistic one: truncating each subhalo at its King radius for the median distance $R \approx 9\,\mathrm{kpc}$ (below) would lower the peak SNR of Sec.~\ref{sec:prospects} from 11.7 to 7.9.} A subhalo with $\rho_s = 1\,M_\odot\,\mathrm{pc^{-3}}$ is then truncated at $r_\mathrm{t} \approx 2.3\,r_s$, just outside the radius of peak circular velocity of the untruncated profile ($2.16\,r_s$), and keeps $M_\mathrm{L}/M_{200} \approx 0.15$ of the would-be virial mass of the subhalo. We adopt $x_\mathrm{t} = 2.3$ for all subhalos in both systems for simplicity. At the line-of-sight-weighted median distance, $R \approx 9\,\mathrm{kpc}$, the King radius would be 1.9 times larger.
The substructure fraction counts only the present-day bound mass $M_\mathrm{L}$ (a fraction 0.72 of which lies inside $r_\mathrm{t}$), i.e.~with a subhalo convergence of $\kappa^\mathrm{I}_\mathrm{L} = f_\mathrm{sub}\kappa^\mathrm{I}$ in truncated subhalos. With $M_\mathrm{L}/M_{200} \approx 0.15$, our fiducial $f_\mathrm{sub} = 0.1$ corresponds to subhalos whose would-be virial mass makes up about two thirds of the convergence at the image positions. 

Subhalos whose angular radius $\gamma_\mathrm{L}$ matches the image angular displacement $\Delta \theta^\mathrm{I} = \widetilde{\mu}^\mathrm{I} \tau$---i.e.~the crossing time of the image across the halo's radius equals the survey time---yield the dominant stochastic lensing effects, for a nearly scale-invariant subhalo spectrum (with $\kappa_\mathrm{L}$ and $\rho_s$ constant), as argued below Eq.~\ref{eq:observable_2}. High-mass halos have stronger lensing effects as $\omega \to 0$ but suffer form factor suppression ($\omega \gamma_\mathrm{L} / \widetilde{\mu}^\mathrm{I} \gg 1$) at observable frequencies $\omega \geq \omega_\mathrm{min} = 2\pi/\tau$ due to their larger size (orange curve in Fig.~\ref{fig:Ctilde}). Low-mass halos escape this suppression, but their lensing effects may simply fall below the astrometric precision floor (cyan curve in Fig.~\ref{fig:Ctilde}). Therefore, halos with scale radii of $r_s = \gamma_\mathrm{L} D_\mathrm{L} \sim  \widetilde{\mu}^\mathrm{I} \tau D_\mathrm{L} \sim  0.05 \, \mathrm{pc} \, (\tau / 10\,\mathrm{yr})$ will yield the strongest signal. Such structures are predicted to have bound masses of $\mathcal{O}(10^{-3}\,M_\odot)$ in $\Lambda$CDM cosmology (using $\rho_s \sim 1\,M_\odot/\mathrm{pc}^3$), some 9 to 11 orders of magnitude lighter than the smallest dark matter structures detected to date~\cite{Powell:2025,DalalKochanek:2002,Gilman:2020,Hsueh:2020}. Along the $\Lambda$CDM scale-density relation, the acceleration SNR on star-free data stays above unity down to $M_\mathrm{L} \approx 3\times10^{-6}\,M_\odot$ (Sec.~\ref{sec:prospects}).

\begin{figure}
    \centering
    \includegraphics[width=0.5\textwidth]{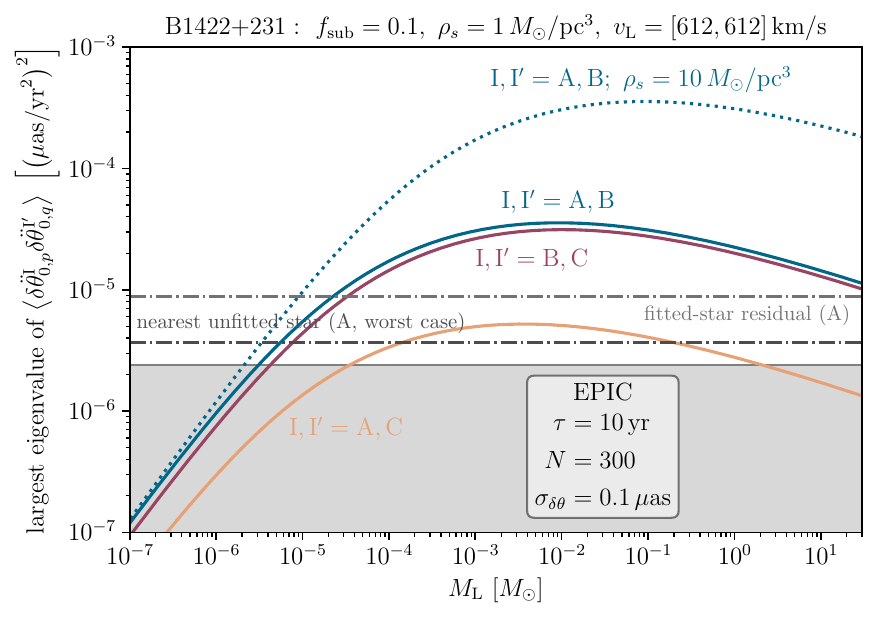}
    \caption{\nblink{code/sensitivity.ipynb} Differential acceleration covariance $\big\langle \delta \ddot \theta_{0,p}^\mathrm{I,I'} \delta \ddot \theta_{0,q}^\mathrm{I,I'} \big\rangle$ (largest eigenvalue) from Eq.~\ref{eq:observable_2} as a function of bound subhalo mass $M_\mathrm{L}$ among the three brightest images A,B,C of quasar B1422+231, for truncated NFW subhalos with a fixed scale density of $\rho_s = 1\,M_\odot / \mathrm{pc}^3$ (solid). The same quantity is shown for the A,B image pair at a higher scale density $\rho_s = 10\,M_\odot / \mathrm{pc}^3$ (dotted blue). 
    The acceleration sensitivity of Eq.~\ref{eq:noise_2} is depicted by the upper boundary of the gray region, for the same EPIC parameters as in Fig.~\ref{fig:Ctilde}. Dash-dotted lines mark the residual stellar-microlensing levels at image A (Sec.~\ref{sec:noise}), obtained by rescaling the campaign of Ref.~\cite{companionstars} to the survey parameters assumed here ($\theta_\mathrm{fit} \approx 17\,\mathrm{\mu as}$): the worst-case acceleration from the nearest unfitted star, and the fitted-star subtraction residual.}
    \label{fig:acc}
\end{figure}

The acceleration statistic in Eq.~\ref{eq:observable_2} provides the strongest forecast over a broad halo-mass range. The PSD $\widetilde{C}^\mathrm{I}_{pq}(\omega)$ has large support at low frequencies, giving sizable corrections to the image separation $\delta \vect{\theta}^\mathrm{I,I'}$ and the differential proper motion $\delta \dot{\vect{\theta}}{}^\mathrm{I,I'}$, which are dominated by the lensing of the largest (sub)halos and degenerate with the macrolensing model and the intrinsic proper motion of the images. However, the differential acceleration covariance of Eq.~\ref{eq:observable_2} is dominated by small subhalos: a parametric estimate for halos with $\gamma_\mathrm{L}/\widetilde{\mu} \gg \tau$ yields:
\begin{align}
\big \langle \big( \delta \ddot \theta_0^\mathrm{I,I'} \big)^2 \big \rangle &\sim B^2 \kappa_\mathrm{L} \theta_\mathrm{E,L}^2 \frac{\widetilde{\mu}^4}{\gamma_\mathrm{L}^4} \ln\frac{\gamma_\mathrm{L}}{\widetilde{\mu}\tau} \nonumber\\
&\sim G \frac{B^6 \kappa_\mathrm{L} v_\mathrm{L}^4}{(1+z_\mathrm{L})^4 D_\mathrm{L}}  \frac{\rho_s^{4/3}}{M_\mathrm{L}^{1/3}} \ln\frac{\gamma_\mathrm{L}}{\widetilde{\mu}\tau}, \label{eq:observable_2_parametric}
\end{align}
where the logarithm counts the frequencies between the crossing rate $\widetilde{\mu}/\gamma_\mathrm{L}$ and $1/\tau$, a consequence of the $1/r$ NFW cusp. Each $e$-fold in that frequency band generates equal contributions from the $1/r$ cusp, as $\widetilde{C} \propto F^2/\omega \propto \omega^{-5}$ multiplied with the $\omega^4$ weight of Eq.~\ref{eq:observable_2} yields a logarithm upon integration. For a mass-independent scale density $\rho_s$, the acceleration signal is therefore dominated by the \emph{lightest} such subhalos, but decouples slowly at larger masses. Halos with a more cored profile (e.g.~the 3D Gaussian profile below Eq.~\ref{eq:F_tNFW}) would lose the logarithmic enhancement in Eq.~\ref{eq:observable_2_parametric}; structures with cuspier profiles (e.g.~prompt cusps) would have a stronger enhancement. We plot the largest eigenvalue of the differential acceleration covariance in Fig.~\ref{fig:acc} for the three brightest image pairs of B1422+231.

The steep scaling $\langle ( \delta \ddot \theta_0 )^2 \rangle \propto B^6 v_\mathrm{L}^4$ in Eq.~\ref{eq:observable_2_parametric} singles out the most highly magnified images, in the fastest-moving lenses, as by far the most promising targets. This strong dependence also makes the forecast sensitive to the macrolensing Jacobian $B^\mathrm{I}$ and the transverse velocity: our fiducial $|\vect{v}_\mathrm{L}| \approx 865\,\mathrm{km/s}$ includes the known CMB-dipole observer term, but the dominant, stochastic group-orbit contribution leaves it uncertain at the order-unity level, and there could be partial cancellations or enhancements among the terms in the sum of Eq.~\ref{eq:mu}. The same intensity-interferometric campaign would, however, resolve the relative proper motions of the images~\cite{Galanis:2023gef} and even those of the microlensing stars~\cite{companionstars}, so the relative proper motions can be measured directly and will not be a significant uncertainty on the signal. At fixed \emph{observed} $\widetilde{\mu}^\mathrm{I}$, the acceleration variance scales only as $B^2$.

Intensity interferometry resolves the magnified angular \emph{shape} of each quasar image, which directly fixes the element-wise ratios of $B^\mathrm{I}_{ij}$. Combined with the flux ratios and the relative image proper motions, this determines all components of $B^\mathrm{I}$ and the bulk proper motion up to the well-known mass-sheet/source-position family of transformations~\cite{FalcoGorensteinShapiro:1985,SchneiderSluse:2014}: a rescaling $\vect{\beta}\to\lambda\vect{\beta}$, $B^\mathrm{I}\to B^\mathrm{I}/\lambda$ leaves image positions, shapes, flux ratios, and image motions invariant. The absolute scale $\lambda$ must be determined externally, e.g.~from time delays. The dominant conversion factors between the substructure signal and our proposed observables can thus be largely calibrated by the data themselves.

\subsection{Instrumental Noise and Stellar Backgrounds} \label{sec:noise}
\paragraph*{Instrumental noise.}
For an astrometric survey with $N = \tau / \Delta t$ observation epochs with relative light-centroiding precision of $\sigma_{\delta \theta}$ (in each direction) equally spaced by $\Delta t$ over a survey time $\tau$, the instrumental noise covariance is:
\begin{align}
    \left.\widetilde{C}^{\mathrm{I,I'}}_{pq}(\omega_n)\right|_\text{noise} 
    &= \delta_{pq} \frac{\sigma_{\delta \theta}^2 \tau}{ N} \label{eq:noise_1} \\
    \left.\left\langle \delta \ddot \theta_{0,p}^\mathrm{I,I'} \delta \ddot \theta_{0,q}^\mathrm{I,I'} \right\rangle\right|_\text{noise}
    &= \delta_{pq} 720 \frac{\sigma_{\delta \theta}^2}{N\tau^4} \label{eq:noise_2}
\end{align}
for the discrete Fourier modes with angular frequencies $\omega_n = 2 \pi n / \tau$ ($n \in [0,1,\dots, N-1]$), and the time-averaged acceleration, respectively (see App.~\ref{app:mu-alpha-est} for a derivation). We plot Eqs.~\ref{eq:noise_1}~and~\ref{eq:noise_2} in Figs.~\ref{fig:Ctilde}~and~\ref{fig:acc}, respectively, as the upper boundaries of the gray regions. The assumed per-epoch precision $\sigma_{\delta \theta} = 0.1\,\mathrm{\mu as}$ could be attained by an EPIC Phase~III facility~\cite{Galanis:2023gef}, and is comparable to the $0.04$--$0.5\,\mathrm{\mu as}$ per-epoch centroid precision forecast for this system \emph{after} simultaneous intensity-interferometric image-model fits in the companion paper~\cite{companionstars}. Since the instrument-limited variance SNR scales as $\sigma_{\delta\theta}^{-2}$, the forecasts below rescale trivially to any other assumed precision. The instrumental noise floor would lead to an acceleration variance of $2.4\times10^{-6}\,\mathrm{\mu as^2\,yr^{-4}}$ for our fiducial EPIC survey parameters. Equations~\ref{eq:noise_1}~and~\ref{eq:noise_2} idealize the noise as white, isotropic, and uncorrelated between epochs and images; a real sub-$\mathrm{\mu as}$ program will likely contend with correlated instrumental systematics, which we cannot foresee here.

\paragraph*{Stellar microlensing.} 
Astrometric microlensing from stars in the lens galaxy G is expected to be the primary astrophysical background. Microlensing by lens-galaxy stars is a mature subfield in the \emph{photometric} domain, long used to probe quasar accretion-disk and X-ray emission-region structure~\cite{wambsganss1990interpretation,webster1991interpreting,rauch1991microlensing,jaroszynski1992microlensed,kochanek2004quantitative,mosquera2011microlensing,moustakas2019astro2020} and to weigh the stellar versus DM content of lens galaxies~\cite{Schechter:2002,Mediavilla:2009,JimenezVicente:2015,Oguri:2014}, down to planet-mass perturbers~\cite{Dai:2018}. Despite their lower convergence $\kappa_*^\mathrm{I}$ compared to the total $\kappa^\mathrm{I}$ for images A,B,C of our benchmark system B1422+231 (App.~\ref{app:B1422}), stellar lenses can have an outsized contribution because they are essentially point-like and more massive than the DM lenses under consideration. In Fig.~\ref{fig:Ctilde}, we plot their naive PSD for $M_* = 0.3\,M_\odot$ and $\kappa_* = 0.05$, roughly the stellar convergence at image A (Tab.~\ref{tab:macro}; cf.~Refs.~\cite{Dogruel:2020,Biggs:2023}). 

However, because stars are much fewer in number and their lensing deflection is not perturbative for close impact parameters, the PSD is not a good measure of their effect. Our companion paper~\cite{companionstars} shows that both the masses and locations of (significant) point-like microlenses can be measured---and their centroid contribution subtracted---out to an angular impact parameter of $\theta_\mathrm{fit} \approx 10 \,\mathrm{\mu as}$ from the main quasar image for $M_* \gtrsim 0.3\,M_\odot$ ($\approx5$--$16\,\mathrm{\mu as}$ for $0.1$--$0.8\,M_\odot$) over its 33-epoch, eight-year campaign, through time-dependent variations of the image's \textit{shape} (the squared visibility measured by intensity interferometry). Combined with photometry, such a multi-year campaign can determine the mass of a stellar microlens at small impact parameter to better than 10\%, localize a close trajectory to $\mathcal{O}(0.1\,\mathrm{\mu as})$, and, for favorable trajectories, constrain its transverse velocity at the order-unity level. The expected number of stars within the fitting region, $N_\mathrm{fit}^\mathrm{I} = \kappa_*^\mathrm{I} \Sigma_\mathrm{cr} \pi \theta_\mathrm{fit}^2 D_\mathrm{L}^2 / M_*$, is $\approx 2.5$ at image A for that campaign and a fiducial mass of $M_* = 0.3 \, M_\odot$. 
Joint inference of \emph{many} stars will likely be limited to the one or two most significant perturbers, with the remainder constrained only in aggregate, since inference power decreases with increasing number of (in this case, stellar microlens) parameters, as forecast for intensity interferometry in different contexts in Refs.~\cite{Dalal:2024,Chen:2025bch}. A poor goodness-of-fit measure for a one- or two-star model would flag those images and epochs whose subtraction cannot be trusted, and should be dropped.

Merely \emph{detecting} the presence of a microlens is much easier than measuring its mass and trajectory, so an image and epoch with no significant perturber can be certified free of stellar contamination out to some radius even where the parameters of a detected star would be poorly constrained: a good $\chi^2$ of the unlensed (macro-magnified) disk model against the measured squared visibilities excludes any point mass close enough to distort the image shape, and thus capable of large astrometric deflections~\cite[Fig.~8]{companionstars}. Because the stellar lenses are few and Poisson-distributed, a subset of images and epochs will be \emph{certifiably star-free} and could approach our instrument-limited forecast. Cluster lenses, such as the one in Sec.~\ref{sec:cluster}, should realize this limit along a substantial fraction of sightlines.

Two residual stellar microlensing noise terms add to the instrumental imprecision in the acceleration channel: the worst-case acceleration from the dominant \emph{unfitted} star just beyond $\theta_\mathrm{fit}$, bounded by $|\ddot \theta_i^\mathrm{I}|^2 < |2 B^\mathrm{I}_{ij} \hat{n}_j \theta_{\mathrm{E},*}^2 (\widetilde{\mu}^\mathrm{I})^2 / \theta_\mathrm{fit}^3 |^2$ with $\hat{n}_j$ a unit vector~\cite[Eq.~3.24]{VanTilburg:2018ykj}, and the residual from imperfect subtraction of the \emph{fitted} stars, with best-case acceleration variance $\approx 2\times10^{-4}\,\mathrm{\mu as^2\,yr^{-4}}$ in Ref.~\cite{companionstars}. Both fall steeply as observations improve, because the fitting radius itself grows: a star is detected once its contribution to the shear $\gamma_* = (\theta_{\mathrm{E},*}/\theta_\mathrm{sep})^2$, measured with per-epoch fractional precision $\propto 1/\mathrm{SNR}_1$, exceeds a fixed threshold $\chi^2 \propto N \gamma_*^2\, \mathrm{SNR}_1^2$, so that
\begin{align}
    \theta_\mathrm{fit} \propto \theta_{\mathrm{E},*}\, \mathrm{SNR}_1^{1/2}\, N^{1/4}. \label{eq:theta_fit}
\end{align}
The unfitted-star variance falls as $\theta_\mathrm{fit}^{-6} \propto \mathrm{SNR}_1^{-3} N^{-3/2}$, much faster than the instrumental floor of Eq.~\ref{eq:noise_2}. The fitted-star residual centroid variance scales as $\propto \mathrm{SNR}_1^{-2}/(N \tau^4)$, and thus tracks the instrumental imprecision at a fixed multiple. Rescaling the campaign of Ref.~\cite{companionstars} to the survey parameters assumed here gives $\theta_\mathrm{fit} \approx 17\,\mathrm{\mu as}$ and acceleration residuals of $ 4\times10^{-6} \,\mathrm{\mu as^2\,yr^{-4}}$ (unfitted, worst case, with the perturber placed exactly at $\theta_\mathrm{fit}$) and $9\times10^{-6}\,\mathrm{\mu as^2\,yr^{-4}}$ (fitted)---within a factor of a few of the instrumental floor of $2.4\times10^{-6}\,\mathrm{\mu as^2\,yr^{-4}}$. These residuals extrapolate the model fits of Ref.~\cite{companionstars}, with its assumptions of isolated point lenses and a simple quasar accretion disk model, to a longer campaign, omitting the degradation from marginalizing over the poorly constrained parameters of a realistic multi-star field. Caustic-network interference at high-$\kappa_*$ images (the ``micro-image swarm'' of Ref.~\cite{2026arXiv260502181M}) could likewise make the residual plateau above these scalings. Most of the information about a star's mass and trajectory is obtained near its closest approach, so one can pin its mass and trajectory during that interval and extrapolate its centroid contribution across the rest of the campaign. An analysis of a realistic multi-star field is left to future work. Stellar microlensing residuals will likely limit the discovery reach of the most ambitious astrometric surveys, while some star-free quasar image pairs retain the instrument-limited sensitivity. 

\paragraph*{Intrinsic photocenter wander.} The quasar's own emission centroid is not perfectly stationary: inhomogeneous accretion-disk variability (localized flares, reprocessing) displaces the flux-weighted centroid by at most the variable-flux fraction times the source size, $|\delta \vect{\beta}_\mathrm{src}| \lesssim f_\mathrm{var}\,\theta_\mathrm{src} \sim 0.01\,\mathrm{\mu as}$ for $f_\mathrm{var}\sim 0.2$ and $\theta_\mathrm{src}\approx 0.04\,\mathrm{\mu as}$ (App.~\ref{app:B1422}). Magnified by $B^\mathrm{I}$, this wander can approach the per-epoch astrometric precision at the most magnified images, and, because the distortion matrices differ between images, it does not cancel in the raw difference $\vect{\theta}^\mathrm{I} - \vect{\theta}^\mathrm{I'}$. 

However, intrinsic source motion is \emph{correlated} among the images, unlike the astrometric microlensing signal from DM substructure (or stars). A two-dimensional source displacement $\delta\vect{\beta}(t)$ is mapped to all images after distortion by the known matrices $B^\mathrm{I}$ and offset by the corresponding inter-image time delay. Because any photocenter wander is common, the reweighted difference $\vect{\theta}^\mathrm{I} - B^\mathrm{I}(B^\mathrm{I'})^{-1}\vect{\theta}^\mathrm{I'}$ (a combination invariant under the mass-sheet rescaling of Sec.~\ref{sec:signals}) cancels it identically while retaining the independent lensing jitter of both sightlines. A third bright image overdetermines $\delta\vect{\beta}(t)$ so that it can be measured outright. The wander is moreover correlated with photometric variability, whereas the lensing-induced jitter is statistically independent between images.  

The predicted delays among the three bright images of B1422+231 are $\lesssim1\,\mathrm{day}$~\cite{Biggs:2023}, and image C trails image B by only $2$--$3$ days in SDSS J1029+2623~\cite{Fohlmeister:2013}. Such delays are compatible with a realistic monitoring cadence, and comparable to the mild microlensing-induced variability of the delays themselves~\cite{TieKochanek:2018}. (Only the wide pair A--(B,C) of the latter system has an inconvenient $744\pm10$-day delay~\cite{Fohlmeister:2013}.) Intrinsic centroid motion is therefore a controllable background, far less severe than the stellar microlensing above, and it mildly favors compact optical continua and systems with at least three bright images.

\begin{figure}
    \centering
    \includegraphics[width=0.5\textwidth]{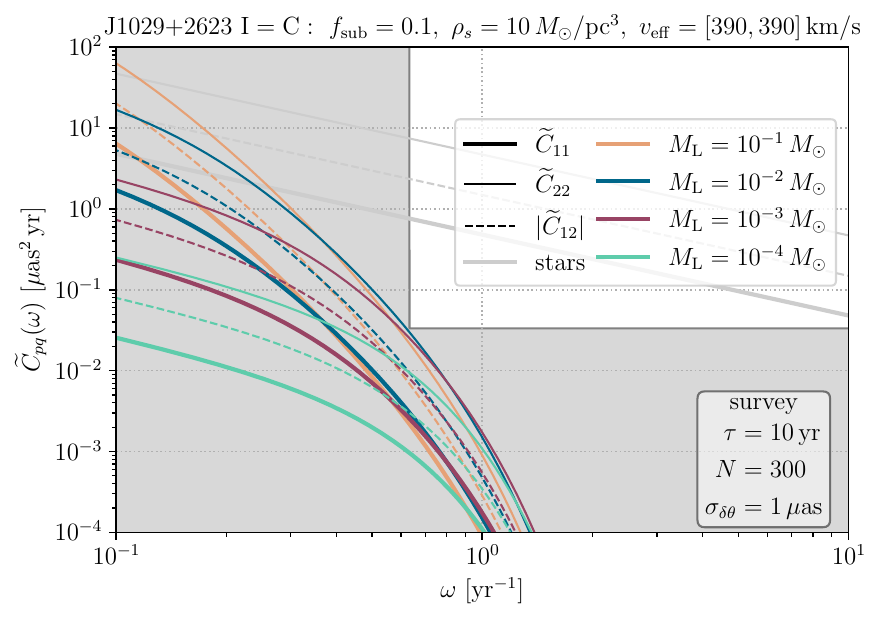}
    \caption{\nblink{code/sensitivity.ipynb} Power spectral density $\widetilde{C}^\mathrm{I}_{pq}$ for the bright image $\mathrm{I}=\mathrm{C}$ of the \emph{cluster}-lensed quasar SDSS J1029+2623, the direct analog of Fig.~\ref{fig:Ctilde}. Bound halo masses $M_\mathrm{L} = \lbrace 10^{-1},10^{-2},10^{-3},10^{-4}\rbrace\,M_\odot$ (orange, blue, maroon, cyan) of truncated NFW subhalos ($r_\mathrm{t} = 2.3\,r_s$) at $\rho_s = 10\,M_\odot/\mathrm{pc^3}$, $f_\mathrm{sub}=0.1$, and the fiducial effective transverse velocity $|\vect{v}_\mathrm{L}^\mathrm{eff}|\approx5.5\times10^2\,\mathrm{km/s}$ (bulk motion plus the moving member-halo term, both magnified by $B^\mathrm{I}$; Sec.~\ref{sec:cluster}). Note the much lower expected stellar microlensing background (light gray; $\kappa_*=0.003$, App.~\ref{app:J1029}) than in the galaxy lens. The gray region is inaccessible to a fiducial $\tau=10\,\mathrm{yr}$, $N=300$ differential-astrometry survey with $\sigma_{\delta\theta}=1\,\mathrm{\mu as}$.}
    \label{fig:Ctilde-J1029}
\end{figure}

\begin{figure}
    \centering
    \includegraphics[width=0.5\textwidth]{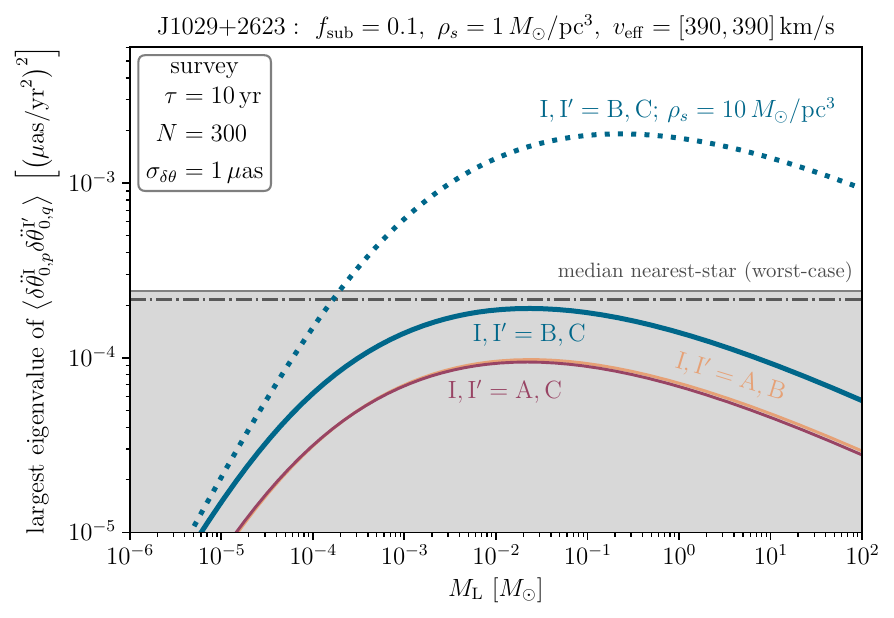}
    \caption{\nblink{code/sensitivity.ipynb} 
    Differential acceleration covariance (largest eigenvalue) for the image pairs of SDSS J1029+2623, the analog of Fig.~\ref{fig:acc}. 
    Solid curves show the three image pairs at $\rho_s = 1\,M_\odot/\mathrm{pc^3}$; the dotted curve is the bright B,C pair at $\rho_s = 10\,M_\odot/\mathrm{pc^3}$. 
    Finite-size effects of the optical source  ($\theta^\mathrm{I}_\mathrm{src}\approx1.3\,\mathrm{\mu as}$) are included through the form factor of Eq.~\ref{eq:finite_source}, which acts as a high-pass filter in halo mass. 
    The sweep rate uses the effective velocity $v_\mathrm{L}^\mathrm{eff}\approx550\,\mathrm{km\,s^{-1}}$ (bulk plus moving member-halo term at the fiducial $f_\mathrm{gal}=0.2$; Sec.~\ref{sec:cluster}).
    The B,C peak signal exceeds that of the galaxy lens B1422+231 (Fig.~\ref{fig:acc}) by about a factor of 5.
    The gray region is the acceleration sensitivity (Eq.~\ref{eq:noise_2}) of a $\tau=10\,\mathrm{yr}$, $N=300$ differential-astrometry survey with $\sigma_{\delta\theta}=1\,\mathrm{\mu as}$, which the bright B,C pair nearly reaches at the fiducial $\rho_s=1\,M_\odot/\mathrm{pc^3}$ (solid blue; peak variance $\mathrm{SNR}\approx0.8$) and clears at 10 times higher halo densities (dotted blue; $\mathrm{SNR}\approx8$).
    The dash-dotted line is the worst-case acceleration from the \emph{median} nearest star: a single, identifiable perturber, absent at any one epoch along half of all sightlines (Sec.~\ref{sec:cluster}).}
    \label{fig:acc-J1029}
\end{figure}

\subsection{Cluster Lens: SDSS J1029+2623} \label{sec:cluster}
Strong lensing by a galaxy \emph{cluster} partly circumvents the stellar microlensing background just discussed and could potentially probe heavier substructure. The three-image quasar SDSS J1029+2623 at $z_\mathrm{S}=2.20$, split by $22.5''$ by a foreground cluster at $z_\mathrm{L}=0.59$~\cite{Inada:2006ph,Oguri:2012vg,Acebron:2022}, is a prototypical example: in the macrolensing model of Refs.~\cite{Acebron:2022,Acebron:2024} (App.~\ref{app:J1029}), the close pair B,C straddles the tangential critical curve and is highly magnified ($|A|\approx22$), while image A is more modestly magnified ($A\approx6.6$). It is one of only about eight cluster lenses (nine quasars, one cluster lensing a pair) found to date, all since 2003~\cite{Inada:2003,Inada:2006ph,Dahle:2013,Shu:2018,Solhaug:2026,Wu:2025curling}; the class should expand rapidly with Rubin, \textit{Euclid}, and \textit{Roman}~\cite{oguri2010gravitationally,Wu:2025curling}. Among the new systems, some star-free lenses will be well suited to setting clean DM limits, and some brighter members may even be accessible to intensity interferometry (Sec.~\ref{sec:discussion}).

The images of SDSS J1029+2623 form about $100\,\mathrm{kpc}$ from the cluster center, where the only stars are the diffuse intracluster light. Its measured surface density in comparable clusters~\cite{Morishita:2017} gives a stellar convergence $\kappa_*^\mathrm{I}\approx0.003$ (App.~\ref{app:J1029}), more than an order of magnitude below that of B1422+231, with a factor-of-few uncertainty. Rarity alone does not eliminate the stellar background in the acceleration channel: the bright images sweep the stellar field at $\widetilde{\mu}^\mathrm{I}\approx1\,\mathrm{\mu as\,yr^{-1}}$, so a star at the typical nearest-neighbor separation of $15$--$20\,\mathrm{\mu as}$ would rival the DM signal (Fig.~\ref{fig:acc-J1029}). Under this fiducial $\kappa_*^\mathrm{I}$ estimate, half of all sightlines are clean at any one epoch, i.e.~no star lies within $16\,\mathrm{\mu as}$; over the campaign, as each image moves by $\widetilde{\mu}^\mathrm{I}\tau\approx11\,\mathrm{\mu as}$, that fraction falls to 37\%. Such star-free pairs could yield powerful limits on the DM power spectrum. 

When a star \emph{is} present, its smooth, nearly quadratic centroid trend over a decade is nearly degenerate with the DM-induced angular acceleration itself. The latter is predominantly carried by the lowest temporal Fourier modes of the centroid, so fitting the secular acceleration trend out of the centroid time series alone (without the intensity-interferometric image-shape information that can identify stars in the galaxy lens) would eliminate the acceleration signal of Eq.~\ref{eq:observable_2} but partially retain the higher-frequency stochastic lensing signal from low-mass halos with crossing times short compared to $\tau$. The excess variance of Eq.~\ref{eq:observable_1} at integer harmonics above the fundamental $2\pi/\tau$ is more resilient, since it cannot be fully reproduced by a quadratic (or low-order) polynomial fit. The accessible bandwidth is set by the finite source: the image sweeps $\widetilde{\mu}^\mathrm{I}\tau/\theta^\mathrm{I}_\mathrm{src}\approx8$ times its own magnified source size over 10 years. If the quasar continuum region is well described by a standard scale-free thin-accretion-disk brightness profile (temperature $T\propto R^{-3/4}$ falling polynomially with radius $R$), then the image visibility-squared of Eq.~\ref{eq:finite_source} has a power-law tail at high frequencies, $|\widetilde{W}^\mathrm{I}|^2\propto\omega^{-5/2}$ (the quasar form factor $\mathcal{F}_\mathrm{q}$ of Ref.~\cite[Eq.~2.42]{Galanis:2023gef}). Folded with the red DM spectrum of $\widetilde{C}\propto1/|\omega|$, the small-halo per-mode variance then decays only as $\propto\omega^{-7/2}$, whereas the Fourier power from a quadratic polynomial falls as $\propto \omega^{-4}$. The finite size of the source suppresses the high harmonics but leaves several usable modes with sensitivity to the small-halo stochastic signal. A robust discovery in a cluster lens requires marginalizing over stellar microlensing, which will likely degrade the ``large''-halo acceleration channel more than the ``small''-halo stochastic lensing signal. We defer a full time-domain forecast of this trade-off to future work.

\paragraph*{Expected motion.} The velocity structure differs qualitatively from that of the galaxy lens. The cluster's internal dispersion is large, $\sigma_v\approx1000\,\mathrm{km\,s^{-1}}$ per axis, but a microhalo's own orbital motion displaces the deflector and enters the image--subhalo sweep rate \emph{unmagnified}, contributing only $\approx0.14\,\mathrm{\mu as\,yr^{-1}}$. One internal contribution \emph{is} macro-magnified, however: the macro-image responds to the motion of every mass component of the deflector, weighted by that component's share of the local deflection gradient, $\dot{\vect{\theta}}{}^\mathrm{I} = B^\mathrm{I}[\vect{\mu} - \sum_c H_c\,\delta\vect{\mu}_c]$ (Secs.~\ref{sec:theory}, App.~\ref{app:J1029}). A virialized, phase-mixed halo has a static potential despite its fast-moving particles, so only \emph{clumped} components contribute. A simplified hierarchy is cluster $\supset$ galaxy-scale halos $\supset$ microhalos, with potentially even more sublevels. The intermediate level---member-galaxy halos orbiting at the cluster dispersion $\sigma_v$---contributes a fraction $f_\mathrm{gal}$ of the convergence-plus-shear at the images (nearby members, plus the known perturber near image B). We estimate $f_\mathrm{gal}\approx0.1$--$0.4$ (fiducial $0.2$) for this dynamically disturbed cluster, raising the effective transverse velocity slightly to $v_\mathrm{L}^\mathrm{eff}=[v_\mathrm{bulk}^2+2(f_\mathrm{gal}\sigma_v)^2]^{1/2}\approx550\,\mathrm{km\,s^{-1}}$ and $\widetilde{\mu}^\mathrm{B,C}\approx1.1\,\mathrm{\mu as\,yr^{-1}}$. The bulk term alone, $\approx470\,\mathrm{km\,s^{-1}}$, would give $\approx1.0\,\mathrm{\mu as\,yr^{-1}}$. Actively merging clusters could have larger effective velocities and correspondingly stronger signals. The peak acceleration covariance of the bright B,C pair, $\approx1.9\times10^{-4}\,\mathrm{\mu as^2\,yr^{-4}}$ at $\rho_s=1\,M_\odot\,\mathrm{pc^{-3}}$ (Fig.~\ref{fig:acc-J1029}), exceeds that of B1422+231 by a factor $\approx5$, peaking at $r_s\approx0.09\,\mathrm{pc}$ and $M_\mathrm{L}\approx0.02\,M_\odot$. 

\paragraph*{Optical astrometry.} SDSS J1029+2623 is too faint ($i\approx18$--$19$) for intensity interferometry, so Figs.~\ref{fig:Ctilde-J1029} and~\ref{fig:acc-J1029} assume differential astrometry with a high-resolution optical imager or amplitude interferometer at a more modest $\sigma_{\delta\theta}=1\,\mathrm{\mu as}$, within a factor of two of the $0.59\,\mathrm{\mu as}$ needed for the fiducial CDM microhalo population to become detectable in a cluster lens. All computations are performed as for B1422+231 but with the cluster macrolens model and the parameters above. We now also include the finite-size form factor for an optical emission region (Eq.~\ref{eq:finite_source}), which at the magnified $\theta^\mathrm{I}_\mathrm{src}\approx1.3\,\mathrm{\mu as}$ lowers the acceleration peak by 16\% and moves it from $M_\mathrm{L}\approx0.013$ to $0.022\,M_\odot$. At this $10\times$ coarser precision, the B,C pair attains a peak variance $\mathrm{SNR}\approx0.8$ at fixed $\rho_s=1\,M_\odot\,\mathrm{pc^{-3}}$ (Tab.~\ref{tab:bottomline}). The subhalo concentrations of Sec.~\ref{sec:pure_CDM} at the image radii ($x_\mathrm{sub}\approx0.13$) give $\rho_s\approx0.4\,M_\odot\,\mathrm{pc^{-3}}$ at the peak, so the fiducial CDM population reaches $\mathrm{SNR}\approx0.35$ (Sec.~\ref{sec:prospects}). Denser-than-fiducial substructure from the enhanced-power scenarios of Sec.~\ref{sec:implications} would be more easily detectable.

If deep imaging confirms the low stellar convergence assumed here, much of the cluster data set could remain instrument limited, making cluster lenses a cleaner route to upper limits. Two systematics dominate its forecast. The first is the macrolens model, which enters as $B^6$: near the critical curve the magnification is model-dependent at the factor-of-two level---an order of magnitude in acceleration variance---and the intensity-interferometric self-calibration of $B^\mathrm{I}$ is unavailable at this flux. Measuring the relative image proper motions cuts this to a factor of a few. Uncertainties could be improved by better measurements of flux ratios, morphology, and time delays, deep imaging, and by ensembles of independent reconstructions~\cite{Oguri:2012vg,Acebron:2022,Fohlmeister:2013}. The complexity already known near images B,C cautions against a smooth-cluster description: a radio--optical flux-ratio anomaly attributed to a $\sim10^{9}\,M_\odot$ perturber near image B~\cite{Kratzer:2011}, and weak evidence for optical microlensing~\cite{Fohlmeister:2013}. Secondly, as for the galaxy lens, neither the bulk transverse motion nor the peculiar motions of the clumped components are measured for this system. We estimate the former statistically and adopt $f_\mathrm{gal}=0.2$ for the latter, itself uncertain by a factor of a few, and an unvirialized or merging cluster could exceed both substantially. Together, these uncertainties can change the forecast SNR by about an order of magnitude or more in either direction.

\paragraph*{Radio interferometry.} VLBI achieves comparable angular resolution and has already resolved the images of strongly lensed radio sources~\cite{Hada:2020}, but the compact radio core/jet is much larger than the optical continuum: $\theta_\mathrm{src}\sim1$--$100\,\mathrm{\mu as}$ unlensed, $\sim30$--$3000\,\mathrm{\mu as}$ magnified. The finite-source high-pass filter from Eq.~\ref{eq:finite_source} then washes out the stochastic signal (over practical time scales), leaving only the acceleration channel. For the most compact ($0.01\,\mathrm{pc}$) cores, we estimate the B,C peak shifts from $M_\mathrm{L}\approx0.02\,M_\odot$ to $\approx30\,M_\odot$ with its amplitude reduced by a factor of $11$. Detecting it requires $\sigma_{\delta\theta}\lesssim0.26\,\mathrm{\mu as}$, well beyond current phase-referenced VLBI ($\sim10\,\mathrm{\mu as}$) and demanding even for next-generation long-baseline or space-based arrays, though it could offer a handle on substructures of $M_\mathrm{L}\sim10$--$10^2\,M_\odot$. An additional complication in this case is that radio emission regions tend to have much larger apparent velocities associated with components of the jet~\cite{Lister:2019,Plavin:2019}. 

Table~\ref{tab:bottomline} summarizes the acceleration-channel forecasts for the two benchmark systems.

\begin{table}[t]
\caption{Summary of the differential angular acceleration channel in the two benchmark systems, for the fiducial $\tau = 10\,\mathrm{yr}$, $N = 300$ campaigns.
Signal variances are the largest eigenvalue of Eq.~\ref{eq:observable_2}, maximized over the bound mass $M_\mathrm{L}$ of truncated NFW subhalos at $\rho_s = 1\,M_\odot\,\mathrm{pc^{-3}}$ ($r_\mathrm{t} = 2.3\,r_s$) and $f_\mathrm{sub}=0.1$. That maximum is located at $M_{\mathrm{L,peak}}$, corresponding to a scale radius $r_{s,\mathrm{peak}} \approx |\widetilde{\vect{\mu}}|\tau D_\mathrm{L}$. The $\Lambda$CDM row instead takes $\rho_s$ from the subhalo concentration relation of Fig.~\ref{fig:SNR} at the image radii, $x_\mathrm{sub}\approx0.03$ and $0.13$, which gives $\rho_s\approx0.8$ and $0.4\,M_\odot\,\mathrm{pc^{-3}}$ at the respective maxima. 
The residual noise after subtracting off the stellar microlensing background is taken to be twice the image-A budget of Sec.~\ref{sec:noise}. 
The variance-ratio SNRs are quoted against the instrumental noise floor only and against the total noise including the stellar microlensing residuals.
The SDSS J1029+2623 column includes the finite-size form factor of the optical emission region and the moving-clump member-halo term at $f_\mathrm{gal}=0.2$. 
Its star count refers to a $10\,\mathrm{\mu as}$-radius reference region.}
\label{tab:bottomline}
\begin{ruledtabular}
\begin{tabular}{lcc}
 & B1422+231 & J1029+2623 \\
\hline
image pair & A,B & B,C \\
$\sigma_{\delta\theta}$ {$[\mathrm{\mu as}]$} & $0.1$ & $1$ \\
peak $\big\langle \delta\ddot\theta_{0}^2 \big\rangle$ $[\mathrm{\mu as^2 \, yr^{-4}}]$ & $3.6\times10^{-5}$ & $1.9\times10^{-4}$ \\
instrumental floor $[\mathrm{\mu as^2 \, yr^{-4}}]$ & $2.4\times10^{-6}$ & $2.4\times10^{-4}$ \\
stellar residual $[\mathrm{\mu as^2 \, yr^{-4}}]$ & $2.5\times10^{-5}$ & --- \\
peak $\mathrm{SNR}$ (clean$\,|\,$stars) & $15 \,|\, 1.3$ & $0.80$ \\
$\Lambda$CDM $\mathrm{SNR}$ (clean$\,|\,$stars) & $12 \,|\, 1.0$ & $0.35$ \\
$r_{s,\mathrm{peak}}$ $[\mathrm{pc}]$ & $7\times10^{-2}$ & $9\times10^{-2}$ \\
$M_{\mathrm{L,peak}}$ $[M_\odot]$ & $1\times10^{-2}$ & $2.2\times10^{-2}$ \\
$\kappa_*$ & $0.05$ & $0.003$ \\
$\theta_\mathrm{fit}$ $[\mathrm{\mu as}]$ & $17$ & --- \\
stars within $\theta_\mathrm{fit}$ & $7.5$ & $0.3$ \\
\end{tabular}
\end{ruledtabular}
\end{table}

\section{DM Structures and Implications} \label{sec:structures}
The astrometric weak lensing signal of Eqs.~\ref{eq:C_tilde_1}~and~\ref{eq:C} is set by the abundance, sizes, and densities of DM structures along the line of sight, encoded either in the small-scale matter power spectrum $P_\delta(k)$ or, equivalently, in the (sub)halo mass function and internal density profiles. The observables of Eqs.~\ref{eq:observable_1}~and~\ref{eq:observable_2} are fed almost exclusively by collapsed structures of physical size $r_\mathrm{L}\sim 10^{-2}$--$1\,\mathrm{pc}$. In $\Lambda$CDM, such (sub)halos have masses $M_\mathrm{L} \sim 10^{-6}$--$1\,M_\odot$ and were seeded by linear density perturbations of comoving wavenumber $k \sim 10^4$--$10^6\,h/\mathrm{Mpc}$\footnote{We label each structure by its bound halo mass $M_\mathrm{L}$ (in our case, 1.4 times the mass enclosed inside the truncation radius) and by the comoving wavenumber $k$ of the \emph{linear} perturbation that seeded it, $M_{200} = (4\pi/3)\,\overline{\rho}_{m,0}\,(\pi/k)^3$ (floating gray axis of Fig.~\ref{fig:matter_power}), where the would-be virial mass is $M_{200}\approx6.5\,M_\mathrm{L}$ for $\rho_s = 1\,M_\odot\,\mathrm{pc^{-3}}$. Virialization shrinks a perturbation far below its comoving Lagrangian radius, so the \emph{physical} wavenumber $\sim 1/r_\mathrm{L}$ at which the collapsed structure feeds the present-day power spectrum is about two decades higher in $\Lambda$CDM (at scales of interest).}, a regime heretofore unconstrained. In an inflationary scenario, these modes left the horizon $\ln(10^7)\approx 16$ $e$-folds after those measured in the CMB (probing correspondingly later inflationary epochs), and re-entered it at temperatures $T\sim 0.5\text{--}50\,\mathrm{MeV}$, so they encode information about both the primordial power spectrum and early dark-sector dynamics that is inaccessible by any other means.

We first define a purely cold DM baseline, then compare it with the benchmark sensitivities and with alternatives that suppress or enhance small-scale structure. Section~\ref{sec:pure_CDM} estimates the structures predicted by \emph{purely cold} DM: pressureless DM inheriting the nearly scale-invariant primordial curvature power spectrum, extrapolated 7 orders of magnitude in comoving wavenumber $k$ beyond CMB scales. Section~\ref{sec:prospects} assesses the prospects for detecting this fiducial population with the campaigns of Sec.~\ref{sec:sensitivity}. Section~\ref{sec:implications} outlines the physics that could produce deviations from this baseline, and what a detection or a null result would imply for each.

\subsection{Purely Cold Dark Matter} \label{sec:pure_CDM}
Our null hypothesis is purely cold DM: a particle that is cold and collisionless by the time the universe cools below $T \sim 30\,\mathrm{MeV}$, well before big bang nucleosynthesis (BBN); that interacts only gravitationally; and that inherits the adiabatic, nearly scale-invariant primordial spectrum extrapolated from CMB scales. Once the background cosmology and primordial spectrum are fixed, this (empirical) model has no additional small-scale parameters. The density of the resulting bound clumps simply follows from the statistics of the initial curvature perturbations $\zeta(k)$ and the standard expansion history. A matter perturbation $\delta(k)$ grows approximately logarithmically while inside the horizon during radiation domination, and linearly during matter domination:
\begin{equation}
    \delta(k,a) \simeq \zeta(k) \ln{\left(\frac{a_{\rm eq}}{a_k}\right)}\frac{a}{a_{\rm eq}}, \label{eq:growth}
\end{equation}
where $a = 1/(1+z)$ is the scale factor, with $a_k \equiv \lbrace a \, \vert \, k/a = H\rbrace$ and $a_{\rm eq}$ denoting its values at horizon crossing of the mode $k$ and at matter-radiation equality, respectively. The extra growth of smaller perturbations during radiation domination, partially offset by the slight red tilt of the primordial curvature power $\Delta^2_\zeta(k) \propto k^{n_s-1}$ with $n_s \approx 0.96$, produces a nearly scale-invariant dimensionless power spectrum $\Delta^2_\delta(k) \equiv k^3 P_\delta(k)/2\pi^2 \sim \mathcal{O}(10)$ at small scales within the context of linear theory (thin solid gray line in Fig.~\ref{fig:matter_power})~\cite{Green:2005,Bertschinger:2006,LoebZaldarriaga:2005}. For purely cold DM, structure formation proceeds hierarchically to arbitrarily large $k$ and small halo masses.

A mode of comoving wavenumber $k$ turns nonlinear when it reaches the spherical-collapse threshold, $\delta(k,a_\mathrm{NL})\simeq \delta_c \equiv  1.686$. This happens for typical ($1\sigma$) upward density fluctuations when $\sigma(M)\,D(a_{\rm NL}) \simeq \delta_c$, where $\sigma(M)$ is the present-day rms density contrast smoothed on the mass scale $M = (4\pi/3)\overline{\rho}_{m,0}(\pi/k)^3$ and $D(a)$ is the linear growth factor normalized to $D(a{=}1)=1$. The density of the resulting bound clump tracks the ambient density of the universe at scale factor $a_\mathrm{NL}$. The smallest halos collapse earliest, so they inherit the higher ambient density of their collapse epoch, and the characteristic (scale) density of a halo scales as $\rho_s \propto \overline{\rho}_m(a_{\rm NL}) \propto (1+z_{\rm coll})^3$. For our fiducial cosmology, $\sigma(M) \approx 12$--$15$ for $M=10^{-6}$--$1\,M_\odot$ using default \texttt{CLASS} settings~\cite{Blas:2011}, with typical collapse redshifts $z_{\rm coll}\sim 8$--$11$, while the densest, rare-peak progenitors collapse at $z\gtrsim 30$. The resulting scale densities, $\rho_s \sim 0.1$--$1\,M_\odot\,\mathrm{pc}^{-3}$ across $M_\mathrm{L} \sim 10^{-6}$--$1\,M_\odot$, roughly track the concentration--mass relations of Refs.~\cite{Moline:2017,Wang:2020,Ludlow:2016,DiemerJoyce:2019} extrapolated to substellar masses.

The associated nonlinear matter power spectrum, shown in Fig.~\ref{fig:matter_power} and detailed in App.~\ref{app:nl-power}, is dominated on these scales by the one-halo term model,
\begin{align}
    P_\delta^{\rm 1h}(k) = \frac{1}{\overline{\rho}_{m,0}^2}\int \dd M\, \frac{\dd n}{\dd M}\, M^2\, |u(k|M)|^2,
\end{align}
with the Sheth--Tormen mass function $\dd n/\dd M$~\cite{PressSchechter:1974,ShethTormen:1999} and NFW profile transforms $u(k|M)$ evaluated at the concentrations above. Published concentration--mass relations for field halos diverge when extrapolated to substellar masses, so we show this term as a band: its lower edge adopts the relation of Ref.~\cite{Wang:2020}, which we regard as pessimistic at these scales (as it is representative of microhalos inside highly underdense voids), and its upper edge the relation of Ref.~\cite{SanchezCondePrada:2014}. Both of these are field-halo relations, appropriate for structures along a generic line of sight, predicting lower concentrations (and thus lower scale densities) than the subhalo relation of Ref.~\cite{Moline:2017}, whose band is plotted in Fig.~\ref{fig:SNR} and which applies to the more concentrated subhalos that survive inside a host halo such as a lens galaxy or cluster. The one-halo term continues the \texttt{halofit} nonlinear spectrum smoothly from $k\sim 30\,h/\mathrm{Mpc}$ down to sub-parsec scales (dark gray band in Fig.~\ref{fig:matter_power}), matches \texttt{halofit}~\cite{Takahashi:2012} to within tens of percent in the overlap region where the two-halo term is negligible, and predicts a nearly flat $\Delta_\delta^2 \sim [2$--$7]\times10^3$ at $k \gtrsim 10^3\,h/\mathrm{Mpc}$ (App.~\ref{app:nl-power}).

The gray band of Fig.~\ref{fig:SNR} plots the scale density $\rho_s$ derived from the concentration--mass relation calibrated on \emph{subhalos}~\cite[Eq.~7]{Moline:2017}. Subhalos are systematically more concentrated than field halos of the same mass, and their median concentration is a function of both mass and the host-centric distance $x_\mathrm{sub} = R_\mathrm{sub}/R_{200}^\mathrm{host}$.\footnote{We ignore the mild dependence of concentration on (low) redshift and assume $z_\mathrm{L}=0$.} We take the halos to be NFW, so that $r_s = r_{200}/c_{200}$ and $\rho_s$ is the density at $r_s$ before truncation at $r_\mathrm{t} = 2.3\,r_s$; the plotted mass is the \emph{bound} mass $M_\mathrm{L} = M_{200}\,m_\mathrm{t}(x_\mathrm{t})/m(c_{200})$, the profile convention of Sec.~\ref{sec:signals}. The three gray lines for $x_\mathrm{sub} = \lbrace1, 0.1, 0.01\rbrace$ bracketing the gray band give $\rho_s \approx 0.11$--$1.5\,M_\odot\,\mathrm{pc}^{-3}$ over $M_\mathrm{L} = 10^{-6}$--$1\,M_\odot$. For B1422+231, the relevant value is $x_\mathrm{sub}\approx0.03$ (black dashed line), since its images lie about $5\,\mathrm{kpc}$ in projection from the lens galaxy, whose isothermal virial radius is $R_{200}\approx310\,\mathrm{kpc}$, and subhalos anywhere along the column contribute, raising the line-of-sight-weighted three-dimensional distance to about twice the projected one (App.~\ref{app:B1422}). For SDSS J1029+2623, whose images sit $\approx100\,\mathrm{kpc}$ from the center of a $\approx1\,\mathrm{Mpc}$ cluster halo, the same estimate gives $x_\mathrm{sub}\approx0.13$.

\subsection{Detection Prospects} \label{sec:prospects}
\begin{figure*}
    \centering
    \includegraphics[width=1\textwidth]{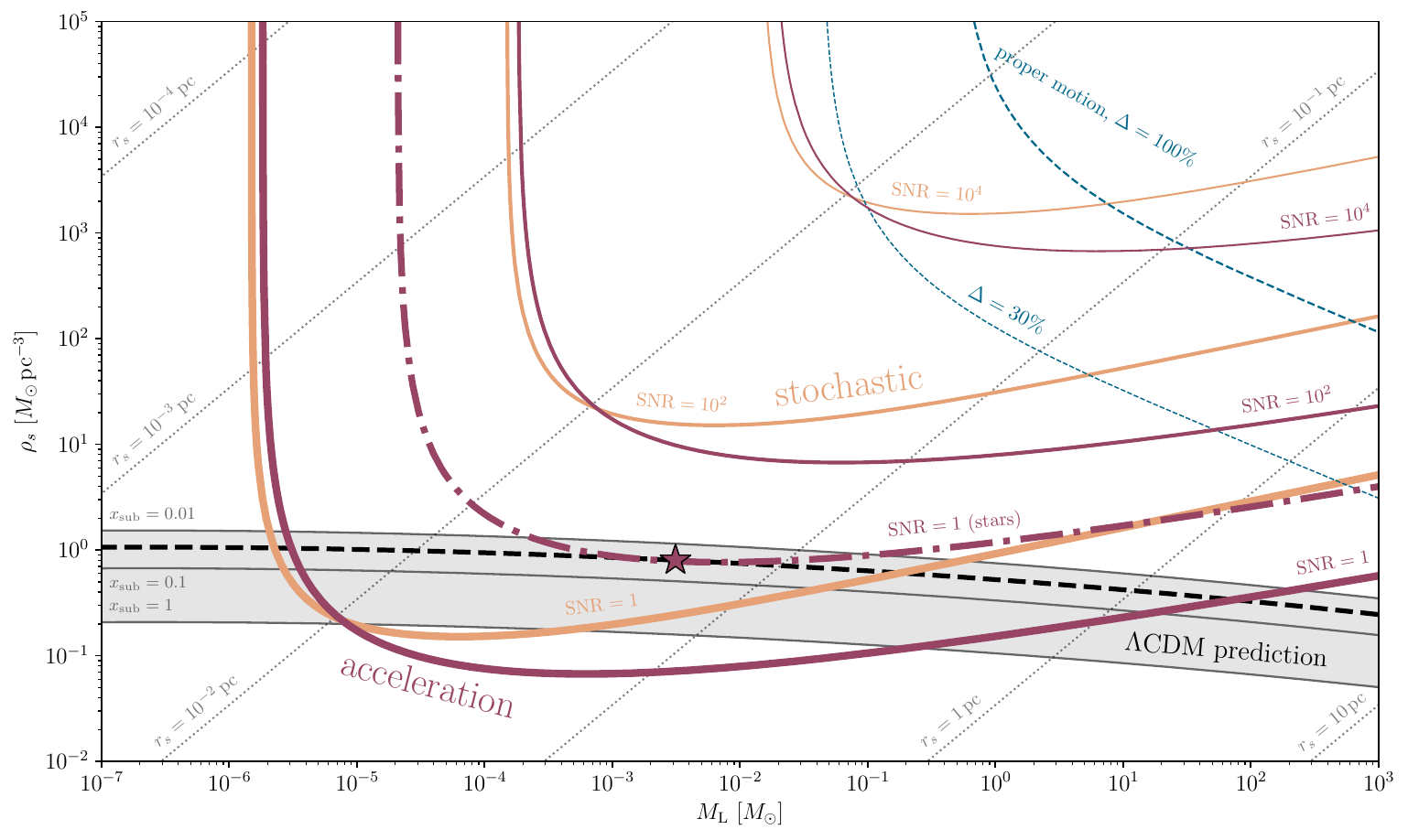}
    \caption{\nblink{code/sensitivity.ipynb} 
    Projected sensitivity of the astrometric weak-lensing signal in the plane of subhalo mass $M_\mathrm{L}$ and scale density $\rho_s$, for the fiducial EPIC survey of Figs.~\ref{fig:Ctilde}~and~\ref{fig:acc} in the galaxy-lensed quasar B1422+231. 
    All contours assume an optimistic substructure fraction $f_\mathrm{sub} = 0.1$ of the convergence at the image positions in surviving microhalos. 
    Stellar microlenses can be identified and subtracted through the intensity-interferometric image-shape fits of Ref.~\cite{companionstars}.
    The solid maroon acceleration contours ($\mathrm{SNR} = 1, 10^2, 10^4$; Eq.~\ref{eq:observable_2}) are instrument-limited, while the thick dot-dashed maroon contour marks the more restrictive $\mathrm{SNR}=1$ once the post-subtraction stellar microlensing residual noise of Sec.~\ref{sec:noise} is included, which burdens primarily this secular acceleration channel. 
    Teal dashed contours show the proper-motion channel at relative precision $\Delta = 30\%$ and $100\%$.
    Orange contours show the stochastic channel (Eq.~\ref{eq:observable_1}, likewise instrument-limited) at the same $\mathrm{SNR}$ levels.  Dotted gray lines indicate constant (physical) scale radius $r_s$.
    Each point of the plane is a \emph{monochromatic} response, assigning that entire substructure convergence $f_\mathrm{sub}\kappa$ to halos of a single bound mass $M_\mathrm{L}$ and scale density. 
    The shaded gray band is the fiducial CDM relation between scale density $\rho_s$ and mass $M_\mathrm{L}$ for NFW subhalos carrying the median concentration $c_{200}(M_{200},x_\mathrm{sub})$ of Ref.~\cite[Eq.~7]{Moline:2017}, truncated at $r_\mathrm{t} = 2.3\,r_s$; the three solid gray lines are host-centric distances $x_\mathrm{sub} = R_\mathrm{sub}/R_{200}^\mathrm{host} = \lbrace1, 0.1, 0.01\rbrace$, the outer two bracketing the band (Sec.~\ref{sec:pure_CDM}). 
    The black dashed line at $x_\mathrm{sub}\approx0.03$ represents the characteristic host-centric distance of images A, B, C of B1422+231 (App.~\ref{app:B1422}).
    Our fiducial survey reaches $\mathrm{SNR}\approx1.0$ against the stellar microlensing residual background (star marker; $\approx12$ on star-free data) and lies above the instrumental $\mathrm{SNR}=1$ contour across $M_\mathrm{L}\approx 3\times10^{-6}$--$70\,M_\odot$, keeping the standard CDM substructure population within reach of the proposed time-domain astrometric weak lensing technique.}
    \label{fig:SNR}
\end{figure*}

\begin{figure*}
    \centering
    \includegraphics[width=1\textwidth]{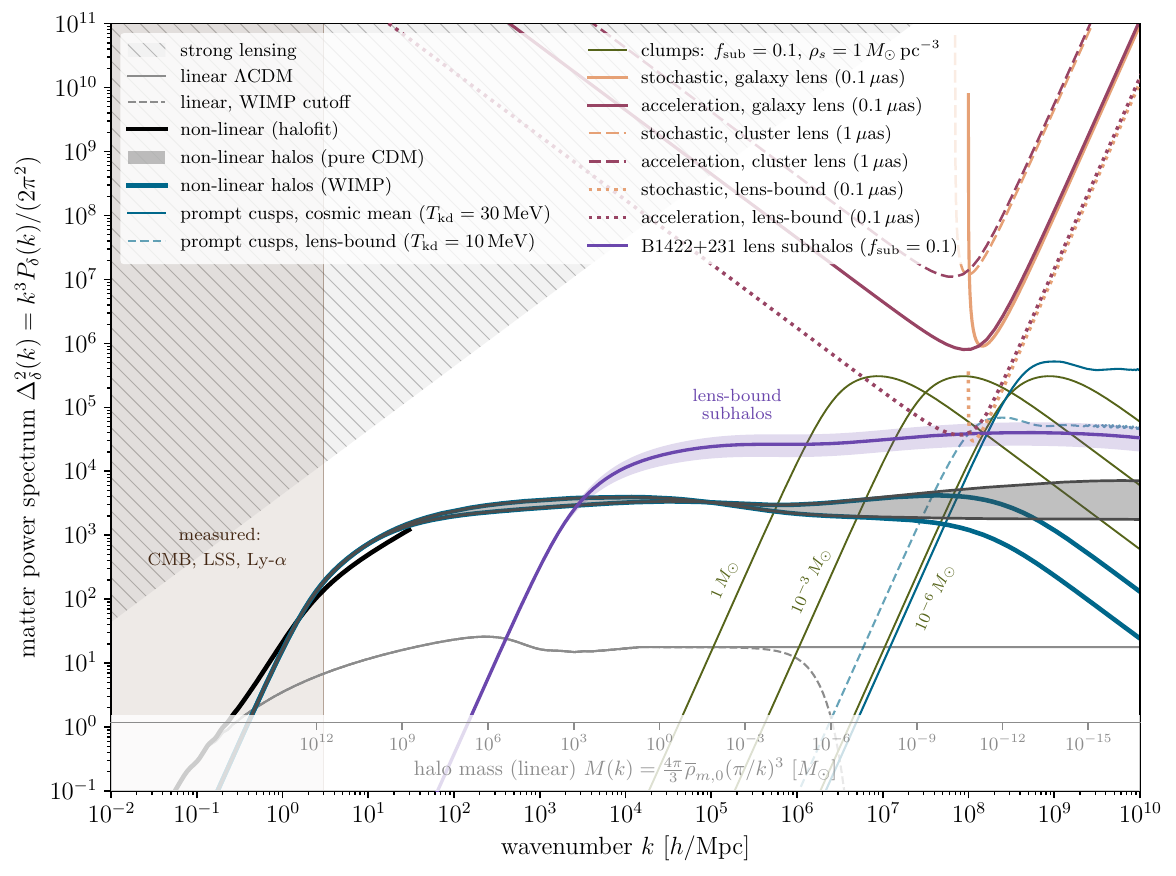}
    \caption{\nblink{code/matter_power.ipynb} 
    Dimensionless matter power spectrum $\Delta^2_\delta(k) = k^3 P_\delta(k)/2\pi^2$ as a function of comoving wavenumber $k$.
    The null hypothesis of purely cold DM has a \emph{linear} power spectrum plotted in thin solid gray; 
    the floating gray axis near the bottom gives the associated (Lagrangian) halo mass $M(k) = (4\pi/3)\,\overline{\rho}_{m,0}\,(\pi/k)^3$ of the comoving \emph{linear} perturbation.
    We estimate the corresponding \emph{nonlinear} spectrum of pure CDM from one-halo terms by the gray band (bracketing the halo concentration--mass relations between those of Refs.~\cite{Wang:2020,SanchezCondePrada:2014}), extending the \texttt{halofit} spectrum (thick black, $k<30\,h/\mathrm{Mpc}$)~\cite{Takahashi:2012} to higher $k$. For a canonical $100\,\mathrm{GeV}$ thermal WIMP with a kinetic decoupling temperature of $T_{\rm kd} = 30\,\mathrm{MeV}$, acoustic damping and free streaming (Sec.~\ref{sec:implications}) cut off the linear spectrum at $k_{\rm cut}\simeq 10^6\,\mathrm{Mpc}^{-1}$ (thin dashed gray); its nonlinear power spectrum from halos (thick blue, band edges) tracks the pure-CDM band at low $k$ and is suppressed near the cutoff, while prompt cusps boost the power at higher $k$ (thin solid blue, cosmic mean)~\cite{DelosWhite:2023}. 
    The thin dashed blue curve shows the same cusps bound to the B1422+231 lens at $T_{\rm kd} = 10\,\mathrm{MeV}$, the lens-bound detection threshold of Sec.~\ref{sec:implications}, in the normalization of the dotted lens-bound thresholds described below (DM convergence at image B, no subhalo enrichment, comoving $k$ at the lens); the lower $T_{\rm kd}$ shifts its left edge at $k\sim 1/r_{\rm cusp}\propto T_{\rm kd}$ into the sensitivity band.
    The brown vertical band at $k \lesssim 3\,h/\mathrm{Mpc}$ marks scales already measured at $\Lambda$CDM-level amplitudes (CMB, LSS, Lyman-$\alpha$ forest). 
    The hatched region marks the strong-lensing regime (for the B1422+231 sightline), where the power in a single $e$-fold at $k$ yields an rms convergence above unity and our perturbative treatment breaks down (App.~\ref{app:power-spectrum}).
    Orange and maroon curves show the projected per-mode sensitivity in the stochastic (Eq.~\ref{eq:observable_1}) and acceleration (Eq.~\ref{eq:observable_2}) channels (the minimum detectable amplitude $\Delta^2_\delta$ in one $e$-fold around a comoving size $\sim \pi/k$) for the two benchmark campaigns of this work, each over $\tau = 10\,\mathrm{yr}$ with $N=300$ epochs: the galaxy-lensed B1422+231 ($B \approx 10.2$, bulk $|\vect{v}_\mathrm{L}| \approx 8.7\times10^2\,\mathrm{km\,s^{-1}}$) at the EPIC precision $\sigma_{\delta\theta} = 0.1\,\mathrm{\mu as}$ (solid), and cluster-lensed SDSS J1029+2623 ($B \approx 22.6$, effective $|\vect{v}_\mathrm{L}| \approx 5.5\times10^2\,\mathrm{km\,s^{-1}}$; Sec.~\ref{sec:cluster}) at $\sigma_{\delta\theta} = 1\,\mathrm{\mu as}$ (dashed). 
    Those curves are for structures spread along the sightline at the cosmic mean density; the dotted curves of the same two colors are the corresponding thresholds for substructure \emph{bound to} the B1422+231 lens (App.~\ref{app:power-spectrum}). 
    The violet band represents the pure-CDM prediction for that lens-bound population using the same halo mass function as the gray band, but with the subhalo concentrations of Ref.~\cite{Moline:2017} over host-centric distances $x_\mathrm{sub} = 0.1$--$0.01$ and the tidal truncation of Sec.~\ref{sec:signals}. 
    Thin olive-green curves show the Poisson power of the \emph{monochromatic} population of Fig.~\ref{fig:SNR}: clumps holding a fraction $f_\mathrm{sub} = 0.1$ of the convergence at image B of B1422+231, all of a single mass $M_\mathrm{L} = \lbrace 1, 10^{-3}, 10^{-6}\rbrace\,M_\odot$ and scale density $\rho_s = 1\,M_\odot\,\mathrm{pc^{-3}}$, with an NFW profile truncated at $r_\mathrm{t} = 2.3\,r_s$. Their peak height is independent of $M_\mathrm{L}$ and falls at $k \approx 2.6/[(1+z_\mathrm{L})\,r_s]$. They are drawn in the cosmic-mean-equivalent normalization of App.~\ref{app:power-spectrum}.
    All curves are instrument-limited. Stellar microlensing backgrounds would degrade the acceleration channel more (Sec.~\ref{sec:noise}): for the galaxy lens, the worst-case post-subtraction residual raises the maroon solid curve by roughly an order of magnitude in $\Delta^2_\delta$, while the star-poor cluster lens ($\kappa_*\approx0.003$) should remain limited by instrumental noise.
}
    \label{fig:matter_power}
\end{figure*}

The lens-bound substructure has the strongest forecast, because the acceleration statistic integrates the full substructure convergence of the overdense lens environment, wherein subhalos are expected to be denser than field halos elsewhere along the line of sight. The forecast of Fig.~\ref{fig:SNR}, Table~\ref{tab:bottomline}, and every \emph{lens-bound} SNR quoted below, rests on an optimistic substructure fraction $f_\mathrm{sub} = 0.1$, i.e.~a tenth of the convergence at the image positions is assumed to reside in surviving microhalos, tidally truncated at $r_\mathrm{t} = 2.3\,r_s$ (Sec.~\ref{sec:signals}). The variance SNR scales linearly with $f_\mathrm{sub}$, and the true survival fraction is uncertain and plausibly much smaller, as we discuss at the end of this subsection. As shown in Fig.~\ref{fig:SNR}, the acceleration channel of Eq.~\ref{eq:observable_2} reaches the fiducial CDM scale-density relation with a peak variance $\mathrm{SNR}\approx12$ at $M_\mathrm{L}\approx3\times10^{-3}\,M_\odot$, and with $\mathrm{SNR}>1$ across $M_\mathrm{L}\approx 3\times10^{-6}$--$70\,M_\odot$ on star-free data. These are monochromatic response curves: each point assigns the entire substructure convergence $f_\mathrm{sub}\kappa$ to a single halo mass. (In the population-level forecast of Fig.~\ref{fig:matter_power}, the small-scale structures are instead distributed across the halo mass spectrum.) 

Stellar contamination limits the galaxy-lens discovery reach. The worst-case post-subtraction stellar residual of Sec.~\ref{sec:noise} caps the sensitivity to monochromatic mass functions at $\mathrm{SNR}\approx1.0$ for $f_\mathrm{sub}=0.1$. A confident \emph{discovery} of the fiducial CDM population in the galaxy lens therefore requires either the certification of clean epochs and images, a higher-than-expected proper motion, or a per-epoch precision better than that assumed here. The last option is available because stellar residuals fall at least as fast as the instrumental noise (cf.~discussion around Eq.~\ref{eq:theta_fit}). 

The $1\,\mathrm{\mu as}$ cluster-lens survey of Sec.~\ref{sec:cluster} trades raw precision for a cleaner environment. Its peak reach is $\mathrm{SNR}\approx0.35$ at $M_\mathrm{L}\approx4\times10^{-3}\,M_\odot$ assuming subdominant stellar microlensing background residuals. This falls short of the fiducial CDM prediction by a factor of 3, but we reiterate that our forecast is highly uncertain, given the unknown moving-clump fraction and effective proper motions of the images. Denser-than-fiducial substructure, $\rho_s\approx10\,M_\odot\,\mathrm{pc^{-3}}$, would be detected at $1\,\mathrm{\mu as}$ with $\mathrm{SNR}\approx8$ over $M_\mathrm{L}\approx2\times10^{-4}$--$3\times10^{4}\,M_\odot$ under the above assumptions.

Figure~\ref{fig:SNR} also shows a proper-motion channel (teal dashed contours), marking where the stochastic lensing contribution to the image proper motion reaches a fraction $\Delta = 30\%$ or $100\%$ of the mean image proper motion $\widetilde{\mu}^\mathrm{I}$ itself. This channel is harder to exploit than the acceleration one, because the mean proper motion is degenerate with the macrolensing model: the first time derivative of the small-scale deflection terms in Eq.~\ref{eq:thetaI} can be absorbed into $\widetilde{\vect{\mu}}{}^\mathrm{I}$, so isolating the substructure contribution requires an external determination of both the macrolensing model and the true relative velocity at precision $\Delta$. The proper motion thus enters at the same order in derivatives of the lensing potential as the image flux ratios, long known to deviate from smooth macromodels at the ten-percent level; B1422+231 in particular furnished some of the earliest evidence for lens substructure~\cite{MaoSchneider:1998}. Within $\Lambda$CDM, both proper motions and flux ratios are dominated by relatively massive structures, whereas the acceleration and stochastic-lensing observables of this work are sourced by the microstructures whose crossing time matches the survey duration.\footnote{\label{fn:FRB_orders}A similar order counting bears on the differential FRB timing proposal of Ref.~\cite{2024PhRvD.110b3516X}, discussed further in Sec.~\ref{sec:discussion}.}

Figure~\ref{fig:matter_power} compares the pure-CDM prediction to the projected per-mode sensitivity for the two populations that source the signal: field halos along the sightline at the cosmic mean density, and substructure bound to the lens itself. The differential astrometry of a strongly lensed quasar is most sensitive to structures of scale radius $r_s\sim \widetilde{\mu}^\mathrm{I}\tau D_\mathrm{L}\sim 0.05\,\mathrm{pc}$ and mass $M_\mathrm{L}\sim 10^{-5}$--$10^{-1}\,M_\odot$. These populate the nonlinear spectrum at $k\sim 1/r_s\sim 10^6$--$10^8\,h/\mathrm{Mpc}$, about two decades above the linear wavenumbers that seeded them under the $\Lambda$CDM null hypothesis.

The contrast between the two populations explains why the line-of-sight curves of Fig.~\ref{fig:matter_power} sit far above the CDM band while the lens-bound ones reach the population predicted for the lens. Structures \emph{along the line of sight}, outside the lens, are guaranteed by the cosmological matter power spectrum, and are included in the integral of Eq.~\ref{eq:C_tilde_1}. Their contribution is nonetheless small, because the sweep length of Eq.~\ref{eq:sweep} weights only a window of order $\chi_\mathrm{L}$ about the lens plane rather than the full $\chi_\mathrm{S}\approx7\,\mathrm{Gpc}$ path to the source: the effective mean-density column is $\approx3\times10^1\,M_\odot\,\mathrm{pc^{-2}}$, an order of magnitude below the matter column through the lens itself, $\kappa^\mathrm{A}\Sigma_\mathrm{cr}/(1+z_\mathrm{L})^2 \approx 4\times10^2\,M_\odot\,\mathrm{pc^{-2}}$. Folded through Eq.~\ref{eq:C_tilde_1}, the full pure-CDM spectrum gives the galaxy-lens campaign a guaranteed variance $\mathrm{SNR}\approx0.005$--$0.014$ in the acceleration channel and $\approx0.002$--$0.005$ per lowest stochastic mode, the ranges spanning the concentration band (instrument-limited; the worst-case stellar residual degrades the acceleration numbers by a further order of magnitude).

The sensitivity to the \emph{lens-bound} population is approximately two orders of magnitude better, for two reasons. One order of magnitude comes from the fact that the sightline crosses an overdense galaxy at the distance where the sweep length of Eq.~\ref{eq:sweep} peaks. Another order of magnitude comes from the higher concentrations of subhalos as compared to field halos~\cite{Moline:2017}. In the forecast of Fig.~\ref{fig:SNR}, we place $f_\mathrm{sub}=0.1$ of the column density in halos of the single best-matched mass, with the more compact, tidally truncated profiles appropriate to subhalos. The cosmic-mean spectrum instead spreads its mass over the entire halo hierarchy, only $\sim0.2\%$ of the mean matter density per $e$-fold of halo mass sitting near the sweet spot, in diffuse NFW profiles whose outskirts contribute little astrometric power. 

The field-halo band in Fig.~\ref{fig:matter_power} omits several positive contributions and therefore underestimates the predicted signal. The largest omission is the substructure bound to the lens galaxy or cluster, since the matter field is \emph{not} generic for a lensed quasar. By construction, the lensed sightlines to the quasar intersect a highly overdense lens galaxy or cluster, whose subhalo abundance, convergence, and concentration exceed the corresponding cosmic-mean values assumed by the line-of-sight sensitivity curves. We account for that population separately, in Fig.~\ref{fig:SNR} and in the lens-bound curves of Fig.~\ref{fig:matter_power}. Another omission is the substructure of the line-of-sight halos themselves, since the one-halo model treats each field halo as a smooth NFW profile, discarding its own subhalos (and subsubhalos), which we expect to add significant power at the wavenumbers that dominate the time-domain astrometric response. We are not aware of any quantitative prediction (or method thereof) for these enhancements in the literature, and defer their estimate to future work. Sharpening this band is, in our view, the single most valuable theoretical input that numerical work could supply.

The leading systematic on both populations is their predicted abundance. The structures that dominate the response sit at the bottom of a hierarchy, as subhalos and subsubhalos of larger hosts in the field and inside the macrolens alike, and, as noted above, our one-halo calculation counts only the top level of that hierarchy. A correct count instead requires following microhalo survival through every level of assembly above them, which is the current bottleneck in the prediction. Inside the macrolens, the harsh tidal field at the image radii compounds the problem, so the survival fraction behind our optimistic $f_\mathrm{sub} = 0.1$ is uncertain and plausibly much smaller than for field halos. The microhalo-survival literature spans early claims of efficient tidal and stellar-encounter disruption~\cite{Zhao:2007} and of substantial survival~\cite{Moore:2005}, with modern, environment-dependent prescriptions in between~\cite{Delos:2019}, while prompt cusps are more resilient~\cite{DelosWhite:2022,DelosWhite:2023,Ando:2026soo}. The acceleration SNR scales linearly with $f_\mathrm{sub}$, so for star-free data on B1422+231 a detection of the fiducial CDM population at its best-matched mass survives down to $f_\mathrm{sub} \approx 0.009$.

\subsection{Beyond Cold Dark Matter} \label{sec:implications}
The purely cold-DM baseline of Sec.~\ref{sec:pure_CDM} is predictive in principle, but the abundance and compactness of the smallest $\Lambda$CDM structures remain uncertain in practice. Departures beyond those uncertainties are diagnostic of new fundamental physics. A measured \emph{deficit} of small-scale power relative to the pure-CDM baseline would imply a cutoff in the transfer function, from the gradient or degeneracy pressure of light DM, or from the free-streaming or collisional damping of a thermal relic that kinetically decoupled (or was ``frozen in'') at late times. Such a deficit would probe the DM particle mass and its elastic couplings to the Standard Model (SM)~\cite{Bechtol:2022koa}. A measured \emph{excess} would point to enhanced primordial power or to dark-sector dynamics at early times. Wavelike DM of low mass adds irreducible interference fluctuations even in the absence of bound structure (App.~\ref{app:scalar-DM-power}), but far below the detection threshold. Several of these effects may be present simultaneously, partially canceling or amplifying one another; for clarity we treat each in isolation below, to illustrate the reach of the technique in each case.

It is useful to translate halo masses into the epoch that produced them. The DM mass inside the Hubble horizon at temperature $T$ is $\simeq 4\times10^{-6}\,M_\odot\,(T/30\,\mathrm{MeV})^{-3}$, so the $10^{-6}$--$1\,M_\odot$ window probed here corresponds to temperatures $T\sim 0.5$--$50\,\mathrm{MeV}$, i.e.~the radiation-dominated universe from shortly after the QCD phase transition down to the eve of BBN. We refer to $T = 30\,\mathrm{MeV}$ as the reference temperature: its horizon holds about an Earth mass of DM, near the bottom of the window probed here, and it is the canonical kinetic-decoupling temperature of a weak-scale WIMP.

\paragraph*{Kinetic decoupling and the damping cutoff.} Any DM candidate with a thermal history departs from the pure-CDM baseline at some scale. Elastic scattering with SM particles keeps the DM kinetically coupled until a decoupling temperature $T_{\rm kd}$, resulting in collisional (acoustic) damping and free streaming and thus the erasure of linear perturbations below some model-dependent scale. At the nonlinear level, the stronger of these two damping effects truncates the halo mass function at low halo masses, and it endows the first peaks to collapse (those at the halo cutoff scale itself) with dense prompt cusps at their centers. We treat the truncated halo population here and the prompt cusps in the next part.

Acoustic damping erases modes inside the comoving horizon at decoupling, $\lambda_{\rm ao}\simeq (a_{\rm kd}H_{\rm kd})^{-1}$, while the DM subsequently free-streams a comoving distance $\lambda_{\rm fs}\simeq v_{\rm kd}\ln(a_{\rm eq}/a_{\rm kd})\,(a_{\rm kd}H_{\rm kd})^{-1}$, with $v_{\rm kd}\simeq(T_{\rm kd}/m_\chi)^{1/2}$ the DM thermal velocity at decoupling, so that
\begin{equation}
    \frac{\lambda_{\rm fs}}{\lambda_{\rm ao}}\simeq \sqrt{\frac{T_{\rm kd}}{m_\chi}}\,\ln\frac{a_{\rm eq}}{a_{\rm kd}}\approx 0.3\left(\frac{m_\chi/T_{\rm kd}}{3\times10^{3}}\right)^{\!-1/2}\!. \label{eq:fs-vs-ao}
\end{equation}
The halo mass cutoff is set by the larger of the two damping scales, $\lambda_{\rm cut} = \max[\lambda_{\rm fs},\lambda_{\rm ao}]$, with $M_{\rm cut}\sim \overline{\rho}_{m,0}\lambda_{\rm cut}^3$.
Acoustic damping therefore sets the cutoff for DM that decouples while very nonrelativistic, while free streaming dominates for warmer relics. Up to the $\mathcal{O}(1)$ factors relating each damping length to a mass, the crossover sits roughly at $m_\chi\simeq 10^{3}\,T_{\rm kd}$. The two calibrated damping masses are~\cite{LoebZaldarriaga:2005,Green:2005,Bringmann:2009}
\begin{align}
    M_{\rm ao}&\simeq 4\times10^{-6}\,M_\odot\left(\frac{T_{\rm kd}}{30\,\mathrm{MeV}}\right)^{\!-3}\!, \label{eq:Mcut-Tkd}\\
    M_{\rm fs}&\simeq 10^{-6}\,M_\odot\left(\frac{m_\chi}{100\,\mathrm{GeV}}\,\frac{T_{\rm kd}}{30\,\mathrm{MeV}}\right)^{\!-3/2}\!\!, \label{eq:Mfs-Tkd}
\end{align}
up to a logarithmic factor in the latter; $M_{\rm ao}$ is just the DM mass inside the comoving horizon at decoupling, and carries no dependence on $m_\chi$.

Our canonical example is a thermal WIMP of mass $m_\chi = 100\,\mathrm{GeV}$ that kinetically decouples at the reference temperature ($m_\chi/T_{\rm kd}\approx 3\times10^3$), so that $M_{\rm cut} = M_{\rm ao}$ exceeds $M_{\rm fs}$ by a factor of a few. Its linear spectrum is exponentially cut off at $k_{\rm cut} \approx 10^6\,\mathrm{Mpc}^{-1}$, corresponding to a minimum halo mass $M_{\rm cut}$ comparable to Earth's mass~\cite{Green:2005,Bertschinger:2006,Diemand:2005,Bringmann:2009} (thin dashed gray and thick blue curves in Fig.~\ref{fig:matter_power}). 

Kinetic decoupling occurs when the momentum-relaxation rate $\Gamma_p\sim (T/m_\chi)\,n_{\rm SM}\,\sigma_{\rm el}$ falls below the Hubble rate. For a contact interaction with the relativistic SM bath, the momentum-transfer cross section scales as $\sigma_{\rm el}\simeq T^2/\Lambda^4$, so that~\cite{LoebZaldarriaga:2005}
\begin{equation}
    T_{\rm kd}\approx 10\,\mathrm{MeV}\left(\frac{\Lambda}{100\,\mathrm{GeV}}\right)\!\left(\frac{m_\chi}{100\,\mathrm{GeV}}\right)^{\!1/4}\!\!, \label{eq:Tkd-Lambda}
\end{equation}
with $\Lambda$ an effective coupling scale.
A detection of bound structure at the fiducial cutoff mass, a few$\,\times10^{-6}\,M_\odot$, requires kinetic decoupling at or above the reference temperature, i.e.~an upper bound on the elastic cross section at $T\sim T_{\rm kd}$, or equivalently a lower bound on the coupling scale $\Lambda\gtrsim 300\,\mathrm{GeV}\,(m_\chi/100\,\mathrm{GeV})^{-1/4}$, which tightens toward lighter DM. Translating this early-universe bound into a present-day cross section limit is model- and target-dependent, but can be quite stringent in some cases. For example, sub-GeV DM coupled to electrons via a heavy mediator at current experimental sensitivity levels~\cite{Essig:2011nj,SENSEI:2023zdf,DAMIC-M:2025luv,DarkSide:2022knj,XENON:2019gfn} would predict cutoff masses well above $1 M_\odot$, far above those probed here.

\paragraph*{Prompt cusps.} A cutoff also qualitatively reshapes the smallest structures. As the first density peaks above the cutoff scale collapse, a power-law cusp $\rho = \mathcal{A}\,r^{-3/2}$ forms quasi-instantaneously at the center of each, with amplitude $\mathcal{A} \simeq 24\,\overline{\rho}_{m,0}\,a_{\rm coll}^{-3/2} R^{3/2}$ and outer extent $r_{\rm cusp}\simeq 0.11\,a_{\rm coll} R$ fixed by the peak's collapse time $a_{\rm coll}$ and comoving size $R \equiv |\delta/\nabla^2\delta|^{1/2}$~\cite{DelosWhite:2022,DelosWhite:2023}. These dense relics survive the hierarchical build-up of larger halos largely intact, and their characteristic mass and size are set by the cutoff itself, so their imprint on the power spectrum is a second diagnostic of $M_{\rm cut}$; for purely cold DM the cusp population recedes to arbitrarily small scales.

In the fiducial WIMP model, cusps initially make up about $2\%$ of the DM, with characteristic masses $M_{\rm cusp}\sim 10^{-6}\,M_\odot$, outer radii $r_{\rm cusp}\sim 5\times 10^{-3}\,\mathrm{pc}$, and an inner phase-space core at $r_{\rm core}\simeq r_{\rm cusp}/500$ set by the primordial coarse-grained phase-space density. Because $\rho \propto r^{-3/2}$ implies $|u(k)|^2 \propto k^{-3}$, the cusp population imprints a \emph{plateau} in $\Delta_\delta^2(k)$ extending from $k\sim 1/r_{\rm cusp}$ to $k\sim 1/r_{\rm core}$ (thin blue line in Fig.~\ref{fig:matter_power}). Anchoring its amplitude to Ref.~\cite{DelosWhite:2023}, the plateau sits at $\Delta_\delta^2 \simeq 160\,f_{\rm surv}(1+z_5)^3/[0.53 + \ln(r_{\rm cusp}/r_{\rm core})] \sim 4\times 10^5$, where $z_5 \simeq 31$ is the redshift by which $5\%$ of peaks have collapsed and $f_{\rm surv}\simeq 0.5$ is the cusp survival fraction. Cusps dominate the matter power spectrum precisely where contributions from larger halos are cut off.

Our benchmark campaigns are forecast to fall short of detecting prompt cusps in the fiducial WIMP model. Carrying a fraction $f_{\rm cusp}\simeq f_{\rm surv}\times2\%\approx1\%$ of the DM convergence $\kappa^\mathrm{I}-\kappa^\mathrm{I}_*$ (80--87\% of $\kappa^\mathrm{I}$ at images A and B) as near-point lenses of $M_{\rm cusp}\sim10^{-6}\,M_\odot$, the lens-bound cusps yield an instrument-noise-only $\mathrm{SNR}\approx0.1$ in both channels for the EPIC campaign on B1422+231, about 4 times that of the line-of-sight cusps ($0.03$ and $0.02$). Along the line of sight, the plateau lies a factor $2$ below that campaign's per-$e$-fold stochastic reach, $\Delta^2_\delta\approx9\times10^5$ near $k\approx1.5\times10^{8}\,h\,\mathrm{Mpc^{-1}}$ (Fig.~\ref{fig:matter_power}). For the lens-bound cusps, in the normalization of the dotted lens-bound curves, it lies a factor $1.7$ above the lens-bound stochastic reach. In both populations, however, its start at $k\sim1/r_{\rm cusp}$ is a factor $2$ beyond the stochastic sensitivity optimum, below which unresolved cusps fall off as $k^3$. 

Lower decoupling temperatures or lower DM particle masses can yield detectable prompt cusps. The collapse time cancels in the cusp mass, $M_{\rm cusp}=(8\pi/3)\mathcal{A}\,r_{\rm cusp}^{3/2}\propto R^3$, so $M_{\rm cusp}\propto M_{\rm cut}\propto T_{\rm kd}^{-3}$ and $r_{\rm cusp}\propto T_{\rm kd}^{-1}$ (assuming acoustic oscillations still set the cutoff, with $m_\chi$ fixed), while $f_{\rm cusp}$ and the plateau height are nearly cutoff-independent. A lower decoupling temperature therefore slides the entire plateau into the sensitivity band at essentially unchanged height (dashed blue in Fig.~\ref{fig:matter_power}, in the normalization of the dotted lens-bound curves): the lens-bound cusps reach $\mathrm{SNR}=1$ at $T_{\rm kd}\approx10\,\mathrm{MeV}$ in both channels, i.e.~$M_{\rm cut}\approx1\times10^{-4}\,M_\odot$. By Eq.~\ref{eq:Tkd-Lambda}, this corresponds to $\Lambda\approx100\,\mathrm{GeV}\,(m_\chi/100\,\mathrm{GeV})^{-1/4}$. The Gaussian description of Eq.~\ref{eq:C_tilde_1} is still sufficient: because the projected mass of an $r^{-3/2}$ cusp rises as $b^{3/2}$, the variance per logarithmic interval in impact parameter $b$ grows toward small $b$, and is cut off not at $r_{\rm cusp}$ but at the magnified source size $\theta^\mathrm{I}_{\rm src}D_\mathrm{L}$. At fixed cusp mass fraction the cusp number density scales as $1/M_{\rm cusp}$, so the weighted number of contributing cusps scales as $T_{\rm kd}^{3}$: it stays above $\sim10$ at the detection threshold, and falls to unity (where the nearest cusp rivals all others combined) at $T_{\rm kd} \approx 2$--$4\,\mathrm{MeV}$.

\paragraph*{Bounds on the DM particle mass.} The existence of gravitationally bound structure down to a cutoff mass $M_{\rm cut}$ sets broadly model-independent lower bounds on the DM particle mass, for both fermionic and bosonic candidates. Fermionic DM obeys Pauli exclusion: its phase space occupation number cannot exceed unity. Applied to the smallest, densest observed halo with velocity dispersion $\sigma^2\sim GM/r_s$ and radius $r_s\sim(M/\rho_s)^{1/3}$, the Tremaine--Gunn bound~\cite{TremaineGunn:1979} on the fermion DM mass $m_\mathrm{f}$ reads
\begin{align}
    m_{\rm f}
    &\gtrsim \frac{\rho_s^{1/8}\,M_{\rm Pl}^{3/4}}{M_{\rm cut}^{1/4}} \label{eq:fermion-bound}
    \approx 400\,\mathrm{keV} \left(\frac{M_{\rm cut}}{10^{-6}\,M_\odot}\right)^{\!-1/4}\!,
\end{align}
at a fiducial scale density $\rho_s = 1\,M_\odot\,\mathrm{pc^{-3}}$, upon which the bound depends very weakly; large internal fermion multiplicities would likewise weaken it only mildly.

Bosonic DM structure growth is inhibited on small scales by gradient pressure, which endows density perturbations with a nonzero effective sound speed, $c_s^2\simeq k^2/(4m^2a^2)$, and hence a Jeans scale. Once a mode enters the horizon, pressure support balances self-gravity on comoving wavelengths below the Jeans length, suppressing the logarithmic growth of structure during radiation domination. With a Jeans mass $M_\mathrm{J}\sim H_{\rm eq}^{1/2}M_{\rm Pl}^2 m^{-3/2}$, requiring $M_\mathrm{J}\lesssim M_{\rm cut}$ yields the fuzzy-DM bound~\cite{Hu:2000}
\begin{equation}
    m_{\rm b}\gtrsim \frac{H_{\rm eq}^{1/3}\,M_{\rm Pl}^{4/3}}{M_{\rm cut}^{2/3}}\approx 10^{-12}\,\mathrm{eV}\left(\frac{M_{\rm cut}}{10^{-6}\,M_\odot}\right)^{\!-2/3}\!, \label{eq:boson-bound}
\end{equation}
with $H_{\rm eq}$ the Hubble rate at matter-radiation equality and $M_{\rm Pl} \equiv G_N^{-1/2} \approx 1.22\times10^{19}\,\mathrm{GeV}$ the Planck mass.
Because both bounds sharpen as $M_{\rm cut}$ decreases ($m_{\rm f}\propto M_{\rm cut}^{-1/4}$, $m_{\rm b}\propto M_{\rm cut}^{-2/3}$), a detection of structure at $10^{-6}\,M_\odot$ would improve the universal lower bound on a fermionic DM mass by nearly four orders of magnitude and that on a bosonic DM mass by nine orders of magnitude above the canonical fuzzy-DM limit (more than six above the dynamical-heating lower bound of Ref.~\cite{Dalal:2022rmp}), essentially excluding bosonic DM whose de Broglie wavelength $2\pi/(m_{\rm b} v)$ at $v\sim10^{-3}$ exceeds the diameter of the Sun, unless the structure formed nongravitationally, as in the large-misalignment mechanism~\cite{Arvanitaki:2019rax}.

\paragraph*{Primordial curvature power spectrum.} Within $\Lambda$CDM, even a coarse measurement of the small-scale matter power at the $\mathcal{O}(10)$ level would constrain the primordial spectrum on scales $\sim 16$ $e$-folds beyond the reach of the CMB. Writing the dimensionless primordial curvature spectrum as $\Delta^2_\zeta(k)=A_s\,(k/k_*)^{n_s-1+\frac{1}{2}\alpha_s\ln(k/k_*)}$, the linear small-scale power inherits a slowly varying effective slope~\cite{LoebZaldarriaga:2005}
\begin{equation}
    \frac{\dd\ln\Delta^2_\delta}{\dd\ln k}=(n_s-1)+\alpha_s\ln\!\frac{k}{k_*}+\frac{2}{\ln(k/k_{\rm eq})}, \label{eq:tilt}
\end{equation}
where the final term is the logarithmic growth of the CDM transfer function. Over the lever arm from the CMB pivot $k_*=0.05\,\mathrm{Mpc}^{-1}$ to the probed scales $k\sim 10^{7}\,\mathrm{Mpc}^{-1}$, $\ln(k/k_*)\approx 19$, so an amplitude measured to $\delta\ln\Delta^2_\delta\approx\ln 10\approx 2.3$ fixes the mean tilt over this range to $\delta(n_s-1)\sim 2.3/19\approx 0.1$. Anchoring the tilt to its CMB value, already determined there to $\sigma(n_s)\approx 0.003$~\cite{ACT:2025}, the same measurement fixes the running to $\delta\alpha_s\sim 2(2.3)/19^2\approx 0.013$. The lever arm enters linearly for the tilt but quadratically for the running, so the former is more than an order of magnitude short of the CMB, while the latter is only a factor of $\sim3$ weaker than the tightest current bound, $\alpha_s=0.0014\pm0.0041$ from CMB and Ly$\alpha$-forest data~\cite{DESILya:2026}. Even at this crude level of precision, a measurement here would be a useful check against \emph{drastic} departures from scale invariance of the primordial spectrum, which remain unexcluded over the intervening 16 $e$-folds.

\paragraph*{Enhanced structure from early dark-sector dynamics.} Several classes of early dark-sector dynamics can enhance small-scale structure, often introducing a characteristic scale tied to the horizon when the relevant dynamics occur.

A global symmetry broken \emph{after} inflation leaves the DM field uncorrelated between horizon patches, and the resulting $\mathcal{O}(1)$ isocurvature fluctuations collapse near matter-radiation equality into dense miniclusters~\cite{HoganRees:1988,KolbTkachev:1993}. Post-inflationary QCD axion miniclusters have canonical masses around $10^{-13}$--$10^{-12}\,M_\odot$ and subsequently assemble hierarchically into a steep mass function of minicluster halos~\cite{Eggemeier:2020}. More than ten percent of the DM collapses into asteroid-mass Bose stars at equality~\cite{Gorghetto:2024vnp}. For lighter axion-like particles, the later onset of oscillations can shift the characteristic minicluster mass directly into the range probed here~\cite{Fairbairn:2017sil}. Collapsing at $z\sim10^2$--$10^3$ rather than $z\sim10$, these structures are orders of magnitude denser than the CDM microhalos of Sec.~\ref{sec:pure_CDM}.

Enhanced structure does not require post-inflationary symmetry breaking. Attractive axion self-interactions at large initial field misalignment can drive a short-lived exponential growth of density perturbations. Under these conditions, an otherwise nearly scale-invariant adiabatic spectrum develops a spike near the horizon scale at the onset of oscillations, which collapses into dense halos, solitons, and potentially oscillons~\cite{Arvanitaki:2019rax}. Nonlinear resonances between axions of near-matching masses can also render density fluctuations nonperturbative, causing similar early collapse~\cite{Cyncynates:2021xzw,Cyncynates:2022wlq}. Attractive long-range dark-sector forces can also drive halo collapse during radiation domination~\cite{Savastano:2019zpr}.

A qualitatively different source is a \emph{first-order} dark-sector phase transition. Stochastic bubble nucleation can seed large curvature perturbations on scales tied to the transition, potentially leading to ultracompact structure~\cite{Liu:2022}. An early matter-dominated era terminated by dark-sector reheating likewise amplifies sub-horizon perturbations, producing abundant early-forming microhalos~\cite{ErickcekSigurdson:2011,Blanco:2019}. Vector DM produced by inflationary fluctuations inherits a peaked primordial spectrum that collapses into dense small-scale substructure~\cite{Graham:2015rva,Gorghetto:2022sue}.

Each one of these mechanisms pushes the abundance and internal density of substellar halos \emph{above} the CDM baseline, and is correspondingly easier to detect (Fig.~\ref{fig:SNR}). Every tenfold enhancement in scale density relaxes the astrometric precision required for a detection by a factor of $\sim3$, bringing these scenarios within reach of more modest campaigns than the fiducial ones considered here. The mass scale at which an excess appears would then date the dark-sector event that produced it.

\section{Discussion} \label{sec:discussion}
We have developed time-domain astrometric weak lensing of multiply imaged quasars as a probe of substellar DM structure. As the photon geodesics forming each macro-image traverse the fine-grained fluctuating gravitational potential of subhalos along the line of sight, they acquire stochastic deflections that manifest as excess astrometric noise with a calculable, red PSD (Eq.~\ref{eq:C}), or as an apparent angular acceleration (Eq.~\ref{eq:est_2_var_main}) for ``large'' halos whose crossing is not time resolved. Differencing the light-centroid positions of image pairs suppresses instrumental systematics, while the strong magnification of the brightest images amplifies the signal.

We calculated the signal and the dominant stellar background for two complementary systems: the galaxy-lensed quadruple B1422+231 and the cluster-lensed triple SDSS J1029+2623. We also estimated the fiducial CDM population of bound halos with $M_\mathrm{L}\sim10^{-6}$--$1\,M_\odot$ and the associated nonlinear matter power spectrum. B1422+231 is a favorable galaxy-lens target because it combines bright, highly magnified images with relatively low stellar convergence. 
At the differential precision forecast for EPIC~\cite{VanTilburg:2023tkl,Galanis:2023gef}, the acceleration channel reaches the fiducial CDM scale-density relation in certifiably star-free data (Figs.~\ref{fig:acc} and~\ref{fig:SNR}); the worst-case post-subtraction stellar residual reduces the variance SNR to approximately unity~\cite{companionstars}.
Cluster lenses offer a complementary route: their image environments may have much lower stellar convergence, although this must be established observationally, and moving unvirialized components can increase the astrometric signal.

Section~\ref{sec:implications} develops the DM science case for measuring halos of $10^{-6}$--$1\,M_\odot$, grown from modes some 16--19 $e$-folds beyond those probed by the CMB. A detection of bound structure down to $M_{\rm cut}\sim10^{-6}\,M_\odot$ raises the universal lower bound on the DM particle mass to $m_{\rm f}\gtrsim400\,\mathrm{keV}$ for fermions and $m_{\rm b}\gtrsim10^{-12}\,\mathrm{eV}$ for bosons (Eqs.~\ref{eq:fermion-bound} and~\ref{eq:boson-bound}), requires kinetic decoupling at $T_{\rm kd}\gtrsim10\,\mathrm{MeV}$, and fixes the running of the primordial tilt to $\delta\alpha_s\sim0.013$. An \emph{excess} would instead be evidence of a dark-sector event---miniclusters, a first-order phase transition, or an early matter-dominated era---and is easier to detect, since every tenfold increase in scale density relaxes the required astrometric precision by a factor of $\sim3$.

Related probes target overlapping masses but use different observables and face different limiting systematics. We discuss each in turn.

Differential timing of repeating FRBs along two sightlines~\cite{2024PhRvD.110b3516X} can be sensitive to the time-delay analog of the signal in Sec.~\ref{sec:observables}.  It requires a challenging setup of phase-referenced receivers that are 20 AU apart to reach enhanced-structure scenarios (beyond $\Lambda$CDM) at masses of $10^{-8}\,M_\odot$.  A second option is to use a strongly lensed, multiply imaged repeating FRB, where the signal may be contaminated by the stellar convergence of the lens and nonlinearities in the macrolens identified in Ref.~\cite{DaiLu:2017} and described in Sec.~\ref{sec:prospects}.

Photometric monitoring of extremely magnified stars crossing micro-caustics behind cluster lenses~\cite{2020AJ....159...49D} is the photometric analog of the stochastic channel of Sec.~\ref{sec:signals}: there too a large number of individually sub-critical halos superpose along the sightline to produce nearly Gaussian surface-density fluctuations, of amplitude $10^{-4}$--$10^{-3}$ on projected scales of $10$--$10^4\,\mathrm{AU}$, i.e.~at halo masses below those probed here. Its systematics are correspondingly distinct: the signal is a residual against a modeled fold-caustic light curve, acquired during rare and short (hours-to-days) transits whose timing must be forecast in advance, and it must be separated from substellar lenses in the cluster, from blended source stars, and from any compact baryonic structure on the same scales.

Pulsar timing residuals~\cite{Baghram:2011is,Dror:2019twh,Ramani:2020hdo,Lee:2020wfn} peak in sensitivity near $10^{-2}\,M_\odot$, but reaching $\Lambda$CDM abundances and densities is difficult on two counts: the \emph{local} subhalo abundance is drastically reduced by Galactic tides acting on the low concentrations that $\Lambda$CDM predicts~\cite{Lee:2020wfn}, and the stochastic gravitational-wave background for which evidence has recently emerged acts as red noise in the signal band, degrading the projected reach by one to three orders of magnitude~\cite{Foster:2026kfg,Cherukupalli:2026cda}.

This work is in a certain sense the complement of astrometric lensing searches for Milky Way substructure~\cite{VanTilburg:2018ykj,Mishra-Sharma:2020ynk,Mondino:2023pnc,Chen:2023xyj}, which exploit the large statistics of the \textit{Gaia} source catalog either by averaging the correlated deflection pattern imprinted across many sources or by searching for rare, high-significance single-lens events. Here the statistical leverage instead comes from the large number of halos along a $\gtrsim \mathrm{Gpc}$ line of sight and from the magnification of the strongly lensed images, and the population sampled is the cosmological one together with the substructure of a lens at moderate redshift, as opposed to the tidally processed present-day population in the solar neighborhood.

A more quantitative forecast requires a full time-domain likelihood over all epochs, images, and sky coordinates, with the macrolensing model (including its mass-sheet scale), image positions, proper motions, and source wander as nuisance parameters. This analysis pipeline should be validated on mock campaigns containing a refined DM (sub)halo spectrum, realistic stellar populations (several simultaneous perturbers per image, binaries, remnants, a full mass function), intrinsic source variability interpolated across the measured time delays, irregular cadence, and (potentially colored or correlated) instrumental noise. Monte Carlo realizations could also quantify the non-Gaussian tails expected when cuspier or compact profiles reduce the effective number of contributing perturbers.
On the theory side, the predicted (sub)halo populations at these small scales are sensitive to tidal stripping, mergers, and accretion during the hierarchical assembly of the lens galaxy~\cite{LaceyCole:1993,Wang:2020,DelosWhite:2023,Moore:2005,Zhao:2007,Delos:2019}, which warrant dedicated numerical and semi-analytic modeling~\cite{Benson:2012,Jiang:2021} both inside and outside the lensing galaxies and clusters. Such modeling would sharpen the predictions for the nonlinear matter power spectrum contributions from both environments.

The target pool of multiply imaged quasars is growing, even if its most valuable members remain scarce: wide-field surveys have enlarged the catalog of lensed quasars to several hundred~\cite{inada2008sloan,Lemon:2023}---the current census contains 364 spectroscopically confirmed systems, of which 87 are quads or triples, with sub-milliarcsecond \textit{Gaia} astrometry and photometry for the individual images~\cite{Ducourant:2026}---with thousands more anticipated~\cite{oguri2010gravitationally}. Intensity interferometry, however, is applicable only to the very brightest systems, most of which are likely already known. The bright sample may still grow modestly as some ambiguous systems are confirmed as true lenses by higher-resolution observations. The wide-separation, cluster-lensed quasars that furnish the clean, star-poor channel are considerably rarer: only about eight such cluster lenses, comprising nine quasars (one cluster lensing a pair), are currently known, all discovered since 2003 and several only recently~\cite{Inada:2003,Inada:2006ph,Dahle:2013,Shu:2018,Solhaug:2026,Wu:2025curling}. Their number should grow sharply as wide-field surveys mature---searches of \textit{Gaia}-based catalogs already yield hundreds of candidates awaiting confirmation~\cite{Wu:2025curling}, and Rubin's LSST alone is forecast to deliver $\sim80$ systems with image separations $\gtrsim10''$~\cite{Solhaug:2026}, with \textit{Euclid} and \textit{Roman} adding more. 

A larger sample would help in three concrete ways. First, because a sizeable fraction of cluster-lens sightlines are expected to be effectively star-free over a decade (Sec.~\ref{sec:cluster}), nondetections in such star-poor lenses should set exceptionally clean upper limits on the small-scale DM power spectrum (or might even see an unexpectedly large signal). Second, while the currently known cluster lenses are too faint for intensity interferometry, the brightest and most strongly magnified members of a much larger sample may reach fluxes comparable to the galaxy lens B1422+231, at which high-precision intensity interferometric observations become possible, along with characterization and subtraction of stellar microlensing contamination. Third, the acceleration signal scales as the fourth power of the effective transverse velocity (Eq.~\ref{eq:observable_2_parametric}), which for a cluster lens is set not by the bulk motion alone but by the coherently moving, clumped mass components that supply part of the local convergence and shear (Sec.~\ref{sec:cluster}). Actively merging, unvirialized clusters have subcomponents streaming at up to the merger velocity of order $10^3\,\mathrm{km\,s^{-1}}$, and that motion is magnified by $B^\mathrm{I}$ along with the images. A factor of a few in velocity is one to two orders of magnitude in acceleration variance. With only about eight cluster lenses known, there is little room to select on this, but a larger catalog has a higher probability of containing at least one bright quasar behind a merger in progress.

Exploiting this growing target pool will require a staged observational program. Immediately, archival and ongoing high-resolution imaging with the \textit{Hubble} and \textit{James Webb Space Telescopes} would refine the macrolensing models, stellar convergences, and time delays that enter our forecasts. Phase-referenced VLBI monitoring of radio-loud systems such as B1422+231 can pilot the differential-astrometry methodology at the $\sim 10\,\mathrm{\mu as}$ level, and perhaps identify exceptionally fast-moving quasar images.

On a longer timescale, $\mathrm{\mu as}$-level differential astrometry on wide-separation cluster lenses (e.g.~those of Refs.~\cite{Inada:2003, Dahle:2013, Shu:2018}) opens the clean, essentially star-free channel of Sec.~\ref{sec:cluster}, close to reaching the fiducial CDM population. This precision regime is a natural target for next-generation space or long-baseline radio interferometry and for purpose-built optical astrometric facilities. The required capabilities extrapolate from facilities now being upgraded on several fronts: the CHARA array is deploying adaptive optics, new beam combiners, and a mobile seventh telescope to extend its baselines~\cite{2024SPIE13095E..02G}; the GRAVITY+ upgrade of the VLTI is extending dual-field, phase-referenced (i.e., differential-astrometric) interferometry to faint sources over the whole sky~\cite{2022Msngr.189...17A}; the next-generation Event Horizon Telescope and space-VLBI concepts such as the \textit{Black Hole Explorer} push (sub)millimeter imaging toward few-$\mathrm{\mu as}$ resolution~\cite{2023Galax..11..107D,2024SPIE13092E..2DJ}; and formation-flying space interferometers, free of the atmosphere altogether, are the subject of active technology roadmaps~\cite{2019BAAS...51g.153M,2025arXiv251018920M}.

The full program of $0.1\,\mathrm{\mu as}$ relative astrometry of the brightest quadruply imaged quasars, reaching subhalos with bound masses of a few $10^{-6}\,M_\odot$ and thus the Earth-mass damping cutoff of a canonical WIMP, is a flagship science case for EPIC~\cite{VanTilburg:2023tkl,Galanis:2023gef,companionstars}. We emphasize that no existing facility delivers sub-microarcsecond relative astrometry on arcsecond scales. The DM reach quantified here provides a concrete science case for such an instrument. Together with the rapidly growing sample of multiply imaged quasars, facilities of this class would probe DM structure and early-universe physics on scales not accessible to current observations.

\acknowledgments
We thank Keiichi Asada, Masha Baryakhtar, Simon Birrer, Neal Dalal, Sten Delos, Marios Galanis, Ue-Li Pen, Julian Mu\~noz, Anna Nierenberg, Jessie Shelton, Sebastian Wagner-Carena, Roland Walter, Neal Weiner, and Huangyu Xiao for insightful discussions. We are grateful to Ana Acebron for sharing data on the macrolensing model of SDSS J1029+2623 in Tab.~\ref{tab:macro-cluster}, and to Abigail Moran for coordinating the companion work~\cite{companionstars}. 
We received helpful comments on this manuscript from Masha Baryakhtar, Cyril Creque-Sarbinowski, Marios Galanis, Junwu Huang, Abigail Moran, Christopher Pezanosky, Elena Pinetti, and Huangyu Xiao.
We acknowledge use of the \texttt{astropy}~\cite{Astropy:2013,Astropy:2018,Astropy:2022}, \texttt{CLASS}~\cite{Blas:2011}, \texttt{halofit}~\cite{Takahashi:2012}, \texttt{lenstronomy}~\cite{Birrer:2018}, \texttt{matplotlib}~\cite{Hunter:2007}, \texttt{numpy}~\cite{Harris:2020}, \texttt{pandas}~\cite{McKinney:2010}, and \texttt{scipy}~\cite{Virtanen:2020} software packages.
\texttt{Claude Code} was used for part of the code development and documentation.
This material is based upon work supported by the National Science Foundation under Grant Nos.~PHY-2622106 and PHY-1818899 and
by the Simons Investigator Grant No.~144924. 

\bigskip

\appendix

\section{Power Spectral Density}  \label{app:psd}

\subsection{Matter Power Spectrum} \label{app:power-spectrum}

The gravitational-potential power spectrum $P_\Phi$ is related to the matter power spectrum $P_\delta$ by
\begin{align}
    P_\Phi(k) = \left(\frac{4\pi G \overline{\rho} a^2}{k^2}\right)^2 P_\delta(k),
\end{align}
with $k$ the 3D comoving wavenumber and $\overline{\rho} = \overline{\rho}_{m,0}(1+z)^3$ the physical DM density\footnote{The fluctuations here are on such small length scales that baryons do not track the DM.} at the epoch along the path, so that $\overline{\rho}a^2 = \overline{\rho}_{m,0}(1+z)$. Consistent with the use of comoving wavenumber $k$, distances along ($\chi$) and transverse to ($\vect{x}_\perp$) the line of sight are comoving throughout this appendix, with $\chi(z) = (1+z)D(z)$ as in Sec.~\ref{sec:theory}. In our calculation of the two-point function, we use Limber's approximation
\begin{align}
    &\langle \Phi(\chi,\vect{x}_\perp) \Phi(\chi',\vect{x}'_\perp) \rangle \\
    &= \int \frac{\dd^3k}{(2\pi)^3} \, P_\Phi(k) e^{-i\left[ k_\parallel (\chi-\chi') + \vect{k}_\perp \cdot (\vect{x}_\perp - \vect{x}'_\perp) \right]} \\
    &\simeq \int \frac{\dd^2 k_\perp}{(2\pi)^2} \, P_\Phi(k_\perp) \delta(\chi-\chi') e^{-i \vect{k}_\perp \cdot (\vect{x}_\perp - \vect{x}'_\perp)}
\end{align}
to obtain the final line.
With this information, the two-point function $C^\mathrm{I}_{pq}(t)$ and the PSD $\widetilde{C}^\mathrm{I}_{pq}(\omega)$ of Eqs.~\ref{eq:2ptC_1} and~\ref{eq:2ptC_2} can be computed as:
\begin{widetext}
    \begin{align}
    C^\mathrm{I}_{pq}(t) &= B_{pi}^\mathrm{I} B_{qj}^\mathrm{I} 4 \iint_0^{\chi_\mathrm{S}} \dd \chi \, \dd \chi' \frac{\chi_\mathrm{S} - \chi}{\chi_\mathrm{S}\chi} \frac{\chi_\mathrm{S} - \chi'}{\chi_\mathrm{S}\chi'} \langle \partial_{\theta_i} \Phi(\chi,0) \partial_{\theta_j} \Phi(\chi',\chi' \widetilde{\mu}^\mathrm{I} t) \rangle  \\
    &= B_{pi}^\mathrm{I} B_{qj}^\mathrm{I} 4 \int_0^{\chi_\mathrm{S}} \dd \chi \, \left(\frac{\chi_\mathrm{S} - \chi}{\chi_\mathrm{S}}\right)^2 \int \frac{\dd^2 k_\perp}{(2\pi)^2} k_\perp^i k_\perp^j P_\Phi(k_\perp) e^{-i \vect{k}_\perp \cdot (\chi \widetilde{\vect{\mu}}{}^\mathrm{I} t)} \\
    \widetilde{C}^\mathrm{I}_{pq}(\omega)
    &= B_{pi}^\mathrm{I} B_{qj}^\mathrm{I} 4 \int_0^{\chi_\mathrm{S}} \dd \chi \, \left(\frac{\chi_\mathrm{S} - \chi}{\chi_\mathrm{S}}\right)^2 \int \frac{\dd^2 k_\perp}{(2\pi)^2} k_\perp^i k_\perp^j P_\Phi(k_\perp) (2\pi) \delta\left[ \omega - \vect{k}_\perp \cdot (\chi \widetilde{\vect{\mu}}{}^\mathrm{I}) \right] \\
    &=B_{pi}^\mathrm{I} B_{qj}^\mathrm{I} \mathcal{R}_{il} \mathcal{R}_{jm} 4 \int_0^{\chi_\mathrm{S}} \dd \chi \, \left(\frac{\chi_\mathrm{S} - \chi}{\chi_\mathrm{S}}\right)^2 (4\pi G \overline{\rho} a^2)^2 \int_0^\infty \frac{\dd k_\perp}{k_\perp}  \int_{-\frac{\pi}{2}}^{\frac{\pi}{2}}  \frac{\dd \phi}{2\pi}
    \begin{pmatrix}
        c_\phi^2 & 0 \\
        0 & s_\phi^2
    \end{pmatrix}_{lm}
    P_\delta(k_\perp) \delta(|\omega| - k_\perp \chi \widetilde{\mu}^\mathrm{I} c_\phi) \\
    &=B_{pi}^\mathrm{I} B_{qj}^\mathrm{I} \mathcal{R}_{il} \mathcal{R}_{jm} \frac{4}{|\omega|} \int_0^{\chi_\mathrm{S}} \dd \chi \, \left(\frac{\chi_\mathrm{S} - \chi}{\chi_\mathrm{S}}\right)^2 (4\pi G \overline{\rho} a^2)^2 \int_{-\frac{\pi}{2}}^{\frac{\pi}{2}}  \frac{\dd \phi}{2\pi}
    \begin{pmatrix}
        c_\phi^2 & 0 \\
        0 & s_\phi^2
    \end{pmatrix}_{lm}
    P_\delta \left(\frac{|\omega|}{\chi \widetilde{\mu}^\mathrm{I} c_\phi} \right). \label{eq:C_power_naive}
\end{align}
\end{widetext}
The steps parallel those of the one-halo-term derivation below (App.~\ref{app:1-halo}). In the second-to-last line, the delta function collapses the 2D $\vect{k}_\perp$ integral onto the ``aligned'' basis with $\hat{\vect{e}}'_1 = \hat{\widetilde{\vect{\mu}}}{}^\mathrm{I}$, in which $\phi$ is the angle of $\vect{k}_\perp$ away from the image proper motion; the rotation matrices $\mathcal{R}$, defined in Eq.~\ref{eq:rot_R}, undo that rotation by the position angle $\zeta^\mathrm{I}$ of $\widetilde{\vect{\mu}}{}^\mathrm{I}$ and return the result to the observer's basis. The last line is Eq.~\ref{eq:C_tilde_1} of the main text, up to the off-plane replacement $\chi \to \mathcal{X}(\chi)$ derived below, and with the two rotations and the projector absorbed into
\begin{align}
    \mathcal{M}_{ij}\big(\phi,\hat{\widetilde{\vect{\mu}}}{}^\mathrm{I}\big) = \mathcal{R}_{il} \mathcal{R}_{jm} \begin{pmatrix}
        c_\phi^2 & 0 \\
        0 & s_\phi^2
    \end{pmatrix}_{lm}, \label{eq:matM}
\end{align}
whose $\phi$ dependence encodes the projection of the potential gradient onto the sweep direction, and whose $\hat{\widetilde{\vect{\mu}}}{}^\mathrm{I}$ dependence enters only through $\zeta^\mathrm{I}$.

\paragraph*{Off-plane perturbers.} Away from the lens plane, both the amplification of a deflection and the sweep rate are reduced. 

Firstly, let $\vect{\alpha}_\mathrm{L}$ denote the reduced deflection of Eq.~\ref{eq:alpha_los} taken to the \emph{lens} plane instead of the source, i.e.~with the replacement $(\chi_\mathrm{S}-\chi)/\chi_\mathrm{S} \mapsto (\chi_\mathrm{L}-\chi)/\chi_\mathrm{L}$ for $\chi < \chi_\mathrm{L}$, and $\vect{\alpha}_\mathrm{L} = 0$ for  $\chi \geq \chi_\mathrm{L}$---this is the astrometric deflection that the ray has already acquired between observer and macrolens. To first order in the perturbation, propagating a ray through the planes of the microhalo lens and the macrolens returns a slightly modified lens equation for small-scale lenses \emph{in front of the macrolens},
\begin{align}
    \delta \theta^\mathrm{I}_{pi} = \alpha_{\mathrm{L} p} + B^\mathrm{I}_{pi} \big(\alpha_i - \alpha_{\mathrm{L}  i}\big), \label{eq:B_offplane}
\end{align}
wherein only the deflection still outstanding at the macrolens plane is magnified. A perturber behind the lens has $\vect{\alpha}_\mathrm{L} = 0$ and has its astrometric deflection magnified in full, whereas one at the observer has $\vect{\alpha}_\mathrm{L} = \vect{\alpha}$ and merely displaces the ray, unamplified.

Secondly, the reduction in the off-plane sweep rate follows from the geometry of the photon geodesics, which are approximately straight between deflections and therefore have a transverse comoving velocity that is piecewise linear in $\chi$, fixed by $\vect{v}_\mathrm{o}$ at the observer, by $\chi_\mathrm{L}\widetilde{\vect{\mu}}{}^\mathrm{I} + \vect{v}_\mathrm{L}$ at the lens plane, and by $\vect{v}_\mathrm{S}$ on the source. Dropping unmagnified velocity contributions (about a 1\% correction for B1422+231) leaves the ``tent function'' of Eq.~\ref{eq:sweep}.

The naive prescription of Eq.~\ref{eq:C_power_naive} over-amplifies perturbative foreground deflectors and over-sweeps background ones. The second effect is the more important of the two, because the sightline weight $\chi^3[(\chi_\mathrm{S}-\chi)/\chi_\mathrm{S}]^2(1+z)^2$ peaks well beyond the lens, while the first effect is a percent-level correction to the naive kernel, which Eq.~\ref{eq:C_tilde_1} omits (see Eq.~\ref{eq:L_S} below). For a scale-invariant spectrum, where the response scales as $\mathcal{X}^3$, taking $\mathcal{X} \to \chi$ would overestimate the contribution of the midpoint between lens and source by $[(\chi_\mathrm{L}+\chi_\mathrm{S})/\chi_\mathrm{L}]^3 \approx 2\times10^2$ (for B1422+231), whereas the over-amplification overestimates that of the plane halfway to the lens by a factor of $4$. When integrated over the total line of sight, taking $\mathcal{X} \to \chi$ would overestimate $\mathcal{L}_\mathrm{S}$ below by a factor of $48$ for B1422+231 and $5.1$ for SDSS J1029+2623.

For a nearly scale-invariant power spectrum with
\begin{align}
    P_\delta(k) = \Delta^2_\delta \frac{2\pi^2}{k^3},
\end{align}
the PSD reduces to:
\begin{align}
    \widetilde{C}^\mathrm{I}_{pq}(\omega) &= B_{pi}^\mathrm{I} B_{qj}^\mathrm{I} \mathcal{J}_{ij} \frac{16\pi}{15} \Delta^2_\delta \frac{\left(\widetilde{\mu}^\mathrm{I}\right)^3}{\omega^4} \mathcal{L}_\mathrm{S}. \label{eq:C_tilde_scale_invariant}
\end{align}
We factored out the line-of-sight integral, keeping $B^\mathrm{I}$ outside:
\begin{align}
    \mathcal{L}_\mathrm{S} \equiv \int_0^{\chi_\mathrm{S}} \dd \chi \, \mathcal{X}^3(\chi) \left(\frac{\chi_\mathrm{S} - \chi}{\chi_\mathrm{S}}\right)^2 (4\pi G \overline{\rho} a^2)^2 \,. \label{eq:L_S}
\end{align}
Incorporating Eq.~\ref{eq:B_offplane} would lower $\mathcal{L}_\mathrm{S}$ by a further $7\%$ and $20\%$ for the sightlines to B1422+231 and SDSS J1029+2623, respectively, for a total factor of $52$ and $6.3$ below the naive prescription; our numerical results include both corrections.

The dependence on the direction of $\widetilde{\vect{\mu}}{}^\mathrm{I}$ is encoded in the matrix:
\begin{align}
    \mathcal{J}_{ij} = \begin{pmatrix}
        1+3 c_{\zeta^\mathrm{I}}^2 & 3 s_{\zeta^\mathrm{I}} c_{\zeta^\mathrm{I}} \\
        3 s_{\zeta^\mathrm{I}} c_{\zeta^\mathrm{I}} & 1 + 3 s_{\zeta^\mathrm{I}}^2
    \end{pmatrix}_{ij}.
\end{align}
The resulting acceleration covariance is
\begin{align}
\langle \hat{\ddot \theta}_{0,p} \hat{\ddot \theta}_{0,q} \rangle 
&= \int \frac{\dd \omega}{2\pi} \widetilde{C}_{pq}(\omega) \omega^4 \mathcal{F}_2(\omega \tau) \\
&= \frac{32 \pi}{21} B_{pi}^\mathrm{I} B_{qj}^\mathrm{I} \mathcal{J}_{ij} \left(\widetilde{\mu}^\mathrm{I}\right)^3 \Delta^2_\delta \frac{\mathcal{L}_\mathrm{S}}{\tau}.
\end{align}

\paragraph*{Sensitivity curves in Fig.~\ref{fig:matter_power}.}
We obtain the orange and maroon curves by inverting Eq.~\ref{eq:C_tilde_1} for the power-spectrum amplitude one comoving wavenumber at a time. At each $k$, we insert a white spectrum truncated there:
\begin{align}
    P_\delta(k') &= 6\pi^2\, \Delta_{\delta,0}^2\, k^{-3}\, \Theta(k-k') \,, \label{eq:white-noise-1} \\
    \Delta_\delta^2(k') &= 3\,\Delta_{\delta,0}^2 \left(\frac{k'}{k}\right)^3 \Theta(k-k') \,. \label{eq:white-noise-2}
\end{align}
The normalization is fixed so that the template amplitude $\Delta_{\delta,0}^2$, reported on the vertical axis, is the total dimensionless variance, with the majority of the power in the top $e$-fold: $\int_{k/e}^{k} \Delta_\delta^2(k')\,\dd\ln k' = (1-e^{-3})\Delta_{\delta,0}^2$. For spectra nearly flat in dimensionless power, such as those of Sec.~\ref{sec:pure_CDM}, the threshold therefore applies to the local value $\Delta_\delta^2(k)$ up to an $\mathcal{O}(1)$ factor.

The truncation makes the angular integral of Eq.~\ref{eq:C_tilde_1} elementary: with $x \equiv |\omega|/(k \chi \widetilde{\mu}^{\mathrm{I}})$, the argument of $P_\delta$ falls below the cutoff only for $|\phi| < \arccos x$, so that $\int ({\dd\phi}/{2\pi})\,c_\phi^2 = (\arccos x+x\sqrt{1-x^2})/{2\pi}$ for $x<1$ and vanishes otherwise. We evaluate the remaining line-of-sight integral numerically in comoving distance $\chi$ from observer to source, at the cosmic mean density, and with appropriate lensing efficiency and off-plane weights given above. For $B^{\mathrm{I}}$, we take the tangential eigenvalue at the brightest image of each system, $\approx \! 10.2$ for image B of B1422+231 and $\approx \! 22.6$ for image C of SDSS J1029+2623; for $\widetilde{\mu}^{\mathrm{I}}$, the proper motion estimates of Secs.~\ref{sec:B1422} and~\ref{sec:cluster}.

The lens-bound curves of Fig.~\ref{fig:matter_power} invert the same expression for a subhalo population confined to the lens plane, where Eq.~\ref{eq:C_tilde_1} is exact with $\mathcal{X} = \chi_\mathrm{L}$. The signal for that halo population is set by its projected surface mass density $\Sigma^\mathrm{I}_\mathrm{L} = f_\mathrm{sub}\kappa^\mathrm{I}\Sigma_\mathrm{cr}$ of Sec.~\ref{sec:one_halo}, which e.g.~for image B of B1422+231 is $9.6\times10^1\,M_\odot\,\mathrm{pc^{-2}}$ ($5.3\times10^1\,M_\odot\,\mathrm{pc^{-2}}$ comoving) for $f_\mathrm{sub} = 0.1$ and $\kappa^\mathrm{B} = 0.46$. We collapse the $\dd \chi$ integral of Eq.~\ref{eq:C_tilde_1} onto the lens plane via the substitution
\begin{align}
    &\int_0^{\chi_\mathrm{S}} \dd \chi \left(\frac{\chi_\mathrm{S} - \chi}{\chi_\mathrm{S}}\right)^2 \big(4\pi G \overline{\rho}_{m,0}(1+z)\big)^2 \big[\cdots\big]_{\mathcal{X}(\chi)} \nonumber \\
    &\qquad \longmapsto \left(\frac{\chi_\mathrm{S} - \chi_\mathrm{L}}{\chi_\mathrm{S}}\right)^2 (4\pi G)^2 \frac{\overline{\rho}_{m,0} \, \Sigma^\mathrm{I}_\mathrm{L}}{f_\mathrm{ST}} \big[\cdots\big]_{\chi_\mathrm{L}}, \label{eq:slab_substitution}
\end{align}
where $\big[\cdots\big]_\mathcal{X}$ denotes the angular integral in the second line of Eq.~\ref{eq:C_tilde_1} at sweep length $\mathcal{X}$. 

The denominator $f_\mathrm{ST} \simeq 0.040$ is the fraction of the cosmic matter density that the Sheth--Tormen mass function places in halos of mass $M_\mathrm{min} < M < M_\mathrm{max}$ and that remains bound after the truncation of Sec.~\ref{sec:signals} (0.21 before truncation), with $M_\mathrm{max} = 10^9\,M_\odot$ the heaviest plausible surviving subhalo at the image radii. We take the same floor for this pure-CDM prediction as in the field-halo band, $M_\mathrm{min} = 10^{-13}\,M_\odot$, but with the subhalo concentrations of Ref.~\cite{Moline:2017} at host-centric distance $x_\mathrm{sub}$ and the truncation $r_\mathrm{t} = 2.3\,r_s$, normalized to the cosmic mean; the masses $M$ of the Sheth--Tormen template are the accreted masses $M_{200}$. The appearance of $\overline{\rho}_{m,0}$ and $f_\mathrm{ST}$ in Eq.~\ref{eq:slab_substitution} is because we report the subhalo density contrast relative to the cosmic mean. Using
\begin{align}
    \frac{\overline{\rho}_{m,0}\, P_\delta(k)}{f_\mathrm{ST}} &= \int \dd \ln M \, \hat{f}(M) \, M_\mathrm{L} \, \big| F(k|M) \big|^2 \nonumber\\
    &\equiv \big\langle M_\mathrm{L} |F|^2 \big\rangle \label{eq:Mu2}
\end{align}
with $\hat{f}$ the normalized bound-mass function, and $M_\mathrm{L}(M)$ and $F(k|M)$ the bound mass and form factor (Eq.~\ref{eq:F_tNFW}) of a subhalo of accreted mass $M$, the RHS of Eq.~\ref{eq:slab_substitution} is simply $(4\pi G)^2 (D_\mathrm{LS}/D_\mathrm{S})^2\,\Sigma^\mathrm{I}_\mathrm{L}\,\langle M_\mathrm{L} |F|^2\rangle$, so Eq.~\ref{eq:C_tilde_1} matches Eq.~\ref{eq:C} with $\kappa^\mathrm{I}_\mathrm{L} = f_\mathrm{sub}\kappa^\mathrm{I}$, in which no cosmic mean density appears at all.

The column on the right-hand side of Eq.~\ref{eq:slab_substitution} is equivalent to a cosmic-mean-density path of $\approx \! 33\,\mathrm{Gpc}$ (several times $\chi_\mathrm{S}$), a reminder that a strongly lensed sightline is anything but generic. At their optima, the lens-bound sensitivity curves accordingly lie a factor $\approx \! 55$ (stochastic) and $\approx \! 22$ (acceleration) below the line-of-sight curves, using otherwise the same system and survey parameters. 

Each curve marks the $\Delta_\delta^2$ at which a single signal statistic equals its noise: for the stochastic channel, $\widetilde{C}^{\mathrm{I}}_{pq}$ of Eq.~\ref{eq:C_tilde_1} at the lowest accessible mode $\omega=2\pi/\tau$ against the white-noise floor of Eq.~\ref{eq:noise_1}; for the acceleration channel, the smeared $\omega^4$ moment of Eq.~\ref{eq:est_2_var_main} against the noise of Eq.~\ref{eq:noise_2}. The stochastic threshold is conservative because we only use the dominant Fourier mode instead of an optimally weighted sum of modes.

\paragraph*{Strong-lensing boundary in Fig.~\ref{fig:matter_power}.}
The hatched region follows from the same line-of-sight kernel and truncated template, with the detection threshold replaced by a breakdown-of-perturbativity threshold. The convergence along a sightline to the source is
\begin{align}
    \kappa(\vect{\theta}) &= \int_0^{\chi_\mathrm{S}} \dd \chi \, W(\chi)\, \delta(\chi\vect{\theta},\chi) \,,  \\
    W(\chi) &\equiv 4\pi G \overline{\rho} a^2 \, \chi \, \frac{\chi_\mathrm{S}-\chi}{\chi_\mathrm{S}} \,.
\end{align}
In the Limber approximation, its variance is $\langle \kappa^2\rangle = \mathcal{K}_\mathrm{S} \int \dd^2 k_\perp/(2\pi)^2 \, P_\delta(k_\perp)$, with
\begin{align}
    \mathcal{K}_\mathrm{S} \equiv \int_0^{\chi_\mathrm{S}} \dd \chi \, W^2(\chi),
\end{align}
the same line-of-sight integral as $\mathcal{L}_\mathrm{S}$ in Eq.~\ref{eq:L_S} with $\mathcal{X}^3(\chi)$ replaced by $\chi^2$. Inserting the truncated white template as in Eq.~\ref{eq:white-noise-1} yields $\langle \kappa^2 \rangle = (3\pi/2)\,\Delta^2_{\delta,0}\,\mathcal{K}_\mathrm{S}/k$, so the sightline ceases to be weakly lensed by the power in the top $e$-fold at $k$ alone once
\begin{align}
    \Delta^2_{\delta,0} \gtrsim \frac{2k}{3\pi\,\mathcal{K}_\mathrm{S}}. \label{eq:strong_lens}
\end{align}
For our fiducial cosmology, $\Lambda_\mathrm{S} \equiv \chi_\mathrm{S}\mathcal{K}_\mathrm{S} = 0.23$ along the B1422+231 sightline ($\chi_\mathrm{S} = 6.9\,\mathrm{Gpc}$) and $0.062$ along that to SDSS J1029+2623 ($5.5\,\mathrm{Gpc}$), giving $\Delta^2_{\delta,0} \simeq 4.5\times10^3\,k\,[h/\mathrm{Mpc}]$ and $1.3\times10^4\,k\,[h/\mathrm{Mpc}]$, respectively. We hatch up to the former (and thus more stringent) boundary.

\subsection{One-Halo Terms} \label{app:1-halo}

To derive Eqs.~\ref{eq:C} and~\ref{eq:Q}, we substitute Eqs.~\ref{eq:alpha_FT1} and~\ref{eq:2ptalpha} into Eq.~\ref{eq:2ptC_1}, resolving the mean number density into the subhalo phase-space distribution $f_\mathrm{L}$ at distance $D_\mathrm{L}$, $n_\mathrm{L}(\vect{\theta}_\mathrm{L}) = \int \dd^3 v \, f_\mathrm{L}(\vect{\theta}_\mathrm{L},\vect{v})$, so that each deflector carries its own drift $\vect{\theta}_\mathrm{L}(t) = \vect{\theta}_\mathrm{L} + \vect{\mu}_\mathrm{L} t$ with unmagnified angular velocity $\vect{\mu}_\mathrm{L} = \vect{v}_\perp/[(1+z_\mathrm{L})D_\mathrm{L}]$ (cf.~Eq.~\ref{eq:mu} and footnote~\ref{fn:mu}):
\begin{widetext}
\begin{align}
    C^\mathrm{I}_{pq}(t) &= B_{pi}^\mathrm{I} B_{qj}^\mathrm{I} \int \dd D_\mathrm{L} \, D_\mathrm{L}^2 \int \dd^2 \theta_\mathrm{L} \int \dd^3 v \, f_\mathrm{L}(\vect{\theta}_\mathrm{L},\vect{v}) \nonumber \\
    &\phantom{=} \times \int \frac{\dd^2 k}{(2\pi)^2} \widetilde{\alpha}_i(\vect{k}) e^{-i \vect{k} \cdot [\vect{\theta}^\mathrm{I}(t') - \vect{\theta}_\mathrm{L}(t')]} \int \frac{\dd^2 k'}{(2\pi)^2} \widetilde{\alpha}_j(\vect{k}') e^{-i \vect{k}' \cdot [\vect{\theta}^\mathrm{I}(t'+t) - \vect{\theta}_\mathrm{L}(t'+t)]}, \label{eq:Cder_0} \\
    &\simeq B_{pi}^\mathrm{I} B_{qj}^\mathrm{I} \int \dd D_\mathrm{L} \, D_\mathrm{L}^2 \int \dd^2 \theta_\mathrm{L} \, n_\mathrm{L} \, \int \frac{\dd^2 k}{(2\pi)^2} \widetilde{\alpha}_i(\vect{k}) e^{-i \vect{k} \cdot [\vect{\theta}^\mathrm{I}(t') - \vect{\theta}_\mathrm{L}]} \int \frac{\dd^2 k'}{(2\pi)^2} \widetilde{\alpha}_j(\vect{k}') e^{-i \vect{k}' \cdot [\vect{\theta}^\mathrm{I}(t'+t) - \vect{\theta}_\mathrm{L}]}, \label{eq:Cder_1} \\
    &= B_{pi}^\mathrm{I} B_{qj}^\mathrm{I} \int \dd D_\mathrm{L} \, D_\mathrm{L}^2 \, n_\mathrm{L} \, \int \frac{\dd^2 k}{(2\pi)^2} \widetilde{\alpha}_i(\vect{k})\widetilde{\alpha}_j^*(\vect{k}) \exp\left\lbrace-i \vect{k} \cdot \widetilde{\vect{\mu}}{}^\mathrm{I} t \right\rbrace; \label{eq:Cder_2} \\
    \widetilde{C}^\mathrm{I}_{pq}(\omega) 
    &= B_{pi}^\mathrm{I} B_{qj}^\mathrm{I} \int \dd D_\mathrm{L} \, D_\mathrm{L}^2 \, n_\mathrm{L} \, \int \frac{\dd^2 k}{(2\pi)^2} \widetilde{\alpha}_i(\vect{k})\widetilde{\alpha}_j^*(\vect{k})  (2\pi) \delta(\omega - \vect{k} \cdot \widetilde{\vect{\mu}}{}^\mathrm{I}) , \label{eq:Cder_3}\\
    &=   B_{pi}^\mathrm{I} B_{qj}^\mathrm{I} \int \dd D_\mathrm{L} \, n_\mathrm{L} \left(\frac{D_\mathrm{LS}}{D_\mathrm{S}}\right)^2 (4 G M_\mathrm{L})^2 (2\pi) \int \dd^2 k \, \frac{k_i k_j}{k^4} F(k\gamma_\mathrm{L})^2 \delta(\omega - \vect{k} \cdot \widetilde{\vect{\mu}}{}^\mathrm{I}) , \label{eq:Cder_4}\\
    &=  B_{pi}^\mathrm{I} B_{qj}^\mathrm{I} \kappa_\mathrm{L} \theta_\mathrm{E,L}^2 \mathcal{R}_{il} \mathcal{R}_{jm} \mathcal{I}_{lm}. \label{eq:Cder_5}
\end{align}
\end{widetext}
Equation~\ref{eq:Cder_0} is exact within the one-halo approximation of Eq.~\ref{eq:2ptalpha}. Only the \emph{relative} displacement $\vect{\theta}^\mathrm{I}(t) - \vect{\theta}_\mathrm{L}(t)$ appears, so each subhalo is swept at the rate $\dd[\vect{\theta}^\mathrm{I} - \vect{\theta}_\mathrm{L}]/\dd t = \widetilde{\vect{\mu}}{}^\mathrm{I} - \vect{\mu}_\mathrm{L}$: the bulk source--lens--observer motion is magnified by $B^\mathrm{I}$ (Eq.~\ref{eq:thetaI}), whereas the deflector's own drift is not.
Equation~\ref{eq:Cder_1} is the approximation made in the main text, in which the internal dispersions are dropped, $|\vect{\mu}_\mathrm{L}| \ll \widetilde{\mu}^\mathrm{I}$, leaving the bulk image motion as the only source of time dependence; the velocity integral is then trivial, $\int \dd^3 v \, f_\mathrm{L} \to n_\mathrm{L}$, and the deflectors are static.
To obtain Eq.~\ref{eq:Cder_2}, we further use the image trajectory in Eq.~\ref{eq:thetaI}, assume that the linear proper motion dominates the stochastic deflection, and neglect variations of $n_\mathrm{L}$ across the relevant angular region.
Retaining the microhalo drifts would replace the frequency-selecting delta function of Eq.~\ref{eq:Cder_3}  by its normalized average over the subhalo velocity distribution, $\langle \delta\big(\omega - \vect{k}\cdot[\widetilde{\vect{\mu}}{}^\mathrm{I} - \vect{\mu}_\mathrm{L}]\big)\rangle_{\vect{\mu}_\mathrm{L}} \equiv n_\mathrm{L}^{-1}\int \dd^3 v \, f_\mathrm{L} \, \delta\big(\omega - \vect{k}\cdot[\widetilde{\vect{\mu}}{}^\mathrm{I} - \vect{\mu}_\mathrm{L}]\big)$. For a zero-mean, isotropic $\vect{\mu}_\mathrm{L}$ with per-axis dispersion $\sigma_{\mu,\mathrm{L}} \ll \widetilde{\mu}^\mathrm{I}$, the term linear in $\vect{\mu}_\mathrm{L}$ averages away, so this smears the spectrum only at relative $\mathcal{O}(\sigma_{\mu,\mathrm{L}}^2/(\widetilde{\mu}^\mathrm{I})^2) \lesssim 1\%$, which we fold into the sweep rate in quadrature, $\langle|\widetilde{\vect{\mu}}{}^\mathrm{I} - \vect{\mu}_\mathrm{L}|^2\rangle = (\widetilde{\mu}^\mathrm{I})^2 + 2\sigma_{\mu,\mathrm{L}}^2$.
Equation~\ref{eq:Cder_3} is the PSD (the Fourier transform of the previous line) from Eq.~\ref{eq:2ptC_2}, which becomes Eq.~\ref{eq:Cder_4} after plugging in the Fourier transform of the single-lens deflection in Eq.~\ref{eq:alpha_FT2}.
Equation~\ref{eq:Cder_5} follows after introducing the lens convergence $\kappa_\mathrm{L}$ and Einstein radius $\theta_\mathrm{E,L}$ in Eq.~\ref{eq:thetaE}. The delta function collapses the 2D $\vect{k}$ integral to a 1D angular one; in an orthonormal system with ``aligned'' basis vectors $\hat{\vect{e}}'_1 = \hat{\widetilde{\vect{\mu}}}{}^\mathrm{I}$ and $\hat{\vect{e}}'_2$, it reads:
\begin{align}
    \mathcal{I}_{lm} 
    &= \frac{2}{|\omega|} \int_{-\pi/2}^{\pi/2} \dd \phi \,
    \begin{pmatrix}
        c^2_\phi & 0 \\
        0 & s^2_\phi
    \end{pmatrix}_{lm} 
    F\left[\frac{\omega \gamma_\mathrm{L}}{\widetilde{\mu}^\mathrm{I} c_\phi} \right]^2.
\end{align}
The matrix $\mathcal{R}$ in Eq.~\ref{eq:Cder_5} represents the rotation by angle $\zeta^\mathrm{I}$ from the ``old'' to the ``aligned'' basis vectors, $\hat{\vect{e}}_j' = \mathcal{R}_{ij}\hat{\vect{e}}_i$, with
\begin{align}
    \mathcal{R}_{ij}  = \begin{pmatrix}
        \cos \zeta^\mathrm{I} & -\sin \zeta^\mathrm{I} \\
        \sin \zeta^\mathrm{I} & \cos \zeta^\mathrm{I}
    \end{pmatrix}
    \label{eq:rot_R}
\end{align}
The matrix multiplication in Eq.~\ref{eq:Cder_5} gives Eqs.~\ref{eq:C} and~\ref{eq:Q}.

\section{Macrolensing Models} \label{app:macrolensing}

\subsection{B1422+231} \label{app:B1422}

We model B1422+231 with a singular isothermal ellipsoid (SIE)~\cite{Kormann:1994,Keeton:2001} plus external shear, implemented in \texttt{lenstronomy}~\cite{Birrer:2018}. Figure~\ref{fig:diagram} shows the resulting configuration. 

\paragraph*{Inputs.} We adopt $z_\mathrm{L}=0.34$ and $z_\mathrm{S}=3.62$, giving physical (non-comoving) angular-diameter distances $D_\mathrm{L}\approx1.00\,\mathrm{Gpc}$, $D_\mathrm{S}\approx1.49\,\mathrm{Gpc}$, and $D_\mathrm{LS}\approx1.20\,\mathrm{Gpc}$ in the fiducial cosmology of Sec.~\ref{sec:intro}; the corresponding comoving distances of Sec.~\ref{sec:theory} are $\chi_\mathrm{L}\approx1.34\,\mathrm{Gpc}$ and $\chi_\mathrm{S}\approx6.89\,\mathrm{Gpc}$. 
The observed image positions $\vect{\theta}^\mathrm{I}$ relative to the brightest image B (at the origin) and flux ratios are listed in Tab.~\ref{tab:macro}; we use the deconvolved \emph{HST}/NICMOS $F160W$ ($H$-band) astrometry and photometry of Ref.~\cite[Tab.~4]{Sluse:2012}. Image~D sits closest to the lens galaxy~G. All position angles quoted in this appendix are measured east of north.
As an initial guess, we take the SIE$+\gamma$ model fit to those data in Ref.~\cite[Tab.~7, $N_\mathrm{lens}=1$]{Sluse:2012}. All four images are also resolved in \textit{Gaia} DR3, and Ref.~\cite{Ducourant:2026} reports a homogeneous SIE$+\gamma$ fit to that absolute astrometry; its positional residuals for this system are relatively large ($53\,\mathrm{mas}$ rms, driven by images B and D), whereas a single SIE$+\gamma$ reproduces the \emph{HST} image and galaxy positions~\cite{Sluse:2012}. 

\paragraph*{Fit.} We solve for the SIE parameters, external shear, and source position $\vect{\beta}_\mathrm{S}$ that reproduce the four image positions exactly (\texttt{lenstronomy} \texttt{Solver4Point} with \texttt{PROFILE\_SHEAR}). The best fit has $\theta_\mathrm{E}=0.776''$, ellipticity components $(e_1,e_2)=(0.065,-0.147)$ (ellipticity $0.28$ with major axis at $-56.9^\circ$), and galaxy center $(0.728,-0.650)''$; external shear $(\gamma_1,\gamma_2)=(-0.045,0.168)$, i.e.~$\gamma_\mathrm{ext}=0.174$ with shear axis at $37.5^\circ$; and source position $\vect{\beta}_\mathrm{S}=(0.527,-0.507)''$. Both orientations reproduce those of Ref.~\cite{Sluse:2012}, whose mass position angle is $-57.0^\circ$ and whose $\theta_\gamma=-52.5^\circ$ points toward the mass producing the shear and is therefore perpendicular to the shear axis. The large external shear reflects the group environment of~G.

\paragraph*{Outputs.} At each image we evaluate the convergence $\kappa$, shear $(\gamma_1,\gamma_2)$, signed magnification $A$, and stellar convergence $\kappa_*$ (Table~\ref{tab:macro}). The inverse magnification tensor used throughout the main text follows as $B^\mathrm{I}=\big[\begin{smallmatrix} 1-\kappa-\gamma_1 & -\gamma_2 \\ -\gamma_2 & 1-\kappa+\gamma_1 \end{smallmatrix}\big]^{-1}$, with $\det B^\mathrm{I}=A$. Images A, B, C are highly magnified with strong, partially aligned shear, whereas D lies well inside the critical curve and is demagnified.

\paragraph*{Host-halo location of the subhalos.} The subhalo concentration relation used for the fiducial CDM band of Fig.~\ref{fig:SNR}~\cite{Moline:2017} depends on the host-centric distance $x_\mathrm{sub}=R_\mathrm{sub}/R_{200}^\mathrm{G}$, which we estimate as follows. The fitted Einstein radius implies an isothermal velocity dispersion $\sigma_v = [\theta_\mathrm{E}D_\mathrm{S}/(4\pi D_\mathrm{LS})]^{1/2}\approx180\,\mathrm{km\,s^{-1}}$, in turn determining the mass $M_{200}=2\sigma^2 R_{200}/G_\mathrm{N}=(4\pi/3)\,200\,\rho_c(z_\mathrm{L})R_{200}^3\approx5\times10^{12}\,M_\odot$ and the virial radius $R_{200}\approx310\,\mathrm{kpc}$ at $z_\mathrm{L}$. The images lie $1.02''$, $0.97''$, $1.07''$, and $0.27''$ from the observed position of~G~\cite[Tab.~4]{Sluse:2012} for A, B, C, D, i.e.~$4.9$, $4.7$, $5.2$, and $1.3\,\mathrm{kpc}$ in projection ($4.85\,\mathrm{kpc}$ per arcsec), so their projected distances are $x_\perp = 0.016$, $0.015$, $0.017$, and $0.004$. Subhalos anywhere along the column contribute, and weighting the line of sight by an NFW subhalo number density with host concentration $c_\mathrm{h}=6$ gives median characteristic distances $x_\mathrm{sub}=0.031$, $0.030$, $0.033$, and $0.013$. We adopt $x_\mathrm{sub}=0.03$ for the bright images A, B, C. The same isothermal model sets the King radius of Eq.~\ref{eq:King} behind our truncation $r_\mathrm{t} = 2.3\,r_s$: inside the smallest projected distance of the bright images, $R = 4.7\,\mathrm{kpc}$ (image B), G has the mean density $\overline{\rho}_\mathrm{host} = 3\sigma_v^2/(2\pi G_\mathrm{N} R^2) \approx 0.17\,M_\odot\,\mathrm{pc^{-3}}$, 3.9 times that inside the median distance $x_\mathrm{sub}R_{200} \approx 9\,\mathrm{kpc}$.

\paragraph*{Stellar component.} The stellar convergence $\kappa_*$ is computed from a de Vaucouleurs profile centered on~G, with total (Salpeter) stellar mass $\log_{10}(M_*/M_\odot)=10.83$ and the measured light-profile shape of Ref.~\cite[Tab.~3]{Sluse:2012}: ellipticity $0.39$ (axis ratio $1${:}$0.61$), effective radius $\theta_\mathrm{eff}=0.41''$, and major-axis position angle $-58.9^\circ$. Our values in the last column of Table~\ref{tab:macro} are somewhat higher (and thus conservative) than those adopted in the microlensing analyses of this system, $\kappa_* = 0.037$, $0.052$, and $0.033$ at A, B, and C~\cite{Dogruel:2020,Biggs:2023}. For the microlensing background of Sec.~\ref{sec:sensitivity}, we take a characteristic microlens mass $M_*=0.3\,M_\odot$.

\paragraph*{Source size.} For the finite-source effects of Sec.~\ref{sec:theory}, we adopt the accretion-disk size derived for this system in the companion paper~\cite{companionstars}, obtained by flux-normalizing a thin-disk blackbody model~\cite{Galanis:2023gef} at vanishing inclination to the observed brightness of image A. The characteristic scale is the source-frame, face-on radius $R_\mathrm{src}\equiv R_{500}\approx6.9\times10^{15}\,\mathrm{cm}$ at which the disk reaches $T_{500}=hc/(\lambda_{500}k_\mathrm{B})$ for $\lambda_{500}=500\,\mathrm{nm}$; because this temperature-defined radius carries an extra factor $(1+z_\mathrm{S})^{4/3}$ relative to the plain geometric size, the unlensed angular size is $\theta_\mathrm{src}=R_\mathrm{src}/[D_\mathrm{S}(1+z_\mathrm{S})^{4/3}]\approx0.04\,\mathrm{\mu as}$. This agrees at the $\sim\!10\%$ level with the optical accretion-disk size of Ref.~\cite{mosquera2011microlensing} once its observed-frame $814\,\mathrm{nm}$, $\cos i=1/2$ estimate is rescaled to rest-frame $500\,\mathrm{nm}$ and face-on orientation ($\theta_\mathrm{src}\propto\lambda^{4/3}$ for $T\propto R^{-3/4}$).

\paragraph*{Flux-ratio anomaly.} The smooth model predicts flux ratios $\mathrm{A{:}B{:}C{:}D}=1{:}1.27{:}0.64{:}0.05$, to be compared with the observed $1{:}1.12{:}0.59{:}0.03$ (Tab.~\ref{tab:macro} magnifications). The discrepancy is due to the well-known flux-ratio anomaly of this system~\cite{MaoSchneider:1998}, conventionally attributed to substructure~\cite{DalalKochanek:2002}. It is not an artifact of the near-IR continuum. A microlens only modulates the flux of a source smaller than its own Einstein radius; a larger source averages over the microlensing magnification pattern and is left unperturbed. For $M_*=0.3\,M_\odot$ in this system, $\theta_{\mathrm{E},*}\approx1.4\,\mathrm{\mu as}$, i.e.~$R_{\mathrm{E},*}=\theta_{\mathrm{E},*}D_\mathrm{S}\approx3\times10^{16}\,\mathrm{cm}\approx0.01\,\mathrm{pc}$ projected onto the source plane, so the accretion disk responsible for the $F160W$ fluxes above ($\theta_\mathrm{src}\approx0.04\,\mathrm{\mu as}$) sits well inside the microlensing regime. The anomaly persists, however, in three tracers emitted far outside $R_{\mathrm{E},*}$, and hence achromatic under stellar microlensing while remaining sensitive to the much larger Einstein radii of $\gtrsim10^{6}\,M_\odot$ substructures: the radio flux ratios~\cite{Patnaik:1992}, from a synchrotron core extended well beyond a stellar Einstein radius~\cite{MaoSchneider:1998}; the $11.7\,\mathrm{\mu m}$ emission flux ratios~\cite{Chiba:2005}, from the hot inner edge of the dust torus at $R_\mathrm{src}\approx2.7\,\mathrm{pc}$; and the narrow-line flux ratios~\cite{Nierenberg:2014}, from a region $\approx60\,\mathrm{pc}$ across.

\begin{table}[t]
\caption{Best-fit macrolensing quantities for B1422+231: image positions $(\theta_x,\theta_y)$ in $\mathrm{arcsec} = \mathrm{''}$ relative to image B, convergence $\kappa$, shear $(\gamma_1,\gamma_2)$, signed magnification $A$, and smooth stellar convergence $\kappa_*$.}
\label{tab:macro}
\begin{ruledtabular}
\begin{tabular}{lccccccc}
I & $\theta_x$ $\mathrm{['']}$ & $\theta_y$ $\mathrm{['']}$ & $\kappa$ & $\gamma_1$ & $\gamma_2$ & $A$ & $\kappa_*$ \\
\hline
A & $0.386$ & $0.317$ & $0.385$ & $0.254$ & $0.410$ & $+6.9$ & $0.049$ \\
B & $0.000$ & $0.000$ & $0.463$ & $-0.097$ & $0.628$ & $-8.7$ & $0.094$ \\
C & $-0.336$ & $-0.752$ & $0.367$ & $-0.405$ & $0.099$ & $+4.4$ & $0.047$ \\
D & $0.947$ & $-0.801$ & $1.715$ & $-0.656$ & $1.770$ & $-0.33$ & $1.20$ \\
\end{tabular}
\end{ruledtabular}
\end{table}

\begin{table}[t]
\caption{Macrolensing quantities for SDSS J1029+2623 from the cluster model of Refs.~\cite{Acebron:2022,Acebron:2024}: image-position offsets $(\theta_x,\theta_y)$ in arcsec relative to image A (with $\theta_x$ along $+$RA), convergence $\kappa$, shear $(\gamma_1,\gamma_2)$, signed magnification $A$, and (estimated) stellar convergence $\kappa_*$.}
\label{tab:macro-cluster}
\begin{ruledtabular}
\begin{tabular}{lccccccc}
I & $\theta_x$ $\mathrm{['']}$ & $\theta_y$ $\mathrm{['']}$ & $\kappa$ & $\gamma_1$ & $\gamma_2$ & $A$ & $\kappa_*$ \\
\hline
A & $0.00$ & $0.00$ & $0.499$ & $0.070$ & $-0.309$ & $+6.6$ & $0.003$ \\
B & $4.05$ & $22.18$ & $0.459$ & $-0.332$ & $0.369$ & $+21.6$ & $0.003$ \\
C & $4.76$ & $20.45$ & $0.505$ & $-0.441$ & $0.310$ & $-21.9$ & $0.003$ \\
\end{tabular}
\end{ruledtabular}
\end{table}

\subsection{SDSS J1029+2623} \label{app:J1029}
For the cluster lens of Sec.~\ref{sec:cluster}, we do not refit the macrolensing model; instead we adopt the local lensing quantities at the three images directly from the cluster-scale reconstruction of Refs.~\cite{Acebron:2022,Acebron:2024}, evaluated at the quasar redshift. That reconstruction is anchored on VLT/MUSE spectroscopy of the cluster members and of the multiply imaged sources~\cite{Acebron:2022}, and subsequently refined by fitting the extended surface brightness of the lensed quasar host galaxy in addition to the pointlike multiple images~\cite{Acebron:2024}.

\paragraph*{Geometry.} We adopt $z_\mathrm{L}=0.588$ and $z_\mathrm{S}=2.199$, giving physical (non-comoving) angular-diameter distances $D_\mathrm{L}\approx1.37\,\mathrm{Gpc}$, $D_\mathrm{S}\approx1.71\,\mathrm{Gpc}$, $D_\mathrm{LS}\approx1.03\,\mathrm{Gpc}$ (comoving $\chi_\mathrm{L}\approx2.17\,\mathrm{Gpc}$, $\chi_\mathrm{S}\approx5.46\,\mathrm{Gpc}$), and a transverse scale of $6.6\,\mathrm{kpc}$ per arcsecond at the lens. The corresponding critical surface density is $\Sigma_\mathrm{cr}\approx2.0\times10^{3}\,M_\odot\,\mathrm{pc}^{-2}$, close to that of B1422+231.

\paragraph*{Inputs and outputs.} The three images span $22.5''$ (A to the close B,C pair), corresponding to a projected $\sim150\,\mathrm{kpc}$ at the lens; the images sit at $\sim90$--$100\,\mathrm{kpc}$ from the cluster center. The convergence $\kappa$ and shear $(\gamma_1,\gamma_2)$ at each image, and the derived signed magnification $A$, are listed in Tab.~\ref{tab:macro-cluster}. Images B and C straddle the tangential critical curve and are highly magnified, making them the prime targets for a differential astrometry search.

\paragraph*{Velocity scale.} The relevant transverse velocity is set by the cluster, not a single galaxy. The image splitting implies an Einstein radius $\theta_\mathrm{E}\approx11$--$18''$, i.e.~an isothermal velocity dispersion $\sigma_v=c\,[\theta_\mathrm{E}\,D_\mathrm{S}/(4\pi D_\mathrm{LS})]^{1/2}\approx0.8$--$1.0\times10^{3}\,\mathrm{km\,s^{-1}}$, which brackets the measured $\theta_\mathrm{E}=15.2''$ ($\sigma_v\approx0.94\times10^{3}\,\mathrm{km\,s^{-1}}$) of Ref.~\cite{Oguri:2012vg}. The \emph{bulk} relative proper motion of Eq.~\ref{eq:mu} combines the known CMB-dipole observer term (transverse component $212\,\mathrm{km\,s^{-1}}$, the dipole apex $35^\circ$ away~\cite{Planck:2018}), the cluster's bulk peculiar velocity ($\sim300\,\mathrm{km\,s^{-1}}$ per axis, from linear theory), and the distance-suppressed source term into $\langle|\vect{\mu}|^2\rangle^{1/2}\approx0.05\,\mathrm{\mu as\,yr^{-1}}$, i.e.~$(1+z_\mathrm{L})D_\mathrm{L}\langle|\vect{\mu}|^2\rangle^{1/2}\approx470\,\mathrm{km\,s^{-1}}$. Magnified by $B^\mathrm{I}$, this alone gives $\widetilde{\mu}^\mathrm{B,C}\approx1.0\,\mathrm{\mu as\,yr^{-1}}$. Two internal contributions modify it. First, a microhalo's own orbital motion enters the sweep rate \emph{unmagnified} (Sec.~\ref{sec:theory}): $\mu_\mathrm{L}=\sqrt{2}\,\sigma_v/[(1+z_\mathrm{L})D_\mathrm{L}]\approx0.14\,\mathrm{\mu as\,yr^{-1}}$, a $\approx1\%$ quadrature correction for the bright images. Second, differentiating the lens equation $\vect{\alpha}=\sum_c \vect{\alpha}_c(\vect{\theta}-\vect{\theta}_c(t))$ for moving macroscopic components gives $\dot{\vect{\theta}}{}^\mathrm{I} = B^\mathrm{I}[\vect{\mu} - \sum_c H_c\,\delta\vect{\mu}_c]$, with $H_c=\nabla\vect{\alpha}_c$ and $\delta\vect{\mu}_c$ each component's peculiar angular velocity. Only \emph{clumped} components contribute since a virialized halo's potential is static. The member-galaxy halos orbiting at $\sigma_v$ supply a fraction $f_\mathrm{gal}$ of the local convergence and shear ($\kappa+\gamma\approx1.0$ at B,C). Since the cluster is not in dynamical equilibrium, we estimate $f_\mathrm{gal}\approx0.1$--$0.4$. An isothermal halo with $\sigma_\mathrm{gal}\approx200\,\mathrm{km\,s^{-1}}$ at $\Delta\theta\approx5''$ alone would supply $\gamma\approx0.07$, about twice the shear at image B from the perturber that explains the flux-ratio anomaly~\cite{Kratzer:2011}. With fiducial $f_\mathrm{gal}=0.2$, $v_\mathrm{L}^\mathrm{eff}=[v_\mathrm{bulk}^2+2(f_\mathrm{gal}\sigma_v)^2]^{1/2}\approx550\,\mathrm{km\,s^{-1}}$ ($\approx500$--$740\,\mathrm{km\,s^{-1}}$ for $f_\mathrm{gal}=0.1$--$0.4$), giving $\widetilde{\mu}^\mathrm{B,C}\approx1.1\,\mathrm{\mu as\,yr^{-1}}$ and a lens-plane sweep velocity $(1+z_\mathrm{L})D_\mathrm{L}\widetilde{\mu}\approx1.1\times10^{4}\,\mathrm{km\,s^{-1}}$, whose uncertainty remains large until the relative image motions are measured directly.

\paragraph*{Tidal truncation.} For Eq.~\ref{eq:King}, we take the projected image distance, $R \approx 95\,\mathrm{kpc}$, and the isothermal cluster with $\sigma_v = 1000\,\mathrm{km\,s^{-1}}$ (the upper end of the range above, which truncates hardest), for which $\overline{\rho}_\mathrm{host}(<R) \approx 0.012\,M_\odot\,\mathrm{pc^{-3}}$, 14 times below that of G at the B1422+231 images. Subhalos with $\rho_s = 1\,M_\odot\,\mathrm{pc^{-3}}$ would then be truncated at $r_\mathrm{t}\approx7.4\,r_s$ and keep $M_\mathrm{L}/M_{200}\approx0.35$ of their would-be virial mass. For ease of comparison, we nevertheless use the B1422+231 value $r_\mathrm{t} = 2.3\,r_s$ for the cluster as well (Sec.~\ref{sec:signals}), at the same value of $f_\mathrm{sub} = 0.1$.

\paragraph*{Stellar component.} Lacking resolved photometry at the image positions, we take the stellar surface density from the intracluster light (ICL) of comparable systems. Resolved ICL profiles of the six Hubble Frontier Fields clusters give $\Sigma_*\approx2$--$6\,M_\odot\,\mathrm{pc^{-2}}$ at the $\sim90$--$100\,\mathrm{kpc}$ image radii~\cite[Fig.~5]{Morishita:2017}; those clusters are far more massive ($M_{500}\approx0.9$--$2.5\times10^{15}\,M_\odot$) than this one ($M_\mathrm{vir}\approx2\times10^{14}\,M_\odot$~\cite{Oguri:2012vg}), so they are a conservative reference. With $\Sigma_\mathrm{cr}=2.0\times10^{3}\,M_\odot\,\mathrm{pc^{-2}}$ we adopt the upper end, $\kappa_*=\Sigma_*/\Sigma_\mathrm{cr}=0.003$ (Tab.~\ref{tab:macro-cluster}), $\gtrsim15\times$ below B1422+231, leaving $N_\mathrm{fit}^\mathrm{I}=\kappa_*^\mathrm{I}\Sigma_\mathrm{cr}\pi\theta_\mathrm{fit}^2 D_\mathrm{L}^2/M_*\approx0.3$ stars (at $M_*=0.3\,M_\odot$) within a $10\,\mathrm{\mu as}$ reference region per image. Stacked photometry of lower-mass clusters gives comparable values at $100\,\mathrm{kpc}$~\cite{Zibetti:2005}. Microlensing of the cluster lens SDSS J1004+4112 instead gives higher values of $\kappa_*\approx0.02$--$0.07$, but at image radii of only $40$--$70\,\mathrm{kpc}$~\cite{ForesToribio:2024b}. Deep resolved photometry of the image environments is a prerequisite for any clean-channel claim. Cluster lenses are not automatically free of local complexity: the persistent radio--optical flux-ratio anomaly of this system points to a $\sim10^{9}\,M_\odot$ perturber near image B~\cite{Kratzer:2011}, and optical monitoring shows weak evidence for microlensing~\cite{Fohlmeister:2013}.

\paragraph*{Source size and the finite-source form factor.} We model the source as a Gaussian of angular size $\theta_\mathrm{src}$, magnified to $\theta^\mathrm{I}_\mathrm{src}=B_\mathrm{t}^\mathrm{I}\,\theta_\mathrm{src}$ at each image, with $B_\mathrm{t}^\mathrm{I}$ the tangential eigenvalue of $B^\mathrm{I}$ ($\approx22.5$ for B,C). The source form factor (Eq.~\ref{eq:finite_source}) multiplies the truncated-NFW one inside the $\phi$ integral of Eq.~\ref{eq:C}, which we evaluate numerically. For the optical continuum, $R_\mathrm{src}\approx1.5\times10^{15}\,\mathrm{cm}$ ($\approx0.6\,$light-day), i.e.~unlensed $\theta_\mathrm{src}\approx0.06\,\mathrm{\mu as}$ and magnified $\theta^\mathrm{I}_\mathrm{src}\approx1.3\,\mathrm{\mu as}$, the acceleration peak (at $\gamma_\mathrm{L}\approx14\,\mathrm{\mu as}$, $M_\mathrm{L}\approx0.022\,M_\odot$ for $\rho_s=1\,M_\odot\,\mathrm{pc^{-3}}$) is lowered by 16\%. Compact radio cores, $R_\mathrm{src}=\lbrace0.01,0.1,1\rbrace\,\mathrm{pc}$ (magnified $\theta^\mathrm{I}_\mathrm{src}\approx\lbrace27,270,2700\rbrace\,\mathrm{\mu as}$), instead suppress all but the largest halos, shifting the surviving acceleration peak to $M_\mathrm{L}\approx\lbrace30,3\times10^{4},9\times10^{5}\rbrace\,M_\odot$ with an amplitude that falls roughly as $1/\theta^\mathrm{I}_\mathrm{src}$. For the most compact core the peak covariance is $\approx1.7\times10^{-5}\,(\mathrm{\mu as/yr^2})^2$, requiring a VLBI light-centroiding precision $\sigma_{\delta\theta}\lesssim0.26\,\mathrm{\mu as}$ for $\mathrm{SNR}=1$ over the fiducial $\tau=10\,\mathrm{yr}$, $N=300$ campaign.

\section{Proper Motion \& Acceleration Estimators} \label{app:mu-alpha-est}
Consider $N=\tau/\Delta t$ observations with uncertainty $\sigma_{\delta\theta}$, equally spaced by $\Delta t$ over $t\in[-\tau/2,\tau/2]$. 
For notational simplicity, we work in one dimension; the two-dimensional generalization is straightforward. Assuming Gaussian errors, the optimal estimators for proper motion $\dot{\theta}_0$ and acceleration $\ddot{\theta}_0$ (both referenced to the midpoint $t=0$ of observations) are those values that minimize the chi-squared statistic:
\begin{align}
    \chi^2 = \sum_{n = -N/2}^{N/2} \frac{1}{\sigma_{\delta \theta}^2} \left\lbrace
    \theta_n - \left[ \theta_0 + \dot{\theta}_0 (n \Delta t) + \ddot{\theta}_0 \frac{(n\Delta t)^2}{2} \right]
    \right\rbrace^2.
\end{align}
At the minimum $\partial_{\theta_0} \chi^2 = \partial_{\dot\theta_0} \chi^2 = \partial_{\ddot \theta_0} \chi^2 = 0$, we have:
\begin{align}
    \hat \theta_0 &= \frac{9}{4} \frac{\sum \theta_n}{N} - 15 \frac{\sum n^2 \theta_n}{N^3}, \label{eq:est_0} \\
    \hat {\dot \theta}_0 &= \frac{12}{ \tau} \frac{\sum n \theta_n}{N^2}, \label{eq:est_1} \\
    \hat {\ddot \theta}_0 &= \frac{30}{\tau^2} \left\lbrace - \frac{\sum \theta_n}{N} + 12 \frac{\sum n^2 \theta_n}{N^3} \right\rbrace; \label{eq:est_2} 
\end{align}
to leading order in the limit of $N \gg 1$. These estimators are unbiased. Since the lensing signal under consideration is stochastic, $\langle \theta_n \rangle = 0$, we also have that Eqs.~\ref{eq:est_0}--\ref{eq:est_2} vanish in expectation value.

Their variances do not vanish. In the continuum limit (valid for $N \gg 1$), we can take:
\begin{align}
\frac{\sum n^p \theta_n}{N^{p+1}} \to \frac{\int_{-\tau/2}^{\tau/2} \dd t \, t^p \theta(t)}{\tau^{p+1}},
\end{align}
and use $\langle \theta(t) \theta(t') \rangle = C(t'-t) = \int \frac{\dd\omega}{2\pi} \widetilde{C}(\omega) e^{-i \omega (t-t')}$ from Eqs.~\ref{eq:2ptC_1} and~\ref{eq:2ptC_2}.

This yields a proper motion variance of:
\begin{alignat}{2}
    \big\langle \hat{\dot \theta}^2_0\big\rangle
    &= \left(\frac{12}{\tau}\right)^2 \iint_{-\tau/2}^{\tau/2} \frac{\dd t \, \dd t'}{\tau^2} \, \int \frac{\dd \omega}{2\pi} \widetilde{C}(\omega) e^{-i\omega (t-t')}  \frac{t \, t'}{\tau^2} \nonumber \\
    &= \int \frac{\dd \omega}{2\pi} \widetilde{C}(\omega) \omega^2 \mathcal{F}_1(\omega \tau) \label{eq:est_1_var}
\end{alignat}
with a smearing form factor defined as:
\begin{align}
    \mathcal{F}_1(a) &\equiv \frac{144 \left(a \cos \left(\frac{a}{2}\right)-2 \sin \left(\frac{a}{2}\right)\right)^2}{a^6} \\
    &\simeq \begin{cases} 
    1 - \frac{a^2}{20} +\mathcal{O}(a^4) & |a| \ll 1 \\
    \frac{144 \cos ^2\left(\frac{a}{2}\right)}{a^4} + \mathcal{O}(a^{-5}) & |a| \gg 1.
    \end{cases}
\end{align}
Similarly, the acceleration estimator's variance is:
\begin{alignat}{2}
    \big\langle \hat{\ddot \theta}^2_0\big\rangle
    &= \left(\frac{30}{\tau^2}\right)^2 \iint_{-\tau/2}^{\tau/2} \frac{\dd t \, \dd t'}{\tau^2} \, \int \frac{\dd \omega}{2\pi} \widetilde{C}(\omega) e^{-i\omega (t-t')}  \\
    & \hspace{12em} \times \left[\frac{144 t^2 t^{\prime 2}}{\tau^4} + 1 - \frac{24 t^2}{\tau^2}\right] \nonumber \\ 
    &= \int \frac{\dd \omega}{2\pi} \widetilde{C}(\omega) \omega^4 \mathcal{F}_2(\omega \tau) \label{eq:est_2_var}
\end{alignat}
and a smearing form factor $\mathcal{F}_2$:
\begin{align}
    \mathcal{F}_2(a) &= \frac{14400 \left[\left(a^2-12\right) \sin \left(\frac{a}{2}\right)+6 a \cos \left(\frac{a}{2}\right)\right]^2}{a^{10}} \\
    &\simeq \begin{cases} 
    1 - \frac{a^2}{28} +\mathcal{O}(a^4) & |a| \ll 1 \\
    \frac{14400 \sin ^2\left(\frac{a}{2}\right)}{a^6} + \mathcal{O}(a^{-7}) & |a| \gg 1.
    \end{cases} 
    \label{eq:F_2_smearing}
\end{align}
These variances have a simple interpretation. The proper-motion and acceleration estimators are first and second time derivatives of the position $\theta(t)$, each derivative bringing a factor $-i\omega$; their variances are therefore the \emph{position} spectrum $\widetilde{C}(\omega)$ of Eq.~\ref{eq:C} weighted by $\omega^2$ or $\omega^4$, the even powers of $\omega$ in Eqs.~\ref{eq:est_1_var} and~\ref{eq:est_2_var}. The form factors $\mathcal{F}_{1,2}(\omega\tau)$ correct for the estimators fitting a finite window $\tau$ rather than differentiating at a point: they tend to unity for $\omega\tau \to 0$, recovering the ideal derivative, and roll off for $\omega\tau \gg 1$, so modes with $|\omega| \gg 1/\tau$ contribute negligibly to the \emph{mean} variances.

The two-dimensional generalizations of Eqs.~\ref{eq:est_1_var} and~\ref{eq:est_2_var} simply amount to adding index subscripts appropriately: $C(t) \to \langle \theta_i(t) \theta_j(t') \rangle =  C_{ij}(t'-t)$ and $\widetilde{C}(\omega) \to \widetilde{C}_{ij}(\omega)$. This replacement is valid if both dimensions are measured at the same regular rate and precision.

The \emph{measurement noise} on the proper motion and acceleration, $\langle \hat{\dot \theta}_{0,i} \hat{\dot \theta}_{0,j} \rangle \equiv \delta_{ij} \sigma_{\dot \theta}^2$ and $\langle \hat{\ddot \theta}_{0,i} \hat{\ddot \theta}_{0,j} \rangle \equiv \delta_{ij} \sigma_{\ddot \theta}^2$, can simply be estimated by squaring Eqs.~\ref{eq:est_1}--\ref{eq:est_2} and using $\langle \theta_n^2 \rangle = \sigma_{\delta \theta}^2$.
The result
\begin{align}
    \sigma_{\dot \theta}^2 &= 12 \, \sigma_{\delta \theta}^2\frac{\Delta t}{\tau^3}, \label{eq:noise_mu} \\
    \sigma_{\ddot \theta}^2 &= 720 \, \sigma_{\delta \theta}^2 \frac{\Delta t}{\tau^5}; 
\end{align}
agrees with that obtained in Refs.~\cite{VanTilburg:2018ykj,Mondino:2023pnc}, and is used in Eqs.~\ref{eq:noise_1}~and~\ref{eq:noise_2} in the main text.

\section{Nonlinear Matter Power Spectrum Predictions} \label{app:nl-power}
Here, we detail our estimates of the nonlinear dimensionless matter power spectrum $\Delta^2_\delta(k)$ plotted in Fig.~\ref{fig:matter_power}, on the sub-parsec scales relevant to the astrometric signal: a halo-model (Press--Schechter) one-halo term, evaluated both without a cutoff (purely cold DM), and with the fiducial WIMP damping cutoff supplemented by prompt cusps.

\subsection{Linear Spectrum, Variance, and Collapse}
We take the linear CDM spectrum $\Delta^2_{\delta,\mathrm{lin}}(k)$ from the Boltzmann code \texttt{CLASS}~\cite{Blas:2011}, extended to small scales with the logarithmic growth of Eq.~\ref{eq:growth}. For the WIMP hypothesis, we impose the damping cutoff multiplicatively,
\begin{align}
    \Delta^2_{\delta,\mathrm{lin}}(k) \to \Delta^2_{\delta,\mathrm{lin}}(k)\, e^{-(k/k_{\rm cut})^2},
\end{align}
with $k_{\rm cut}\simeq 1.06\times 10^6\,\mathrm{Mpc}^{-1}$ for our fiducial $100\,\mathrm{GeV}$, $T_{\rm kd}=30\,\mathrm{MeV}$ thermal relic~\cite{Green:2005,Bertschinger:2006,DelosWhite:2023}. The variance of the field smoothed on a mass scale $M$ with a real-space top-hat of comoving radius $R = (3M/4\pi\overline{\rho}_{m,0})^{1/3}$ is
\begin{align}
    \sigma^2(M) = \int_0^\infty \frac{\dd k}{k}\, \Delta^2_{\delta,\mathrm{lin}}(k)\, W^2(kR), \label{eq:sigmaM}
\end{align}
with the real-space top-hat $W(x) = 3(\sin x - x\cos x)/x^3$. The rms density contrast $\sigma(M)$ rises slowly to $\approx 15$ near the WIMP cutoff mass $M_{\rm cut}\simeq (4\pi/3)\overline{\rho}_{m,0}(\pi/k_{\rm cut})^3 \approx 4\times 10^{-6}\,M_\odot$. A halo of mass $M$ turns nonlinear at the scale factor $a_{\rm coll}$ where $\sigma(M)\,D(a_{\rm coll}) = \delta_c = 1.686$, with $D(a)$ the linear growth factor normalized to $D(1)=1$; its characteristic density then tracks the mean density at collapse, $\rho_s \propto \overline{\rho}_m(a_{\rm coll}) \propto (1+z_{\rm coll})^3$. 

\subsection{Field Halos}
On small scales, the nonlinear power spectrum is dominated by the one-halo term,
\begin{align}
    P^{\rm 1h}_\delta(k) = \frac{1}{\overline{\rho}_{m,0}^2}\int \dd M\, \frac{\dd n}{\dd M}\, M^2\, |u(k|M)|^2, \label{eq:P1h}
\end{align}
where $u(k|M)$ is the density-profile transform normalized to $u\to1$ as $k\to0$. We adopt the Sheth--Tormen mass function~\cite{PressSchechter:1974,ShethTormen:1999},
\begin{align}
    \frac{\dd n}{\dd \ln M} = \frac{\overline{\rho}_{m,0}}{M}\, f(\nu)\, \left|\frac{\dd \ln \sigma}{\dd \ln M}\right|,
\end{align}
with first-crossing distribution $f(\nu) = A\sqrt{2/\pi}\,\tilde\nu\,(1+\tilde\nu^{-2p})\,e^{-\tilde\nu^2/2}$, peak height $\nu = \delta_c/\sigma(M)$, $\tilde\nu = \sqrt{a}\,\nu$, and $(A,a,p)=(0.3222,0.707,0.3)$. Halos are taken to be NFW, $\rho \propto (r/r_s)^{-1}(1+r/r_s)^{-2}$, truncated at $r_{200} = c\,r_s$. The concentration $c(M)$, calibrated in simulations only down to $\sim 10^{-6}\,M_\odot$ at $z=0$ in zoom-in configurations~\cite{Wang:2020} and otherwise extrapolated over many decades in mass, is a significant modeling uncertainty. We  bracket it between the relation of Ref.~\cite{Wang:2020} (the lowest concentrations in the literature at these masses) and the relation of Ref.~\cite{SanchezCondePrada:2014}, which yields roughly twice the concentration near the cutoff mass. In a WIMP scenario, we suppress both of these relations above $M_{\rm cut}$ by the cutoff-suppression factor of Ref.~\cite{Wang:2020}. Since the amplitude of the one-halo term at wavenumber $k$ scales as the squared central density of the halos resolved at that scale, $\propto [c^2/m(c)]^2$, the resulting band widens toward high $k$, reaching a factor of several at the smallest scales. The corresponding transform is the standard NFW form factor
\begin{align}
    u(k|M) = \frac{1}{m(c)}\Big[ &\sin(kr_s)\big(\mathrm{Si}((1{+}c)kr_s) - \mathrm{Si}(kr_s)\big) \nonumber\\
    &+ \cos(kr_s)\big(\mathrm{Ci}((1{+}c)kr_s) - \mathrm{Ci}(kr_s)\big) \nonumber\\
    &- \frac{\sin(ckr_s)}{(1{+}c)kr_s}\Big], \label{eq:u_NFW}
\end{align}
with $m(c) = \ln(1+c) - c/(1+c)$. Evaluated this way, $\Delta^2_\delta = k^3 P^{\rm 1h}_\delta/2\pi^2$ matches the \texttt{halofit} result~\cite{Takahashi:2012} to within tens of percent at $k\sim 10$--$30\,h/\mathrm{Mpc}$, where the two concentration relations agree, and extends it to sub-parsec scales (dark gray band in Fig.~\ref{fig:matter_power}).

The near-flatness of the pure-CDM one-halo term has a simple origin. At wavenumber $k$ the integrand of Eq.~\ref{eq:P1h} peaks on the marginally resolved halos, those with $k\,r_s(M)$ of order a few: lighter halos act as coherent point masses [$u(k|M)\to 1$] but carry too little $M^2\,\dd n/\dd M$, while heavier ones are suppressed by their $r^{-1}$ central cusp, $|u(k|M)|^2 \propto (k r_s)^{-4}$. Those dominant halos have $r_{200} = c\,r_s \sim c/k$ and hence $k^3 M \propto c^3(M)$, so the explicit $k^3$ in $\Delta^2_\delta$ cancels against the $k^{-3}$ scaling of halo mass, leaving an amplitude that depends on $k$ only through the slow mass dependence of the concentration and of the mass function. 

\subsection{Prompt Cusps}
As a peak in the linear density field collapses, a power-law cusp forms quasi-instantaneously at its center~\cite{DelosWhite:2023},
\begin{align}
    \rho_{\rm cusp}(r) = \begin{cases} \mathcal{A}\, r^{-3/2}, & r_{\rm core} < r < r_{\rm cusp}, \\[2pt] \mathcal{A}\, r_{\rm core}^{-3/2}, & r < r_{\rm core}, \end{cases}
\end{align}
with amplitude $\mathcal{A} \simeq 24\,\overline{\rho}_{m,0}\,a_{\rm coll}^{-3/2}R^{3/2}$ and outer extent $r_{\rm cusp}\simeq 0.11\,a_{\rm coll}R$ fixed by the peak's comoving size $R\equiv|\delta/\nabla^2\delta|^{1/2}$ and collapse time $a_{\rm coll}$, and an inner core at $r_{\rm core}\simeq r_{\rm cusp}/500$ set by the maximal coarse-grained phase-space density (Liouville's theorem). The cusp mass within $r_{\rm cusp}$ is $M_{\rm cusp} = (8\pi/3)\mathcal{A}\,r_{\rm cusp}^{3/2}$. The transform of a single cusp,
\begin{align}
    \tilde\rho(k) = \frac{4\pi}{k}\int_0^\infty \dd r\, r\,\rho_{\rm cusp}(r)\sin(kr),
\end{align}
is analytic; the $r^{-3/2}$ part yields Fresnel integrals $S(x) = \int_0^x \sin(\pi t^2/2)\,\dd t$,
\begin{align}
    \tilde\rho(k) = \frac{4\pi \mathcal{A}}{k}\Bigg[& \sqrt{\frac{2\pi}{k}}\bigg( S\Big(\sqrt{\tfrac{2k r_{\rm cusp}}{\pi}}\Big) - S\Big(\sqrt{\tfrac{2k r_{\rm core}}{\pi}}\Big)\bigg) \nonumber\\
    &+ \frac{\sin(kr_{\rm core}) - kr_{\rm core}\cos(kr_{\rm core})}{k^2\, r_{\rm core}^{3/2}}\Bigg]. \label{eq:cusp_FT}
\end{align}
This has three regimes: for $k \ll 1/r_{\rm cusp}$ the cusp is unresolved and $\tilde\rho \to M_{\rm cusp}$; for $1/r_{\rm cusp}\ll k\ll 1/r_{\rm core}$ the $r^{-3/2}$ profile gives $\tilde\rho\to 4\pi \mathcal{A}\sqrt{\pi/2}\,k^{-3/2}$; and for $k\gtrsim 1/r_{\rm core}$ the constant-density core cuts off the transform. Equation~\ref{eq:cusp_FT} is also what we use for the cusp \emph{signal} estimates of Sec.~\ref{sec:implications}, in place of the truncated-NFW form factor of Sec.~\ref{sec:signals}. The distinction is immaterial while the cusps are unresolved, but not once $k\,r_{\rm cusp}\gtrsim1$: in halo form-factor terms, the intermediate regime reads $F = \tilde\rho/M_{\rm cusp}\to 3\sqrt{\pi/8}\,(k\,r_{\rm cusp})^{-3/2}$, a shallower power law than the $(k r_s)^{-2}$ of the NFW $1/r$ cusp. The cusp population contributes a one-halo term $P_\delta^{\rm cusp}(k) = \overline{\rho}_{m,0}^{-2}\int \dd n\, |\tilde\rho(k)|^2$. On the intermediate plateau, $|\tilde\rho|^2 = 8\pi^3 \mathcal{A}^2 k^{-3}$, so
\begin{align}
    \Delta^2_{\delta,\rm cusp}(k)\big|_{\rm plateau} = \frac{4\pi}{\overline{\rho}_{m,0}^2}\int \dd n\, \mathcal{A}^2. \label{eq:cusp_plateau}
\end{align}
At $k\lesssim 1/r_{\rm cusp}$, the cusps act as point sources and $\Delta^2_{\delta,\rm cusp}\propto k^3$, while above $k\sim 1/r_{\rm core}$ it is cut off by the core.

We fix the plateau amplitude using the cusp annihilation rate of Ref.~\cite{DelosWhite:2023}. The annihilation $J$-factor of one cusp is $J_{\rm cusp} = 4\pi \mathcal{A}^2[0.531 + \ln(r_{\rm cusp}/r_{\rm core})]$, and its volume-average per unit DM mass is $J_{\rm cusps}/M_{\rm DM} = \overline{\rho}_{m,0}^{-1}\int\dd n\, J_{\rm cusp} \simeq 160\,f_{\rm surv}(1+z_5)^3\,\overline{\rho}_{m,0}$, with $z_5\simeq 31$ the redshift by which $5\%$ of peaks have collapsed and $f_{\rm surv}\simeq 0.5$ the survival fraction. Eliminating $\int\dd n\, \mathcal{A}^2$ between this relation and Eq.~\ref{eq:cusp_plateau} gives a plateau independent of the cusp abundance,
\begin{align}
    \Delta^2_{\delta,\rm cusp}\big|_{\rm plateau} = \frac{160\,f_{\rm surv}(1+z_5)^3}{0.531 + \ln(r_{\rm cusp}/r_{\rm core})} \approx 4\times 10^5.
\end{align}
For the thin blue curve in Fig.~\ref{fig:matter_power} we multiply this amplitude by the single-cusp shape $|\tilde\rho(k)|^2 k^3/(8\pi^3 \mathcal{A}^2)$, averaged over a log-normal spread of width $\sigma_{\ln r_{\rm cusp}} = 0.6$ about $r_{\rm cusp}\sim 5\times10^{-3}\,\mathrm{pc}$.

\subsection{Irreducible Scalar Dark Matter Power} \label{app:scalar-DM-power}
Interference fringes in a bath of virialized scalar DM waves produce irreducible density fluctuations. These unbound fluctuations can perturb stellar trajectories and photon geodesics. Examples include gravitational heating of stars in dwarf galaxies and star clusters~\cite{BarOr:2018pxz,Marsh:2018zyw,Dalal:2022rmp}, and metric perturbations in pulsar-timing~\cite{Khmelnitsky:2013lxt} and astrometric~\cite{Mishra-Sharma:2020ynk} observables.

In this appendix, we calculate the power spectrum $P_\delta(k)$ of \emph{unbound} scalar DM fluctuations, along the lines of Ref.~\cite{Mishra-Sharma:2020ynk}. Consider a minimally coupled free scalar field $\phi(x) = \int \dd^3k \, (2\pi)^{-3} (2k^0)^{-1/2}(a_{\vect{k}} e^{-i k \cdot x} + a_{\vect{k}}^\dagger e^{+i k \cdot x})$ with mass $m$ and energy density $\rho(x) = [\dot{\phi}(x)^2 + (\nabla \phi(x))^2 + m^2 \phi(x)^2]/2$. Here, $x = (t,\vect{x})$ and $k = (k^0, \vect{k})$ are four-vectors, with $k^0 = \sqrt{m^2 + \vect{k}^2}$. DM can be taken to be in a mixed state with number density operator expectation value $\langle a_{\vect{k}}^\dagger a_{\vect{k}'} \rangle = n_0 f(\vect{k}) (2\pi)^3 \delta^3(\vect{k} - \vect{k}')$, where $n_0$ is the number density of particles and the momentum distribution is normalized to unity $\int \dd^3 k \, (2\pi)^{-3} f(\vect{k}) = 1$. If the latter has support only over nonrelativistic wavenumbers $|\vect{k}| \ll m$, as appropriate for DM, the mean energy density is simply $\langle \rho(x) \rangle \simeq m n_0$.

Reference~\cite[Eq.~13]{Mishra-Sharma:2020ynk} gives $\langle \rho(x) \rho(x') \rangle$ under these assumptions. At equal times and in the nonrelativistic limit, the fractional-density correlation for $\delta(x) \equiv \rho(x)/\langle \rho(x) \rangle -1$ is
\begin{align}
    &\langle \delta(\vect{x}) \delta(\vect{x}') \rangle = \frac{\langle \rho(\vect{x}) \rho(\vect{x}') \rangle}{\langle \rho \rangle^2} - 1 \\
    &\simeq  \int \frac{\dd^3 k}{(2\pi)^3}\frac{\dd^3 k'}{(2\pi)^3} f(\vect{k}) f(\vect{k}') \cos\left[(\vect{k}-\vect{k}') \cdot (\vect{x}-\vect{x}') \right]. \nonumber
\end{align}
The momentum distribution of virialized scalar DM can be roughly approximated by a Maxwell--Boltzmann distribution with mean velocity $\bar{\vect{v}}$ and velocity dispersion $\sigma_v$:
\begin{align}
    f(\vect{k}) = \left(\frac{2 \pi}{m^2 \sigma_v^2}\right)^{3/2} \exp \left\lbrace -\frac{(\vect{k} - m \bar{\vect{v}})^2}{2 m^2 \sigma_v^2} \right\rbrace. \label{eq:fk}
\end{align}
The power spectrum $P_\delta$ can be found by comparison to the two-point function $\langle \tilde{\delta}(\vect{k}) \tilde{\delta}(\vect{k}')^* \rangle = P_\delta(\vect{k}) (2\pi)^3 \delta^3(\vect{k}-\vect{k}')$ of the Fourier transform $\tilde{\delta}(\vect{k}) \equiv \int \dd^3 x \, e^{-i\vect{k}\cdot \vect{x}} \delta(\vect{x})$. After a change of variables and an elementary Gaussian integral, the power spectrum for the $f(\vect{k})$ ansatz from Eq.~\ref{eq:fk} is:
\begin{align}
    P_\delta(\vect{k}) \simeq \frac{\pi^{3/2}}{m^3 \sigma_v^3} \exp\left\lbrace -\frac{\vect{k}^2}{4 m^2 \sigma_v^2} \right\rbrace,
\end{align}
independent of the direction of $\vect{k}$. This spectrum is white for $k \ll k_{\mathrm{cut}}$ and is exponentially cut off at $k_{\mathrm{cut}} = 2m\sigma_v$, giving order-unity fractional density variations near the scale $1/k_{\mathrm{cut}}$. The corresponding dimensionless power spectrum $\Delta^2_\delta(k) \equiv k^3 P_\delta(k)/2\pi^2 = (4/\sqrt{\pi})\,x^3 e^{-x^2}$, with $x \equiv k/2m\sigma_v$, peaks at $x = \sqrt{3/2}$, i.e.~$k = \sqrt{6}\,m\sigma_v \approx 1.2\,k_{\mathrm{cut}}$, where it reaches
\begin{align}
    \max_k \Delta^2_\delta(k) = \frac{4}{\sqrt{\pi}}\left(\frac{3}{2}\right)^{3/2} e^{-3/2} \approx 0.93,
\end{align}
confirming that the interference fringes induce at most order-unity density contrast per mode. This maximal power is independent of the DM mass $m$; only the scale $1/k_{\mathrm{cut}} = 1/2m\sigma_v$ at which it is attained depends on $m$. Comparing to the per-mode sensitivity of Fig.~\ref{fig:matter_power}, which lies orders of magnitude above unity, this irreducible wave-DM contribution lies below the sensitivity of the astrometric technique proposed here.

\bibliography{q-lensing}

\end{document}